\documentclass[12pt]{article}

\usepackage{latexsym}
\usepackage{amssymb,amsfonts,amsmath}
\usepackage{graphicx} 
\usepackage{indentfirst}
\usepackage{bbm}
\usepackage{amssymb}
\usepackage{verbatim}
\usepackage{amsmath, amsthm,amssymb}
\usepackage{mathrsfs}
\usepackage{hyperref}
\usepackage{amsfonts}
\usepackage{dsfont}
\usepackage{cite}
\usepackage{xcolor}
\usepackage{slashed}
\parskip=\medskipamount

\usepackage[mathscr]{euscript}

\newcommand{\de}{{\nabla}}

\numberwithin{equation}{section}

\newcommand {\cB}{{\cal B}}
\newcommand {\cC}{{\cal C}}
\newcommand {\cD}{{\cal D}}

\newcommand {\cF}{{\cal F}}
\newcommand {\cG}{{\cal G}}
\newcommand {\cH}{{\cal H}}

\newcommand {\cK}{{\cal K}}
\newcommand {\cL}{{\cal L}}

\newcommand {\cN}{{\cal N}}
\newcommand {\cO}{{\cal O}}

\newcommand {\cR}{{\cal R}}

\newcommand {\cU}{{\cal U}}
\newcommand {\cV}{{\cal V}}

\newcommand{\scT}{{\mathscr{T}}}
\newcommand{\bbD}{\mathbb D}

\newcommand{\RM}{R(M)}
\newcommand{\RD}{R(\mathbb D)}

\newcommand{\RS}{R(S)}
\newcommand{\RK}{R(K)}
\def\a{\alpha}
\def\b{\beta}

\def\d{\delta}
\def\e{\epsilon}
\def\f{\phi}
\def\g{\gamma}
\def\G{\Gamma}

\def\l{\lambda}

\def\o{\omega}

\def\q{\theta}
\def\r{\rho}
\def\s{\sigma}
\def\t{\tau}

\def\D{\Delta}
\def\F{\Phi}

\def\L{\Lambda}
\def\O{\Omega}

\def\S{\Sigma}

\def\ri{{\rm i}}
\def\re{{\rm e}}

\newcommand{\rd}{\mathrm d}
\newcommand{\hm}{{{m}}}
\newcommand{\hn}{{{n}}}
\newcommand{\hp}{{{p}}}
\newcommand{\hq}{{{q}}}
\newcommand{\ha}{{{a}}}
\newcommand{\hb}{{{b}}}
\newcommand{\hc}{{{c}}}
\newcommand{\hd}{{{d}}}
\newcommand{\he}{{{e}}}
\newcommand{\hM}{{{M}}}
\newcommand{\hN}{{{N}}}
\newcommand{\hA}{{{A}}}
\newcommand{\hB}{{{B}}}
\newcommand{\hC}{{{C}}}
\newcommand{\hal}{{{\a}}}
\newcommand{\hbe}{{{\b}}}
\newcommand{\hga}{{{\g}}}
\newcommand{\hde}{{{\d}}}
\newcommand{\hrh}{{{\rho}}}

\newcommand{\ve}{\varepsilon}

\newcommand{\pa}{\partial}
\newcommand{\hf}{\frac12}

\newcommand{\vf}{\varphi}

\newcommand{\be}{\begin{equation}}
\newcommand{\ee}{\end{equation}}
\newcommand{\bea}{\begin{eqnarray}}
\newcommand{\eea}{\end{eqnarray}}
\newcommand{\non}{\nonumber}
\newcommand{\ba}{\begin{array}}
\newcommand{\ea}{\end{array}}

\newcommand{\bsubeq}{\begin{subequations}}
\newcommand{\esubeq}{\end{subequations}}

\newcommand{\gD}{{\mathbb D}}

\def\double #1{#1{\hbox{\kern-2pt $#1$}}}

\newcommand{\eps}{\varepsilon}

\newcommand{\eol}{\notag \\}

\newcommand{\loco}{\vert}
\newcommand{\doubar}{{{\loco}\!{\loco}}}

\expandafter\ifx\csname url\endcsname\relax
  \def\url#1{\texttt{#1}}\fi
\expandafter\ifx\csname urlprefix\endcsname\relax\fi
\providecommand{\eprint}[2][]{\url{#2}}
\begin{document}
%%%%%%%%%%%%%%%%
%%%%%%%%%%%%%%%%

\begin{titlepage}
\begin{flushright}
September, 2022
\end{flushright}
\vspace{5mm}

\begin{center}
{\Large \bf 
Hyper-Dilaton Weyl Multiplets\\ 
of $5{\rm D}$ and $6{\rm D}$ Minimal Conformal Supergravity 
%in ${\rm D}=5,6$ 
}
\end{center}

\begin{center}

{\bf
Jessica Hutomo,
Saurish Khandelwal,\\
Gabriele Tartaglino-Mazzucchelli, and Jesse Woods
} \\
\vspace{5mm}

\footnotesize{
{\it 
School of Mathematics and Physics, University of Queensland,
\\
 St Lucia, Brisbane, Queensland 4072, Australia}
}
\vspace{2mm}
~\\
\texttt{j.hutomo@uq.edu.au; s.khandelwal@uq.edu.au;\\ g.tartaglino-mazzucchelli@uq.edu.au;
jesse.woods@uq.net.au}\\
\vspace{2mm}

\end{center}

\begin{abstract}
\baselineskip=14pt

By extending the recent analysis of arXiv:2203.12203 for
${\mathcal{N}}=2$ conformal supergravity in four dimensions, we define new hyper-dilaton Weyl multiplets for five-dimensional $\cN=1$, and six-dimensional $\cN=(1,0)$ conformal supergravities. These are constructed by coupling the five- and six-dimensional standard Weyl multiplets to on-shell hypermultiplets and reinterpreting the systems as new multiplets of conformal supergravity. In the five-dimensional case, we also construct a new hyper-dilaton Poincar\'e supergravity by coupling to an off-shell vector multiplet compensator. As in four dimensions, a $BF$-coupling induces a non-trivial scalar potential for the five-dimensional dilaton that admits AdS$_5$ vacua.
\end{abstract}
\vspace{5mm}

\vfill
\end{titlepage}

%%%%%%%%%%%%

\newpage
\renewcommand{\thefootnote}{\arabic{footnote}}
\setcounter{footnote}{0}

\tableofcontents{}
\vspace{1cm}
\bigskip\hrule

\allowdisplaybreaks
%%%%%%%%%%%%%%%%%%%%%%%%%%%%%%%%%
\section{Introduction}

A key ingredient to efficiently engineer off-shell supergravity-matter couplings is the fact that  Poincar\'e gravity can be formulated as conformal gravity coupled to a compensating scalar field \cite{Deser,Zumino}.
This approach plays an equally important role both in the superconformal tensor calculus 
and in superspace supergravity 
formalisms --- see
\cite{freedman,Lauria:2020rhc} and\cite{SUPERSPACE,Buchbinder-Kuzenko} for reviews and  references. 
In the supersymmetric case, conformal gravity is turned into an off-shell representation of the local superconformal algebra
containing the vielbein as one of its independent fields. Such multiplet is referred to as the \emph{Weyl} multiplet of conformal supergravity.
The scalar compensator is also lifted to an off-shell locally superconformal multiplet. Depending on the amount of supersymmetry, due to the existence of several possible choices of compensating multiplets, it is possible to obtain several different off-shell Poincar\'e supergravity theories. Moreover, the fact that the Weyl multiplets themselves are in general not unique adds to the richness of the off-shell representations.

The first instance where variant representations of the Weyl multiplets were presented is six-dimensional (6D) minimal $\cN=(1,0)$ supergravity \cite{Bergshoeff:1985mz} --- see also \cite{Sokatchev:1988aa,CVanP,Bergshoeff:2012ax,OzkanThesis,Linch:2012zh,BKNT16,BNT-M17,Butter:2018wss} for further references on 6D conformal supergravity. In this case, it was noted that the so-called \emph{standard Weyl} multiplet could be turned into a \emph{dilaton Weyl} multiplet by reinterpreting the system described by an on-shell tensor multiplet in a standard Weyl multiplet background as a new conformal supergravity multiplet. Such a \emph{tensor-dilaton} Weyl multiplet plays an important role since, once coupled to a second off-shell conformal compensating multiplet, is the one used to construct the simplest versions of two-derivatives Poincar\'e supergravity theories. Extending the idea of \cite{Bergshoeff:1985mz}, dilaton Weyl multiplets have been discovered also for five-dimensional (5D) $\cN=1$ \cite{Bergshoeff:2001hc} and, more recently, for four-dimensional (4D) $\cN=2$ conformal supergravities in \cite{Butter:2017pbp} and \cite{Gold:2022bdk}.

In 5D $\cN=1$ supergravity the known dilaton Weyl multiplet is constructed by coupling an on-shell vector multiplet to the standard Weyl multiplet \cite{Bergshoeff:2001hc}. See  \cite{Kugo:2000hn,Fujita:2001kv,Kugo:2002vc,Bergshoeff:2002qk,Bergshoeff:2004kh,KT-M08,BKNT-M14} for more discussions of 5D conformal supergravity and its matter couplings.
The 4D $\cN=2$ analogue of this type of \emph{vector-dilaton} Weyl multiplet was constructed in \cite{Butter:2017pbp}. It is natural to expect that different on-shell multiplets containing a scalar field playing the role of a dilaton could be used to engineer other multiplets of conformal supergravity. In fact, for the 4D $\cN=2$ case, in \cite{Gold:2022bdk} a so-called \emph{hyper-dilaton} Weyl multiplet was constructed by using an on-shell hypermultiplet. The scope of this paper is to present the extension of the analysis of \cite{Gold:2022bdk} to 5D $\cN=1$ and 6D $\cN=(1,0)$ supergravities.

Besides a mathematically oriented interest in classifying variant Weyl multiplets, it is worth to explore new options to define off-shell Poincar\'e supergravities with an eye on their broad range of applications.
For example, in our opinion, for theories with eight supercharges it remains an open problem to have a simple, though general, off-shell approach for gauged supergravities with no physical matter hypermultiplets. In the presence of physical charged hypermultiplets (with no central charge) one will need to use representations containing an infinite number of auxiliary/matter fields \cite{Galperin:1984av,Karlhede:1984vr,Lindstrom:1987ks,Lindstrom:1989ne,Lindstrom:2009afn,Galperin:2001seg,Bagger:1987rc,K06,KT-M_5D2,KT-M_5D3,KT-M08,Kuzenko:2008ep,Kuzenko:2009zu,BKNT-M14,Sokatchev:1988aa,Linch:2012zh,Butter:2014gha,Butter:2014xua,Butter:2015nza},  but with only physical vector multiplets it might be beneficial to use simpler approaches, if possible. The exploration of options to fill this gap was one of the main motivations for the recent construction in \cite{Gold:2022bdk}.

Interesting applications of off-shell approaches to (gauged) supergravity includes the construction of locally supersymmetric higher-derivative invariants
\cite{BSS1,LopesCardoso:1998tkj,Mohaupt:2000mj,Hanaki:2006pj,CVanP,Bergshoeff:2012ax,Butter:2013rba,Butter:2013lta,Kuzenko:2013vha,OP131,OP132,OzkanThesis,BKNT-M14,Kuzenko:2015jxa,BKNT16,Butter:2016mtk,BNT-M17,NOPT-M17,Butter:2018wss,Butter:2019edc,Hegde:2019ioy,Mishra:2020jlc}.
This topic has recently attracted a renewed attention due to advances in the study of black-hole entropy and next to leading order AdS/CFT calculations
--- see the very recent works
\cite{Bobev:2020egg,Bobev:2021oku,Bobev:2021qxx,Hristov:2022lcw,Bobev:2022bjm,Cassani:2022lrk} 
and references therein.
Vector-dilaton Weyl multiplets have been a main ingredient to construct off-shell higher-derivative supergravities in five and six dimensions, see
\cite{BSS1,CVanP,Bergshoeff:2012ax,OP131,OP132,OzkanThesis,BKNT-M14,NOPT-M17,Butter:2018wss,Mishra:2020jlc}.
We hope new hyper-dilaton Weyl multiplets might play an interesting role to extend some of these results to gauged supergravity.
The construction of alternative Weyl multiplets could also play an interesting role to develop alternative approaches to localisation of quantum field theories on curved space-times  --- see \cite{Pestun:2016zxk} for a recent extensive review.  In this context, off-shell supersymmetry  has been a central ingredient for localisation and new Weyl multiplets could offer alternative starting points.

This paper is organised as follows. In section \ref{5DMultipletSuperspace} we use the conformal superspace approach to 5D $\cN=1$ supergravity \cite{BKNT-M14} to review the locally superconformal multiplets used in our analysis. Specifically, we introduce the 5D standard Weyl multiplet and its geometric superspace construction, the on-shell hypermultiplet, the linear  (or $\cO(2)$) multiplet, and the Abelian vector multiplet. Section \ref{SWM} is devoted to a description of the standard Weyl multiplet in terms of component fields in the notations of our paper. Section \ref{hyper+HDWM} describes the construction of the new 32+32 5D $\cN=1$ hyper-dilaton Weyl multiplet. In section \ref{section-3}, we couple the hyper-dilaton Weyl multiplet to a vector multiplet compensator to recover a new 40+40 Poincar\'e supergravity. This can be thought of as a 5D analogue of the 4D off-shell $\cN=2$ supergravity introduced by M\"uller in 1986 \cite{Muller_hyper:1986ts} and redefined by using superconformal techniques in \cite{Gold:2022bdk}. A distinctive feature of the dilaton Poincar\'e supergravity is the fact that the off-shell multiplet is irreducible, while the two-derivative supergravity action leads to the minimal on-shell Poincar\'e supergravity multiplet coupled to an extra physical matter multiplet containing the dilaton. Interestingly, as in the 4D analysis of \cite{Gold:2022bdk}, in subsection \ref{chiral analysis} we show how it is possible to engineer a non-trivial scalar potential for the dilaton and obtain an AdS$_5$ vacua in a framework different than standard gauged supergravity.
In sections \ref{6DMultipletSuperspace}, \ref{6DCSWM}, and \ref{6Dhyper+HDWM}  we closely repeat the 5D analysis of sections \ref{5DMultipletSuperspace}, \ref{SWM}, and \ref{hyper+HDWM} in the case of 6D $\cN=(1,0)$ conformal supergravity. The paper also contains two technical appendices, A and B, where we collect conformal superspace identities from \cite{BKNT-M14} and \cite{BKNT16,BNT-M17} used in our paper.

\section{Superconformal multiplets in 5D $\cN=1$ superspace}
\label{5DMultipletSuperspace}

This section reviews the salient details of several superconformal multiplets pertinent to this work. We first describe the standard Weyl multiplet of conformal supergravity in 5D $\cN=1$ conformal superspace before moving on to the discussion of various matter multiplets: the on-shell hypermultiplet, together with the off-shell linear and Abelian vector multiplets. Here we make use of the formulation and results of \cite{BKNT-M14}. 
We also refer the reader to the following list of papers for other work on flat and curved superspace and off-shell multiplets in five dimensions \cite{KL,KT-M5D1,Howe5Dsugra,HL,KT-M_5D2,KT-M_5D3,KT-M08,K06}.

\subsection{The standard Weyl multiplet}
\label{SWM-superspace}
In five dimensions, the standard Weyl multiplet of $\cN=1$ conformal supergravity \cite{Bergshoeff:2001hc} contains $32+32$ physical components and is associated with the gauging of the superconformal algebra $\rm F^2(4)$. Associated respectively with local translations, $Q$-supersymmetry, ${\rm SU(2)}_R$ symmetry, and dilatations are the vielbein $e_\hm{}^\ha$,
the gravitino $\psi_\hm{}_\hal^i$,
the ${\rm SU(2)}_R$ gauge field $\phi_\hm{}^{ij}$, and a dilatation gauge field $b_\hm$. There are three composite connections which are associated with the remaining gauge symmetries: these are
the spin connection  $\omega_\hm{}^{\ha\hb}$, the $S$-supersymmetry connection
$\phi_\hm{}_\hal^i$, and the special conformal connection
$\mathfrak{f}_\hm{}^\ha$ that are algebraically determined in terms of the other fields by imposing constraints on some of the
curvature tensors. To achieve an off-shell representation of the 5D $\cN=1$ local superconformal algebra, three covariant auxiliary fields are introduced: a real antisymmetric
tensor $w_{\ha\hb}$, a fermion $\chi_\hal^i$, and a real auxiliary scalar $D$. In this subsection, we show how to embed this in conformal superspace. The component structure of the multiplet is given in section \ref{SWM}. 

The 5D $\cN=1$ conformal superspace is parametrised by
local bosonic $(x^{\hm})$ and fermionic $(\theta_i)$ coordinates 
$z^{\hM} = (x^{\hm},\q^{{\mu}}_i)$, 
where $\hm = 0, 1, 2,3, 4$, ${\mu} = 1, \cdots , 4$ and $i = 1, 2$.
By gauging the
full 5D $\cN=1$ superconformal algebra, we introduce
covariant derivatives $ {\nabla}_{\hA} = (\nabla_{\ha} , \nabla_{\hal}^i)$ which take the form
\bsubeq
\bea\label{eq:covD}
\de_{\hA} 
= E_{\hA} - \o_{\hA}{}^{\underline{b}}X_{\underline{b}} 
&=& E_{\hA} - \hf \O_{\hA}{}^{\ha \hb} M_{\ha \hb} - \Phi_{\hA}{}^{ij} J_{ij} - B_{\hA} \mathbb{D} - \mathfrak{F}_{\hA}{}^{\hB} K_{\hB}~,
\\ &=& E_{\hA} - \hf \O_{\hA}{}^{\ha \hb} M_{\ha \hb} - \Phi_{\hA}{}^{ij} J_{ij} - B_{\hA} \mathbb{D} - \mathfrak{F}_{\hA}{}^{\a i} S_{\a i} - \mathfrak{F}_{\hA}{}^{a} K_{a} ~.
\eea
\esubeq
Here $E_{\hA} = E_{\hA}{}^{\hM} \partial_{\hM}$ is the inverse super-vielbein,
$M_{\ha \hb}$ are the Lorentz generators, $J_{ij}$ are generators of the
${\rm SU(2)}_R$ $R$-symmetry group,
$\mathbb D$ is the dilatation generator, and $K_{\hA} = (K_{\ha}, S_{\hal i})$ are the special superconformal
generators.
The super-vielbein one-form is $E^{\hA} =\rd z^{\hM} E_{\hM}{}^{\hA}$ with $E_{\hM}{}^{\hA} E_{\hA}{}^{\hN} =\d_{\hM}^{\hN}$,
 $E_{\hA}{}^{\hM} E_{\hM}{}^{\hB}=\d_{\hA}^{\hB}$.
Associated with each structure group generator $X_{\underline{a}} = (M_{\ha\hb},J_{ij},\bbD, S_{\hal i}, K_\ha)$ is the following connection super one-form 
$\omega^{\underline{a}} = (\O^{\ha\hb},\F^{ij},B,\mathfrak{F}^{\hal i},\mathfrak{F}^{\ha})= \rd z^\hM \omega_\hM{}^{\underline{a}} = E^{\hA} \o_{\hA}{}^{\underline{a}}$.

To describe the standard Weyl multiplet in conformal superspace,
one constrains the algebra of covariant derivatives
\begin{align}
[ \nabla_\hA , \nabla_\hB \}
	&= -\mathscr{T}_{\hA\hB}{}^\hC \nabla_\hC
	- \frac{1}{2} {\mathscr{R}(M)}_{\hA\hB}{}^{\hc\hd} M_{\hc\hd}
	- {\mathscr{R}(J)}_{\hA\hB}{}^{kl} J_{kl}
	\non \\ & \quad
	- {\mathscr{R}(\mathbb{D})}_{\hA\hB} \mathbb D
	- {\mathscr{R}(S)}_{\hA\hB}{}^{\hga k} S_{\hga k}
	- {\mathscr{R}(K)}_{\hA\hB}{}^\hc K_\hc~,
	\label{nablanabla}
\end{align}
to  be   completely determined   in terms of the symmetric super-Weyl tensor superfield $W_{\hal \hbe}$, which is a superconformal primary with conformal dimension 1
\be
W_{\hal \hbe} = W_{\hbe \hal} \ , \quad
K_{\hA} W_{\hal \hbe} = 0 \ ,
 \quad \mathbb D W_{\hal \hbe} = W_{\hal \hbe} \ ,
\ee
and obeys the Bianchi identity
\be 
\nabla_\hga^k W_{\hal \hbe} = \nabla_{(\hal}^k W_{\hbe \hga )} + \frac{2}{5} \eps_{\hga (\hal} \nabla^{\hde k} W_{\hbe ) \hde} \ . \label{WBI}
\ee
The relation $W_{\hal \hbe} = 1/2 ({\S}^{\ha \hb})_{\hal \hbe}W_{\ha \hb}$ means that the super-Weyl tensor is equivalent to an antisymmetric rank-2 tensor superfield $W_{\ha \hb} = -W_{\hb \ha}$.
In  \eqref{nablanabla}
$ \mathscr{T}_{\hA \hB}{}^C$ is the torsion, and $ \mathscr{R}(M)_{\hA \hB}{}^{\hc \hd}$,
$ \mathscr{R}(J)_{\hA \hB}{}^{kl}$, $ \mathscr{R}(\mathbb D)_{\hA \hB}$, $ \mathscr{R}(S)_{\hA \hB}{}^{ {{\g}}k}$, and $ \mathscr{R}(K)_{\hA \hB}{}^{\hc}$
are the curvatures associated with Lorentz, ${\rm SU(2)}_R$,
dilatation, $S$-supersymmetry, and special conformal
boosts, respectively. Their expressions in terms of $W_{\hal\hbe}$ and its descendant superfields of
dimension 3/2
\bsubeq \label{descendantsW-5d}
\begin{gather}
W_{\hal \hbe \hga}{}^k := \nabla_{(\hal}^k W_{\hbe \hga )} \ , \quad X_\hal^i := \frac{2}{5} \nabla^{\hbe i} W_{\hbe\hal} \ ,
\end{gather}
and of dimension 2
\bea
W_{\hal \hbe \hga \hde} := \nabla_{(\hal}^k W_{\hbe \hga \hde) k} \ , \quad 
X_{\hal \hbe}{}^{ij} := \nabla_{(\hal}^{(i} X_{\hbe)}^{j)} 
\ , \quad
Y := \ri \nabla^{\hga k} X_{\hga k} \ .
\eea
\esubeq
are collected in appendix \ref{conformal-identities}.
%%%
Note that in this paper we will make use of conformal superspace
with a redefined vector covariant derivative. This corresponds
to choosing the ``traceless'' frame conventional constraints
employed for the first time
in subsection 4.4 and appendix C of \cite{BKNT-M14}.
The component and superspace  structure corresponding to this choice are summarised in section \ref{SWM} and appendix \ref{conformal-identities}.

The superfields
$W_{\hal \hbe {\g}}{}^{k}$,
$X_{\hal}{}^i$,
$W_{\hal \hbe {\g} {\d}}$,
$X_{\hal \hbe}{}^{ij}$, and $Y$
 satisfy nontrivial Bianchi identities given in \cite{BKNT-M14}. Thus, only these five superfields and their vector derivatives appear upon taking successive spinor derivatives on $W_{\hal \hbe}$. Eq.~\eqref{S-on-X_Y-a} gives the action of the $S$-generators on these independent descendants that prove to be all annihilated by $K_{\ha}$.

The conformal supergravity gauge group $\cG$ is generated by
{\it covariant general coordinate transformations},
$\delta_{\rm cgct}$, associated with a local superdiffeomorphism parameter $\xi^{\hA}$ and
{\it standard superconformal transformations},
$\delta_{\cH}$, associated with the following local superfield parameters:
the dilatation $\s$, Lorentz $\L^{\ha \hb}=-\L^{\hb \ha}$,  ${\rm SU(2)}_R$ $\L^{ij}=\L^{ji}$,
 and special conformal transformations $\L^{\hA}=(\eta^{\hal i},\L^{\ha}_{K})$.
The covariant derivatives transform as
\bea
\d_\cG \nabla_{\hA} &=& [\cK , \nabla_{\hA}] \ ,
\label{TransCD}
\eea
where $\cK$ denotes the first-order differential operator
\bea
\cK = \xi^{\hC}  {\nabla}_{\hC} + \hf  {\L}^{\ha \hb} M_{\ha \hb} +  {\L}^{ij} J_{ij} +  \s \mathbb D +  {\L}^{\hA} K_{\hA} ~.
\eea
A covariant (or tensor) superfield $U$ transforms as
\be
\d_{\cG} U =
(\d_{\rm cgct}
+\d_{\cH}) U =
 \cK U
 ~. \label{trans-primary}
\ee
The superfield $U$ is said to
be \emph{superconformal primary} of dimension $\D$ if $K_{\hA} U = 0$ (it suffices to require that $S_{\a i} U = 0$) and $\mathbb D U = \D U$.

\subsection{The on-shell hypermultiplet}
\label{HDWM-superspace}
The on-shell realisation for the hypermultiplet contains $4+4$ degrees of freedom, exactly as in the four-dimensional case
\cite{Fayet:1975yi,FS2}. In conformal superspace, it is described by a Lorentz scalar superfield $q^{i\underline{i}}$ subject to the constraint
\bea
\de_{\hal}^{(i} q^{j) \underline{j}} = 0~, 
\label{onshell-sspace}
\eea
which is equivalent to
\bea
\de_{\hal}^{i}q^{k \underline{k}} = -\hf \ve^{ik} \rho_{\hal}^{\underline{k}}~, \qquad \rho_{\hal}^{\underline{k}}:= \de_{\hal}^{j}q_{j}{}^{\underline{k}}~. \label{onshell-sspace-1}
\eea
Here, the index $\underline{i}=\underline{1},\underline{2}$ denotes an SU(2) flavour 
index. The superfield $q^{i \underline{i}}$ is a Lorentz scalar and superconformal primary,
\bea
M_{\ha \hb} q^{i \underline{i}} =0~, \qquad K_A q^{i \underline{i}} = 0~, \qquad J^{jk} q^{i \underline{i}} = \ve^{i(j} q^{k)\underline{i}}~. \label{hyper-primary}
\eea
Eqs.~\eqref{onshell-sspace}, \eqref{hyper-primary}, and the relation \eqref{alg-S-spinor} tell us that $ \mathbb{D} q^{i \underline{i}} = \frac{3}{2} q^{i \underline{i}}$.

The independent descendants of $q^{i \underline{i}}$ are obtained by acting on it with spinor derivatives. We obtain several implications of the (anti-)commutation relations \eqref{new-frame_spinor-spinor}, \eqref{new-vectspinor}, along with the constraints \eqref{onshell-sspace-1}, and \eqref{hyper-primary}: 
\bsubeq \label{twospinors-on-q}
\bea
\de_{\hal}^i \de_{\hbe}^j q_{j}{}^{\underline{i}} &=& \de_{\hal}^i \rho_{\hbe}^{\underline{i}} = -4 \ri \de_{\hal \hbe} q^{i \underline{i}}~, \\
\de^{\hal i} \rho_{\hal}^{\underline{i}} &=& 0~,
\eea
\esubeq
with $\de_{\hal \hbe}:= (\G^{\ha})_{\hal \hbe} \de_{\ha}$. 
Next, we shall consider
\bea
\{ \de_{\hal}^k, \de_{\hbe k} \} \rho^{\hbe \underline{i}}
&=&  4 \ri \de_{\hal \hbe}\rho^{\hbe \underline{i}} -2 \ri \,W_{\hal}{}^{\hbe} \rho_{\hbe}^{\underline{i}} +18 \ri \,X_{\hal}^{k}\, q_{k}{}^{\underline{i}}~, \label{dd-rho-right}
\eea
where we have made use of \eqref{new-frame_spinor-spinor} and the $S$-supersymmetry transformation
\bea
S_{\hal i} \rho_{\hbe}^{\underline{i}} = 12 \ve_{\hal \hbe} \,q_{i}{}^{\underline{i}} \qquad \Longrightarrow \qquad K_{\ha}\, \rho_{\hbe}^{\underline{i}} = 0~. \label{S-rho}
\eea
On the other hand, by virtue of \eqref{twospinors-on-q}, we also have that
\bea
\{ \de_{\hal}^k, \de_{\hbe k} \} \rho^{\hbe \underline{i}} &=& 4 \ri \de^{\hbe}_k \de_{\hal \hbe} q^{k \underline{i}}
= 4 \ri \, [\de^{\hbe}_k, \de_{\hal \hbe}] q^{k \underline{i}} - 4 \ri \de_{\hal \hbe} \rho^{\hbe \underline{i}}~. \label{dd-rho-left}
\eea
Applying the commutation relation \eqref{new-vectspinor}, we can then equate \eqref{dd-rho-right} and \eqref{dd-rho-left} to obtain
\bea
(\nabla_{\ha}
\rho^{\underline{i}}\,
\G^{\ha})^{\hal}
&=&
-\frac{3}{4}
({\rho}^{\underline{i}}\,
\S^{\hb \hc})^{\hal} W_{\hb \hc}
-\frac{3}{2}X^{\hal k}q_{k}{}^{\underline{i}}~. \label{vect-rho-sspace}
\eea
We can then hit both sides of \eqref{vect-rho-sspace} with $\de_{\hal}^i$ and make use of \eqref{twospinors-on-q}, \eqref{new-vectspinor}, and the identity \eqref{Wdervs-a}. This results in the equation 
\bea
\Box q^{i \underline{i}} 
&=&
\frac{3 \ri}{16}X^{i}\rho^{\underline{i}} 
+ \frac{3}{64} \big( Y-  W^{{a} {b}} W_{{a} {b}} \big)\,q^{i\underline{i}}~, \qquad \Box := \nabla^{\ha}\nabla_{\ha}~. \label{boxq-sspace}
\eea
Both \eqref{vect-rho-sspace} and \eqref{boxq-sspace} describe on-shell conditions for the hypermultiplet's fields when $W_{ab}=0$. However, as we will discuss in more detail later, in a non-trivial curved background these equations can be reinterpreted as conditions linking $q^{i\underline{i}}$ and $\rho_\a^{\underline{i}}$ with fields of the standard Weyl multiplet.

By restricting to $\xi^a\equiv0$, the local superconformal $\d= \d_{Q} + \d_{\cH}$ transformations of the covariant superfields $q^{i \underline{i}}$ and $\r_{\a}^{\underline{i}}$ can be derived using the relations \eqref{onshell-sspace-1}, \eqref{twospinors-on-q}, and \eqref{S-rho}. This leads to
\bsubeq \label{d-hyper-sspace}
\bea
\delta q^{i\underline{i}} 
&=& 
\frac{1}{2}\xi^{ i}\rho^{\underline{i}}
+\L^{i}{}_{k}q^{k\underline{i}} 
+ \frac{3}{2}\s q^{i\underline{i}}
~,
\\
\delta \rho_{\hal}^{\underline{i}} 
&=&
-4\ri(\G^{\ha} {\xi}_i)_{\hal} 
\nabla_{\ha} q^{i\underline{i}}
 +\hf \L_{\ha \hb} (\S^{\ha \hb} \rho^{\underline{i}})_{\hal}    
+2 \s \rho_{\hal}^{\underline{i}} 
-12 \eta_{\hal}^i q_i{}^{\underline{i}}~.
\eea
\esubeq
As we will describe later, these will lead to the analogue transformations of the component fields in the hypermultiplet.

\subsection{The $\cO(2)$ multiplet}

The linear multiplet \cite{FS2,deWit:1980gt,deWit:1980lyi,deWit:1983xhu,N=2tensor,Siegel:1978yi,Siegel80,SSW,deWit:1982na,KLR,LR3}, 
 or $\cO(2)$ multiplet, can be described in 5D $\cN=1$ conformal superspace \cite{BKNT-M14} in terms of the superfield $G^{ij}= G^{ji}$, with $(G^{ij})^{*}= \ve_{ik} \ve_{jl} G^{kl}$ and satisfies the defining constraint
\bea
\de_{\a}^{(i} G^{jk)} = 0~.
\eea
Here $G^{ij}$ is a superconformal primary dimension-3 superfield,
\bea
K_A G^{ij}=0~, \qquad \mathbb{D} G^{ij} = 3 G^{ij}~.
\eea
To elaborate on the component structure of the superfield $G^{ij}$, we list the following useful identities:
\bsubeq \label{O2spinorderivs}
\begin{align}
\nabla_\hal^i G^{jk} &= 2 \eps^{i(j} \varphi_\hal^{k)} \ , \\
\nabla_\hal^i \varphi_\hbe^j &= - \frac{\ri}{2} \eps^{ij} \eps_{\hal\hbe} F + \frac{\ri}{2} \eps^{ij} \cH_{\hal\hbe} + \ri \nabla_{\hal\hbe} G^{ij} \ , \\
\nabla_\hal^i F &= - 2 \nabla_\hal{}^\hbe \varphi_\hbe^i - 3 W_{\hal\hbe} \varphi^{\hbe i} - \frac{3}{2} X_{\hal j} G^{ij} \ , \\
\nabla_\hal^i \cH_{\ha} &= 4 (\S_{\ha\hb})_\hal{}^\hbe \nabla^\hb \varphi_\hbe^i - \frac{3}{2} (\G_\ha)_\hal{}^\hbe W_{\hbe \hga} \varphi^{\hga i}
-\frac{1}{2} (\G_\ha)_\hga{}^\hbe W_{\hbe \hal} \varphi^{\hga i} \ ,
\end{align}
\esubeq
where we have defined the independent descendant superfields
\bsubeq \label{O2superfieldComps}
\begin{align}
\varphi_\hal^i &:= \frac{1}{3} \nabla_{\hal j} G^{ij} \ , \\
F &:= \frac{\ri}{12} \nabla^{\hga i} \nabla_\hga^j G_{ij} = - \frac{\ri}{4} \nabla^{\hga k} \varphi_{\hga k} \ ,\\
\cH_{abcd} &:= \frac{\ri}{12} \eps_{abcde} (\G^{e})^{\a \b} \nabla_{\a}^i \nabla_{\b}^j G_{ij} 
\equiv \eps_{\ha\hb\hc\hd\he} \cH^\he \ .
\end{align}
\esubeq
It can be checked that $\cH^{a}$ obeys the differential condition 
\be 
\nabla_\ha \cH^\ha = 0~, \qquad {\cH}^{\ha}:= -\frac{1}{4!}\ve^{\ha \hb \hc \hd \he} \cH_{\hb \hc \hd \he}~.
\ee
The descendants \eqref{O2superfieldComps} prove to be annihilated by $K_a$ and to satisfy
\bsubeq \label{O2-S-actions}
\begin{align} S_\hal^i \varphi_\hbe^j &= - 6 \eps_{\hal \hbe} G^{ij} \ , \\
S_\hal^i F &= 6 \ri \varphi_\hal^i \ , \\
S_\hal^i \cH_\hb &= - 8 \ri (\G_\hb)_\hal{}^\hbe \varphi_\hbe^i \ .
\end{align}
\esubeq
We refer the reader to \cite{BKNT-M14} for a superform description of the $\cO(2)$ multiplet. 

\subsection{The Abelian vector multiplet}
\label{vector-superspace}

In conformal superspace \cite{BKNT-M14}, an Abelian vector multiplet is described by a superfield $W$, which is superconformal primary of dimension 1,
 $K_A W=0$ and
 $\bbD W= W$. Moreover, it is real, 
$(W)^* = W$, and obeys the Bianchi identity
\bea 
\nabla_{{\a}}^{(i} \nabla_{{\b}}^{j)} W 
= \frac{1}{4} \eps_{{\a} {\b}} \nabla^{{\g} (i} \nabla_{{\g}}^{j)} 
W \ .
\label{vector-Bianchi}
\eea
 
It is useful to introduce the following descendant superfields constructed from spinor derivatives of $W$:
\bsubeq \label{components-vect}
\begin{align}
\l_\hal^i := - \ri\nabla_\hal^i W \ , \qquad
 X^{ij} := \frac{\ri}{4} \nabla^{\hal (i} \nabla_\hal^{j)} W 
= - \frac{1}{4} \nabla^{\hal (i}  \l_\hal^{j)} \ .
\end{align}
These superfields, along with
\be 
\cF_{\hal\hbe} := - \frac{\ri}{4} \nabla^k_{(\hal} \nabla_{\hbe) k}  W - W_{\a \b} W
= \frac{1}{4} \nabla_{(\hal}^k \l_{\hbe) k} 
- W_{\a \b} W
~,
\ee
\esubeq
satisfy the following identities:
\bsubeq \label{VMIdentities}
\begin{align}
\nabla_\hal^i \l_\hbe^j 
&= - 2 \eps^{ij} \big(\cF_{\hal \hbe} + W_{\hal\hbe}  W\big) 
- \eps_{\hal\hbe} X^{ij} - \eps^{ij} \nabla_{\hal\hbe} W \ , \\
\nabla_\hal^i \cF_{\hbe\hga} 
&=
 - \ri \nabla_{\hal (\hbe} \l_{\hga)}^i 
 - \ri \eps_{\hal (\hbe} \nabla_{\hga )}{}^\hde \l_\hde^i 
- \frac{3\ri}{2} W_{\hbe\hga} \l_\hal^i - W_{\hal\hbe \hga}{}^i  W \non\\
&~~~+ \frac{\ri}{2} W_{\a (\b} \l_{\g)}^i
-\frac{3 \ri}{2} \ve_{\a (\b} W_{\g)}{}^{\d} \l_{\d}^i\ , \\
\nabla_\hal^i X^{jk} &= 2 \ri \eps^{i (j} 
\Big(
\nabla_\hal{}^\hbe \l_\hbe^{k)} 
- \hf W_{\hal\hbe} \l^{\hbe k)} 
+ \frac{3\ri}{4} X_\hal^{k)} W
\Big) 
\ .
\end{align}
\esubeq
We also note the relation $\cF_{\a \b} = \hf (\S^{ab})_{\a \b} \, \cF_{ab}$. 
The $S$-supersymmetry generator acts on these descendants as
\begin{align}
S_\hal^i \l_\hbe^j &= - 2 \ri \eps_{\hal\hbe} \eps^{ij}  W \ , \qquad
S_\hal^i \cF_{\hbe\hga} = 4 \eps_{\hal (\hbe}  \l_{\hga)}^i \ , \qquad
S_\hal^i X^{jk} = - 2 \eps^{i (j}  \l_\hal^{k)} \ ,
\end{align}
while all the fields are annihilated by the $K_a$ generators.

For the construction of Poincar\'{e} supergravity models later in subsection \ref{chiral analysis}, it is important to note that given a system of $n$ Abelian vector multiplets $W^I$, with $I = 1, 2, \dots n$, we can construct the following composite $\cO(2)$ multiplet and its descendant superfields \cite{BKNT-M14}:
\bsubeq \label{O2composite-N}
\bea
G_I^{ij}
&=& C_{IJK} \Big( 2 W^J X^{ij\, K}
- \ri \l^{\a J\,(i } \l_{\a}^{j) K}\Big)~,\\
%%%%%%%%%%%%%%%%%%%%%%%
\vf_{\a\, I}^i &=& C_{IJK} \bigg(
\ri X^{ij\, J} \l_{\a j}^K 
-2 \ri \cF_{\a \b}^{J} \l^{\b i K}
- \frac{3}{2} X_{\a}^i W^J W^K
-2 \ri W^J \de_{\a \b} \l^{\b i K}\non\\
&&- \ri \de_{\a \b} W^J \l^{\b i K}
- 3 \ri W_{\a \b} W^J \l^{\b i K}
\bigg)~, \\
%%%%%%%%%%%%%%%%%%%%%%%%%%%%%%%%%%%%
F_I &=& C_{IJK} \bigg( X^{ij J}X_{ij}^{K}
- \cF^{ab J} \cF_{ab}^{K}
+ 4 W^J \de^{a} \de_{a} W^K
+ 2 (\de^a W^J) \de_{a} W^K \non\\
&&+ 2 \ri (\de_{\a}{}^{\b} \l_{\b}^{iJ}) \l^{\a K}_{i}{} -6 W^{ab} \cF_{ab}^J W^K 
-\frac{39}{8} W^{ab} W_{ab} W^J W^K
+ \frac{3}{8} Y W^J W^K \non\\
&&+ 6 X^{\a i}\l_{\a i}^J W^K
-3 \ri W_{\a \b} \l^{\a i J} \l^{\b K}_{i}
\bigg)~, \\
%%%%%%%%%%%%%%%%%%%%%%%%%%%%%%%%%%%%%%
\cH_{a I} &=& C_{IJK} \bigg(-\hf \ve_{abcde} \cF^{bc\, J} \cF^{de \,K}
+ 4 \de^{b} \big( W^J \cF_{ba}^K + \frac{3}{2} W_{ba} W^J W^K\big)\non\\
&&+ 2 \ri (\S_{ba})^{\a \b} \de^{b} (\l_{\a}^{iJ} \l_{\b i}^K)
\bigg)~,
\eea
\esubeq
where $C_{IJK} = C_{(IJK)}$ is a completely symmetric constant.
These are the superspace analogue of the composite linear multiplets constructed in \cite{Bergshoeff:2001hc}.

\section{The standard Weyl multiplet in components}
\label{SWM}

We begin by identifying the various component fields of the 5D $\cN=1$ standard Weyl multiplet \cite{Bergshoeff:2001hc} within the superspace geometry described in subsection \ref{SWM-superspace}. Let us start with the vielbein $(e_{\hm}{}^{\ha})$ and gravitino $(\psi_{\hm}{}^{i}_{ \hal})$. These appear as the coefficients of $\rd x^\hm$ of the super-vielbein $E^\hA = (E^\ha, E^\hal_i) = \rd z^\hM\, E_\hM{}^\hA$,
\bea\label{singlebar}
e_{\hm}{}^{\ha} (x):= E_{\hm}{}^{\ha}(z)|~, \qquad \psi_{\hm}{}^{i}_{ \hal}(x) := 2 E_{\hm}{}^{i}_{\hal} (z)|~,
\eea
where a single vertical line next to a superfield denotes the usual component projection to $\q = 0$, i.e. $V(z) | := V(z) |_{\q=0}$. This operation can be written in a coordinate-independent way using the so-called double-bar projection
\begin{align}\label{doublebar}
e{}^{\ha} = \rd x^{\hm} e_{\hm}{}^{\ha} = E^{\ha}\doubar~,~~~~~~
\psi_{\hal}^{i} = \rd x^{\hm} \psi_{\hm}{}^{i}_{ \hal} =  2 E_{\hal}^{i} \doubar ~,
\end{align}
where the double-bar denotes setting $\q = \rd \q = 0$.
Analogously, the remaining fundamental and composite one-forms are obtained by taking the projections of the corresponding superspace one-forms,
\begin{align}
	\phi^{ij} := \Phi{}^{ij} \doubar~, \quad
	b := B\doubar ~, \quad
\omega^{\ha \hb} := \Omega{}^{\ha \hb} \doubar ~, \quad  
\phi^{\hal i} := 2 \,\mathfrak F{}^{\hal i}\doubar~, \quad
\mathfrak{f}{}^{\ha} := \mathfrak{F}{}^{\ha}\doubar~. 
\end{align}
The covariant auxiliary matter fields are contained within the super-Weyl tensor $W_{\hal \hbe}$ and its independent descendants, 
\bsubeq \label{comps-W}
\bea
w_{\hal \hbe} &:=&  W_{\hal \hbe}\loco \ ,
\\
\chi_{\hal}^{i} &:=& \frac{3 \ri}{32}X_{\hal}^{i} \loco = \frac{3\ri}{80}\de^{\hbe i} W_{\hbe \hal}\loco~,
\\
D&:=& -\frac{3}{128}Y \loco=
-\frac{3\ri}{320}\de_{\hal}^k\de_{\hbe k}W^{\hal \hbe}\loco
~.
\eea
\esubeq
The other components of $W_{\hal \hbe}=\hf(\S^{\ha\hb})_{\hal\hbe}W_{\ha\hb}$ are given by $W_{\ha \hb}{}_{\hal}^{i} \loco=(\S_{\ha\hb})^{\hbe\hga}W_{\hbe \hga}{}_{\hal}^{i} \loco$ and by $X_{\ha \hb}{}^{ij} \loco=(\S_{\ha\hb})^{\hal\hbe}X_{\hal \hbe}{}^{ij} \loco$.
These will turn out to be composite and expressed in terms of the component curvatures.

Taking the double-bar projection of $\de = E^{\hA} \de_{\hA}$,
we define the component vector covariant derivative $\de_{\ha}$ to coincide with the projection of the superspace derivative $\de_{\ha} \loco$,
\begin{align}
e_{\hm}{}^{\ha} \nabla_{\ha} = \pa_{\hm}
	- \frac{1}{2} \psi_{\hm}{}^{\hal}_i \nabla_{\hal}^i \loco
	- \frac{1}{2} \omega_{\hm}{}^{\ha \hb} M_{\ha \hb}
	- b_{\hm} \mathbb D
	- \phi_{\hm}{}^{ij} J_{ij}
	- \frac{1}{2} \phi_{\hm}{}^{\hal i} S_{ \hal i}
	- {\mathfrak f}_{\hm}{}^{\ha} K_{\ha}
	~.
\end{align}
Here, the projected spinor covariant derivative $\nabla_{\hal}^i\loco$
corresponds to the generator of $Q$-supersymmetry. It is defined such that
if $\cU = U\vert$, then $Q_{\hal}^i \cU:=\nabla_{\hal}^i| \cU := (\nabla_{\hal}^i U) \loco$.
For the other generators, e.g.
$M_{\ha \hb}\cU =(M_{\ha \hb} U)\loco$, there is no ambiguity in identifying the bar projection; hence, local diffeomorphisms,
$Q$-supersymmetry transformations, and so forth descend naturally
from their corresponding rule in superspace.

The component supercovariant curvature tensors are given by
\begin{align}
[\nabla_\ha, \nabla_\hb]
	&= - R(P)_{\ha \hb}{}^\hc \nabla_{\hc}
	- R(Q)_{\ha \hb}{}^{\hal}_i \nabla_\hal^i\loco
	- \frac{1}{2} \RM_{\ha \hb}{}^{\hc \hd} M_{\hc \hd}
	- R(J)_{\ha\hb}{}^{ij} J_{ij}
	\eol & \quad
	- \RD_{\ha\hb} \gD 
	- \RS_{\ha\hb}{}^{\hga k} S_{\hga k}
	- \RK_{\ha\hb}{}^\hc K_\hc~.
\end{align}
We have introduced $R(P)_{\ha \hb}{}^{\hc} = \scT_{\ha \hb}{}^\hc \loco$, 
$R(Q)_{\ha \hb}{}_\hal^i = \scT_{\ha\hb}{}_\hal^i \loco$,
and
$ {R}(M)_{\ha \hb}{}^{\hc \hd}$, $ {R}(J)_{\ha \hb}{}^{ij}$, $ {R}(\mathbb D)_{\ha \hb}$, $ {R}(S)_{\ha \hb}{}^{\hga k}$, and $ {R}(K)_{\ha \hb}{}^{\hc}$ coinciding with the lowest components of the corresponding superspace curvature tensors given in appendix \ref{conformal-identities}.

The constraints on the superspace curvatures determine how to supercovariantise
a given component curvature by taking the double-bar projection of the
superspace torsion and curvature two forms. This leads to \cite{BKNT-M14}
\begin{subequations}\label{eq:hatCurvs}
\begin{align}
R(P)_{\ha\hb}{}^\hc &= 2 \,e_\ha{}^\hm e_\hb{}^\hn \cD_{[\hm} e_{\hn]}{}^\hc
	- \frac{\ri}{2} \psi_{\ha  j} \G^\hc \psi_\hb{}^j\ , \\
R(Q)_{\ha \hb}{}_{\hal}^i &=
	e_\ha{}^\hm e_\hb{}^\hn \cD_{[\hm} \psi_{\hn]}{}_\hal^i
	+ \ri (\G_{[\ha } \phi_{\hb]}{}^i)_\hal
	\eol & \quad
	+ \frac{1}{8} w_{\hc\hd} \Big(
		3 (\S^{\hc\hd} \G_{[\ha })_\hal{}^\hbe - (\G_{[\ha } \S^{\hc\hd})_\hal{}^\hbe
		\Big) \psi_{\hb]}{}_\hbe^i~, \\
R(M)_{\ha\hb}{}^{\hc\hd} &= \cR(\omega)_{\ha\hb}{}^{\hc\hd}
	+ 8 \,\delta_{[\ha }{}^{[\hc} {\mathfrak f}_{\hb]}{}^{\hd]}
	- 2 \,\psi_{[\ha  j} \Sigma^{\hc\hd} \phi_{\hb]}{}^{j}
	\eol & \quad
	+ \frac{16\ri}{3} \delta_{[\ha }{}^{[\hc}  \psi_{\hb]i} \G^{\hd]} \chi^i
	- \ri \psi_{[\ha  i} \Big(\G_{\hb]} R(Q)^{\hc\hd  i} +
		2\G^{[\hc} R(Q)_{\hb]}{}^{\hd] i} \Big)
	\eol & \quad
	+ \frac{\ri}{2} \psi_{\ha j} \psi_\hb{}^j w^{\hc\hd}
	- \frac{\ri}{4} (\psi_{\ha  j} \G_\he \psi_{b}{}^j) \tilde w^{\hc\hd\he} \ , \\
R(J)_{\ha\hb}{}^{ij} &= \cR(\phi)_{\ha\hb}{}^{ij}
	- 3 \,\psi_{[\ha }^{(i} \phi_{\hb]}{}^{j)}
	- 8\ri \,\psi_{[\ha }^{(i} \G_{\hb]} \chi^{j)} \ , \\
R(\bbD)_{\ha\hb} &= 2\, e_\ha{}^\hm e_\hb{}^\hn \pa_{[\hm} b_{\hn]}
	+ 4 \,{\mathfrak f}_{[\ha\hb]}
	+ \psi_{[\ha  j} \phi_{\hb]}{}^j
	+ \frac{8\ri}{3} \,\psi_{[\ha  j} \G_{\hb]} \chi^j
	 \ .
\end{align}
\end{subequations}
In the above we have introduced the spin, dilatation, and
${\rm SU}(2)_R$ covariant derivative
\begin{align}
\cD_\hm = \pa_\hm
	- \frac{1}{2} \omega_\hm{}^{\hb \hc} M_{\hb \hc}
	- b_\hm \mathbb D
	- \phi_\hm{}^{ij} J_{ij}~,  \qquad
\cD_{\ha} = e_{\ha}{}^{\hm} \cD_{\hm}~,
\end{align}
along with the curvatures
\bsubeq
\begin{align}
\cR(\omega)_{\ha\hb}{}^{\hc\hd} &:= 2 e_\ha{}^\hm e_\hb{}^\hn \Big(\partial_{[\hm} \omega_{\hn]}{}^{\hc\hd} - 2 \omega_{[\hm}{}^{\hc\he} \omega_{\hn]}{}_\he{}^\hd\Big)  \ , \\
\cR(\phi)_{\ha\hb}{}^{ij} &:= 2\,e_\ha{}^\hm e_\hb{}^\hn \Big( \pa_{[\hm} \phi_{\hn]}{}^{ij}
	+ \phi_{[\hm}{}^{k (i} {\phi}^{j)}{}_{\hn]k }\Big)~.
\end{align}
\esubeq
The component curvatures turn out to obey ``traceless'' conventional constraints \cite{BKNT-M14}
\begin{align} \label{componentConstTRv2}
R(P)_{\ha\hb}{}^\hc &= 0~, \qquad
(\G^\ha)_\hal{}^\hbe R(Q)_{\ha \hb}{}_{\hbe}^i = 0~, \qquad
R(M)_{\ha\hb}{}^{\hc\hb} = 0~,
\end{align}
which allow us to solve for the composite connections as follows:
\begin{subequations} \label{componentComposite}
\begin{align}
\omega_{\ha\hb\hc} &= 
	\omega(e)_{\ha\hb\hc} + \frac{\ri}{4} (
	\psi_{\ha k} \G_\hc \psi_\hb{}^k
	+ \psi_{\hc k} \G_\hb \psi_\ha{}^k
	- \psi_{\hb k} \G_\ha \psi_\hc{}^k)
	+ 2 b_{[\hb} \eta_{\hc ] \ha} ~, \\
\ri \,\phi_{\hm}{}^i
	&= \frac{2}{3} (\G^{[\hp} \delta_\hm{}^{\hq]} + \frac{1}{4} \G_\hm \S^{\hp\hq})
		\Big(
		\cD_{[\hp} \psi_{\hq]}{}^i
		+ \frac{1}{8} w_{\hc\hd}
			\big(3 \S^{\hc\hd} \G_{[\hp} \psi_{\hq]}{}^i
			- \G_{[\hp} \S^{\hc\hd} \psi_{\hq]}{}^i
		\big)\Big)~, \\
{\mathfrak f}_\ha{}^\hb &=
	- \frac{1}{6}\cR(\omega)_{\ha \hc}{}^{\hb\hc}
	+ \frac{1}{48} \delta_\ha{}^\hb \cR(\omega)_{\hc\hd}{}^{\hc\hd}
	- \frac{\ri}{6} \psi_{\hc j} \G^{[\hb} R(Q)_\ha{}^{\hc]j}
	- \frac{\ri}{12} \psi_{\hc j} \G_\ha R(Q)^{\hb\hc j}
	\eol & \quad
	+ \frac{1}{3} \psi_{[\ha  j} \S^{\hb\hd} \phi_{\hd]}{}^j
	- \frac{1}{24} \delta_\ha{}^\hb (\psi_{\hc j} \S^{\hc\hd}  \phi_\hd{}^j)
	- \frac{2\ri}{3} (\psi_{\ha  j} \G^\hb \chi^j)
	\eol & \quad
	- \frac{\ri}{12} \psi_{\ha j} \psi_\hc{}^j w^{\hb\hc}
	+ \frac{\ri}{24} (\psi_{\ha j} \G_\he \psi_\hd{}^j) \tilde w^{\hb\hd\he}
	\eol & \quad
	+ \frac{\ri}{192} \delta_\ha{}^\hb
		\Big(2 (\psi_{\hc j} \psi_\hd{}^j) w^{\hc\hd} -
		(\psi_{\hc j} \G_\he \psi_\hd{}^j) \tilde w^{\hc\hd \he}\Big)~,
\end{align}
\end{subequations}
where $\o(e)_{\ha \hb \hc} = -\hf (\cC_{\ha \hb \hc} + \cC_{\hc \ha \hb } -\cC_{\hb \hc \ha})$ is the usual torsion-free spin connection in terms of the anholonomy coefficient $\cC_{\hm \hn}{}^{\ha} := 2 \pa_{[\hm} e_{\hn]}{}^{\ha}$. We have also defined
\bea
\tilde{w}_{abc} = \hf \ve_{abcde} w^{de}~.
\eea
The curvature $R(\mathbb{D})_{\ha \hb}$ now vanishes due to eqs.~\eqref{componentComposite}.  
We stress that in the traceless frame $\phi_m{}^i$ has no dependence upon the matter field $\chi^i$ and $\mathfrak{f}_a{}^b$ has no dependence upon $D$. This choice minimises the dependence of the covariant derivatives upon some matter ``auxiliary'' fields and will simplify part of the analysis in the coming sections.

The supersymmetry transformations of the fundamental gauge connections of the Weyl
multiplet can be derived directly from the transformations of their corresponding
superspace one-forms.
We restrict
to all local superconformal transformations except local 
translations (covariant general coordinate transformations).
We denote such transformations by $\d = \d_{Q} + \d_{\cH}$ and define the operator
\bea
\delta
= 
\xi^{\hal}_iQ_{\hal}^i
+\hf \l^{\ha \hb}M_{\ha \hb} 
+ \l^{ij}J_{ij} 
+ \l_{\mathbb{D}}\mathbb{D} 
+ \l^{\ha} K_{\ha}
+ \eta^{\hal i} S_{\hal i}
~.
\eea
Here, the local component parameters are respectively defined as the $\q=0$ components of the corresponding superfield parameters, $\xi^{\hal}_i := \xi^{\hal}_i \loco$, $\l^{\ha \hb} := \L^{\ha \hb} \loco$, $\l^{ij} := \L^{ij} \loco$, $\l_{\mathbb{D}} := \sigma \loco$, $\l^{\ha} := \L^{\ha}_K \loco$, and $\eta^{\hal i} := \eta^{\hal i} \loco$.
The local superconformal transformations of 
the independent connection fields of the standard Weyl multiplet
are given by \cite{BKNT-M14}
\bsubeq\label{transf-standard-Weyl}
\bea
\delta e_{\hm}{}^{\ha} 
&=& 
\ri\, (\xi_i\G^{\ha}{\psi}_{\hm}{}^i)
- \lambda_{\mathbb{D}}{e}_{\hm}{}^{\ha}
+ \lambda^{\ha}{}_{\hb}e_{\hm}{}^{\hb}
~,
\label{d-vielbein}
\\
%%%%%%
\delta \psi_\hm{}_\hal^i &=&
	2 \cD_\hm \xi_\hal^i
	- \frac{1}{4} w_{\hc\hd} \Big(
		(\G_\hm \S^{\hc\hd})_\hal{}^\hbe 
		- 3 (\Sigma^{\hc\hd} \G_\hm)_\hal{}^\hbe \Big) \xi_\hbe^i
	+ 2 \ri \,(\G_\hm \eta^i)_\hal \non\\
&&+\frac{1}{2}\lambda^{\ha \hb}(\S_{\ha \hb} \psi_{\hm}{}^{i})_{\hal}
+ \lambda^{i}{}_{j} \,\psi_{\hm}{}_{\hal}^{j}
- \frac{1}{2}\lambda_{\mathbb{D}}\,{\psi}_{\hm}{}^{i}_{\hal}
~,
\label{d-gravitino}
\\
%%%%%%
 \delta  \phi_{\hm}{}^{ij} &=& \pa_{\hm}\l^{ij}
-2 \phi_{\hm}{}^{(i}{}_{k} \l^{j) k}
+3 \xi^{(i} \phi_{\hm}{}^{j)} 
-3 \eta^{(i} \psi_{\hm}{}^{j)}
+8 \ri \xi^{(i} \G_{\hm} \chi^{j)}~,
\label{d-SU2}
%%%%%%
\\
\delta b_{\hm}
&=& \partial_{\hm} \lambda_{\mathbb{D}} 
-\frac{8\ri}{3}\,\xi_i \G_{\hm} \chi^{i} 
-\xi_i   \phi_m{}^i
-\psi_{\hm}{}^i\eta_i  
- 2\lambda_{\hm}
~.
\label{d-dilatation}
\eea
%%%%%
In like fashion, one can derive the transformations $\d \o_{\hm}{}^{\ha \hb}, \d \phi_{\hm \hal}{}^i$, and $\d \mathfrak{f}_{\hm \ha}$, which we omit since these fields are composite. For the covariant matter fields, the transformations are given by \cite{BKNT-M14}
\bea
 \delta w_{\ha \hb} &=& 
2 \ri \, \xi_{i} R(Q)_{\ha \hb}{}^{i} 
- \frac{32 \ri}{3} \xi_{i} \S_{\ha \hb} \chi^i
- 2 \l_{[\ha}{}^{\hc} w_{\hb] \hc} 
+ \l_{\mathbb{D}} w_{\ha \hb}~,
  \label{d-W}
%%%%%%%
\\
\delta  \chi^{\hal i}
&=& \hf \xi^{\hal i} D
-\frac{1}{16} (\xi_{j} \S^{\ha \hb})^{\hal} R(J)_{\ha \hb}{}^{ij}
-\frac{3}{128} (\de_{\ha} w_{\hb \hc}) \Big( 3 (\xi^{i} \G^{\ha} \S^{\hb \hc})^{\hal} + (\xi^{i} \S^{\hb \hc} \G^{\ha})^{\hal} \Big)\non\\
&&+ \frac{3}{256} w_{\ha \hb} w_{\hc \hd} \ve^{\ha \hb \hc \hd \he} (\xi^i \G_{\he})^{\hal}
+ \frac{3\ri}{16} (\eta^{i} \S^{\ha \hb})^{\hal} w_{\ha \hb}
 \non\\
&&-\hf \l^{\ha \hb}(\chi^{i}\S_{\ha \hb})^{\hal}
+\l^{i}{}_{j}\chi^{\hal j}
+\frac{3}{2} \l_{\mathbb{D}} \chi^{\hal i}
~,
\label{d-chi}
%%%%%%%%%
\\
\delta D
&=& 2 \ri \, \xi_{i}\G^{\ha}\de_{\ha} \chi^{i} +  \ri w_{\ha \hb} (\xi_{i} \S^{\ha \hb} \chi^i)
+2 \eta_{i} \chi^{i}
+2 \l_{\mathbb{D} }D
~,
\label{d-D}
\eea
\esubeq
where
\bsubeq\label{nabla-on-W-and-Chi}
\bea
\de_{\ha}w_{\hb \hc} &=& 
\cD_{\ha}w_{\hb \hc}
-\ri \psi_{\ha i} R(Q)_{\hb \hc}{}^{i}
+ \frac{16 \ri}{3} \psi_{\ha i} \S_{\hb \hc} \chi^i~,\\
%%%%%%%%
\de_{\ha} \chi^{\hal i}
&=& 
\cD_{\ha}\chi^{\hal i} 
-\frac{1}{4} \psi_{\ha}{}^{\hal i} D 
-\frac{3\ri}{32} (\phi_{\ha}{}^i \S^{\hb \hc})^{\hal}w_{\hb \hc}
+\frac{1}{32} (\psi_{\ha j} \S^{\hb \hc})^{\hal} R(J)_{\hb \hc}{}^{ij} \non\\
&+& \frac{3}{256} (\de_{\hb} w_{\hc \hd}) \Big( 3 (\psi_{\ha}{}^{i} \G^{\hb} \S^{\hc \hd})^{\hal}+ (\psi_{\ha}{}^{i} \S^{\hc \hd} \G^{\hb})^{\hal} \Big)
-\frac{3}{512} w_{\hb \hc} w_{\hd \he} \ve^{\hb \hc \hd \he {f}} (\psi_{\ha}{}^i \G_{{f}})^{\hal}~.~~~~~~~~~
\eea
\esubeq
%

%%%%%%%%%%%%%%%%%%%%%%%%%%%%%%%%%%%%%%%%%%%%%%%%%%%%%%%%
%%%%%%%%%%%%%%%%%%%%%%%%%%%%%%%%%%%%%%%%%%%%%%%%%%%%%%%%

\section{The hyper-dilaton Weyl multiplet in 5D}
\label{hyper+HDWM}
The aim of this section is to construct a new $32+32$
hyper-dilaton Weyl multiplet of off-shell 
$\cN=1$ conformal supergravity in five dimensions. The analysis closely follows the 4D $\cN=2$ case of \cite{Gold:2022bdk}.

In constructing such a hyper-dilaton Weyl multiplet, our starting point is the component structure of the on-shell hypermultiplet. This can be readily extracted from the previous superspace realisation (see subsection \ref{HDWM-superspace}) via the bar projection. As was shown before, taking successive spinor derivatives of $q^{i \underline{i}}$ leads to $\rho_{\hal}^{\underline{i}}$ or the vector derivatives of $q^{i \underline{i}}$ and $\rho_{\hal}^{\underline{i}}$. Hence, the independent components of the on-shell hypermultiplet are simply the Lorentz scalar field $q^{i \underline{i}} \loco$ which is superconformal primary,  and the spinor field $\rho_{\hal}^{\underline{i}} \loco$. 

In what follows, we will associate the same symbol for the covariant component fields and the corresponding superfields, when the interpretation is clear from the context. 
The superfields $q^{i \underline{i}}$ and $\rho_{\a}^{\underline{i}}$ are all annihilated by $K^a$; hence, all their bar projections are $K$-primary fields. The local superconformal transformations of the component fields follow directly from the projections of \eqref{d-hyper-sspace}, which give
\bsubeq\label{hyper-susy}
\bea
\delta q^{i\underline{i}} 
&=& 
\frac{1}{2}\xi^{ i}\rho^{\underline{i}}
+\lambda^{i}{}_{k}q^{k\underline{i}} 
+ \frac{3}{2}\lambda_{\mathbb{D}}q^{i\underline{i}}
~,
\label{d-qii}
\\
\delta \rho_{\hal}^{\underline{i}} 
&=&
-4\ri(\G^{\ha} {\xi}_i)_{\hal} 
\nabla_{\ha} q^{i\underline{i}}
 +\hf \lambda_{\ha \hb} (\S^{\ha \hb} \rho^{\underline{i}})_{\hal}    
+2 \lambda_{\mathbb{D}}\rho_{\hal}^{\underline{i}} 
-12 \eta_{\hal}{}^i q_i{}^{\underline{i}}~,
\label{d-rho}
\eea
\esubeq
where 
\bea
\nabla_{\ha} q^{i\underline{i}}
=
\cD_{\ha} q^{i\underline{i}}
 -\frac{1}{4}\psi_{\ha}{}^{i}
 \rho^{\underline{i}}~.
 \label{DD'q}
\eea
The algebra of the local transformations \eqref{hyper-susy} closes only when the following equations of motion for the fields $q^{i\underline{i}}$ and $\rho_{\hal}^{\underline{i}}$ are imposed:
\bsubeq \label{on-shell-hyper}
\bea
(\nabla_{\ha}
\rho^{\underline{i}}\,
\G^{\ha})^{\hal}
&=&
-\frac{3}{4}
({\rho}^{\underline{i}}\,
\S^{\hb \hc})^{\hal} w_{\hb \hc}
+16 \ri\,{\chi}^{\hal k}q_{k}{}^{\underline{i}}~,
\label{on-shell-hyper-1}
\\
\Box q^{i \underline{i}} 
&=&
2\chi^{i}\rho^{\underline{i}} 
-2D \,q^{i\underline{i}} -\frac{3}{64} w^{{a} {b}} w_{{a} {b}} q^{i \underline{i}}~, \qquad \Box := \nabla^{\ha}\nabla_{\ha}~.
\label{on-shell-hyper-2}
\eea
\esubeq
The above equations are obtained by bar-projecting \eqref{vect-rho-sspace}, \eqref{boxq-sspace}, and using the definitions \eqref{comps-W}. The expressions for 
$\nabla_{\ha}\rho^{\hal \underline{i}}$ 
and $\Box q^{i\underline{i}}$ in terms of the derivatives 
$\cD_{\ha}$ are given by
\bsubeq\label{vect_der_hyper5D}
\bea
\nabla_{\ha} \rho^{\hal\underline{i}}
&=& 
\cD_{\ha}\rho^{\hal \underline{i}} 
+2\ri (\psi_{\ha k}\G^{\hb})^{\hal} \left( 
\cD_{\hb} q^{k\underline{i}} 
-\frac{1}{4}{\psi_{\hb}}^{k}
\rho^{\underline{i}} 
\right) 
+6\phi_{\ha}{}^{\hal k}q_{k}{}^{\underline{i}}
~,
%%%%%%
%%%%%%
\\
\Box q^{i\underline{i}} 
&=& 
\cD^{\ha}\cD_{\ha} q^{i\underline{i}}  
- 3 {\mathfrak{f}_{\ha}}^{\ha}  q^{i \underline{i}}
- \frac{1}{4}\rho^{\underline{i}} \cD_{\ha}{\psi^{\ha i}}
- \frac{1}{2}{\psi^{\ha i}}\cD_{\ha}\rho^{\underline{i}}
\nonumber\\
&&
+ \frac{\ri}{4}\phi_{\ha}{}^i \G^{\ha} \rho^{\underline{i}} 
+\frac{\ri}{2}
({\psi_{\ha}}^{(i}\G^{\hb} {\psi}^{k)\ha})
\cD_{\hb} q_k{}^{\underline{i}} 
+4\ri ({\psi_{\ha}}^{ i} \G^{\ha} \chi_{k}) q^{k \underline{i}} 
\non\\
&&
-\frac{1}{8} w^{\hc \hd} (\psi_{\hc}{}^i 
\G_{\hd}
\rho^{\underline{i}})
+ \frac{3}{2} ({\psi^{\ha i}}{\phi_{\ha k}})
q^{k\underline{i}}
- \frac{\ri}{8}
({\psi_{\ha}}^{(i}
\G^{\hb}
{{\psi}^{k)}{}^{\ha} })
({\psi_{\hb k}}\rho^{\underline{i}}) ~.
\eea
\esubeq

Equations \eqref{on-shell-hyper} can then be interpreted as algebraic equations for the 
standard Weyl multiplet that determine
the auxiliary fields $\chi^{\hal i}$ and $D$ in terms of 
$q^{i\underline{i}}$ and $\rho_{\hal}^{\underline{i}}$, 
together with the other independent fields of the standard Weyl multiplet.
Assuming that $q^{i\underline{i}}$ is an invertible matrix,
\bea
q^2:=q^{i\underline{i}}q_{i\underline{i}}=\ve_{ij}\ve_{\underline{i}\underline{j}}q^{i\underline{i}}q^{j\underline{j}}
=2\det{q^{i\underline{i}}}\ne0
~,
\eea
then the following relations hold
\bsubeq\label{compositeSSbD}
\bea
{\chi}^{{\hal} i} 
&=&
\frac{\ri}{8} q^{-2}q^{i\underline{i}}\Big[
- (\nabla_{\ha}
\rho_{\underline{i}}\,
\G^{\ha})^{\hal}
-\frac{3}{4} w_{\hc \hd} 
(\rho_{\underline{i}}\S^{\hc \hd})^{\hal}
\,\Big] 
~,
\label{chi}
\\
%%%%%%%%%%%%
D 
&=&
- \frac{1}{2} q^{-2}q_{i\underline{i}} \Box q^{i\underline{i}}
+\frac{\ri}{16} q^{-2} \Big[
- (\nabla_{\ha}
\rho_{\underline{i}}\,
\G^{\ha})^{\hal}
-\frac{3}{4} w_{\hc \hd} 
(\rho_{\underline{i}}\S^{\hc \hd})^{\hal}
\,\Big]  \rho_{\a}^{\underline{i}} 
-\frac{3}{128}w_{\ha \hb}w^{\ha \hb}  ~.
~~~~~~
\label{D_def}
\eea
\esubeq

Note that so far, we have only used one of the four 
equations that are equivalent
to \eqref{on-shell-hyper-2} to solve for $D$ in eq.~\eqref{D_def}.
It is simple to show that the remaining independent
three equations are equivalent to the following 
\be
\de^a (q^{i(\underline{i}}\de_aq_i{}^{\underline{j})})=0
~.
\label{2-4-constraints}
\ee
As in the 4D $\cN=2$ case \cite{Gold:2022bdk},
this equation is solved by turning 
the SU(2)$_R$ connection $\phi_m{}^{kl}$
into a composite field.
Let us describe how.

Following the analysis of the $\cN=2$ hyper-dilaton Weyl multiplet in 4D \cite{Gold:2022bdk}, here we also note that, associated to an on-shell hypermultiplet, there is always a triplet of composite linear multiplets \cite{FS2,deWit:1980gt,deWit:1980lyi,deWit:1983xhu}. 
The covariant component fields of the 5D $\cN=1$ off-shell linear (or $\cO(2)$) multiplet are defined in terms of the bar projections of \eqref{O2superfieldComps}:
an SU$(2)_R$ triplet of Lorentz scalar fields $G^{ij} = G^{ij} \loco$\,; 
a spinor field $\varphi_{\hal i}  = \varphi_{\hal i} \loco$\,;
a scalar field $F  = F \loco$;
and a covariant closed anti-symmetric four-form field strength $H_{abcd} := \cH_{abcd} \loco$. 
The latter is equivalent to a conserved dual vector 
$H^{\ha}:= -1/{4!}\,\ve^{\ha \hb \hc \hd \he}H_{\hb \hc \hd \he}$.\footnote{The Levi-Civita tensor with world indices 
is defined as $\ve^{mnpqr} := \ve^{\ha\hb\hc\hd\he} e_\ha{}^{m} e_\hb{}^{n} e_\hc{}^{p} e_\hd{}^{q} e_\he{}^{r}$, such that
$\ve_{abcde}$ and $\ve^{abcde}$ are normalised as $\ve_{01234} = -\ve^{01234}= 1$.}
At the component level it holds that
\bea
H^a = h^a + 2 \psi_{b i} \S^{ab} \vf^i+ \frac{\ri}{2} \ve^{abcde} \psi_{bi}\S_{cd}\psi_{ej} G^{ij}
~.
\eea
The covariant conservation equation for ${H}_{\ha}$ is
\bea
\de^{\ha} {H}_{\ha} 
&=& 
0~.
\label{covariant-current}
\eea
The constraint implies the existence of a gauge three-form potential, $b_{\hm \hn \hp}$, 
and its exterior derivative $h_{\hm \hn \hp \hq}:= 4 \pa_{[\hm}b_{\hn \hp \hq]}$, with $b_{mnp} := \cB_{mnp} \loco$.

In the standard Weyl multiplet background, the local superconformal transformations of the covariant fields can be derived using the relations \eqref{O2spinorderivs} and \eqref{O2-S-actions}, which lead to
\bsubeq
\label{linear-multiplet}
\bea
\d G_{ij}
&=&
-2 \xi_{(i} \varphi_{j)}
- 2\lambda_{(i}{}^{k}  G_{j) k} 
+3 \lambda_{\mathbb{D}} G_{ i j}
~,
%%%%%%
\\
\d \varphi_{\hal i}
&=&
-\frac{\ri}{2}\xi_{\hal i} F 
-\frac{\ri}{2}{H}_{\ha} (\G^{\ha} {\xi}_{i})_{\hal} 
- \ri   
(\G^{\ha} {\xi}^{j})_{\hal} 
\de_{\ha} G_{i j}
+6 \eta_{\hal}{}^{j}   G_{ij}
\nonumber\\ 
&& 
+\frac{1}{2} \lambda^{\ha \hb}
(\S_{\ha \hb}\varphi_{i})_{\hal} 
-\lambda_i{}^j \varphi_{j \hal}
+\frac{7}{2}\lambda_{\mathbb{D}}\varphi_{\hal i}~,
%%%%%%
\\
\d F
&=&
2 {\xi}^{i} \G^{\ha} \de_{\ha} \varphi_i
- \frac{3}{2}({\xi}^{i} \S^{\ha \hb} {\varphi}_i)
w_{\ha \hb}
+16 \ri ({\xi}^{i} {\chi}^{ j}) G_{i j} 
\non\\
&&+6 \ri\, \eta^{i} \varphi_{i} +4 \lambda_{\mathbb{D}} F ~,\\
%%%%%%
\d {H}_{\ha}
&=&
-4 \xi^{i} \S_{\ha \hb} \de^{\hb} \varphi_{ i}
+\frac{3}{2} (\xi^{i} \G^{\hb}{\varphi}_{i})w_{\ha \hb}
+ \hf \ve_{\ha \hb \hc \hd \he} w^{\hb \hc} (\xi^{i} \S^{\hd \he}{\varphi}_{i})
\non\\
&&
+ \lambda_{\ha}{}^{\hb}\,{H}_{\hb}
+4 \lambda_{\mathbb{D}} \,{H}_{\ha}
-8 \ri
\eta^{i}\G_{\ha} {\varphi}_i
~,
\eea
\esubeq
where
\bsubeq
\bea
\de_{\ha} G_{ij} &=& \cD_{\ha} G_{ij} + \frac{3}{4} \psi_{\ha(i} \phi_{j)}~, \\
\de_{\ha} \vf_{{\a} i} &=& \cD_{\ha} \vf_{{\a} i} 
-\frac{\ri}{4} \psi_{\ha {\a} i } F 
-\frac{\ri}{4} (\G^{\hb} \psi_{\ha i})_{{\a}} H_{\hb}
+\frac{\ri}{2}  (\G^{\hb} \psi_{\ha}{}^{j})_{{\a}} \de_{\hb} G_{ij} - 3 \phi_{\ha {\a}}{}^{j} G_{ij}~.
\eea
\esubeq
The locally superconformal transformations of $b_{\hm \hn \hp}$ are 
\bea
\d b_{\hm \hn \hp}
&=&
2 \, \ve_{\ha \hb \hc \hd \he} e_{\hm} {}^{\ha} e_{\hn}{}^{\hb} e_{\hp}{}^{\hc} (\xi_i\S^{\hd \he} \varphi^i)
- 12 \ri (\psi_{[\hm}{}^{i} \S_{\hn \hp]} \xi^j) G_{ij}
+3 \pa_{[\hm}l_{\hn \hp]}~,
\label{d-bmn}
\eea
where we have also included 
the gauge transformation
$\d_l b_{\hm \hn \hp}= 3 \pa_{[\hm}l_{\hn \hp]}$
leaving 
$h_{\hm \hn \hp \hq}$
and
${H}^{\ha}$ invariant.
For convenience, we have summarised the dilatation weights
of the fields of the $\cO(2)$ multiplet in Table 
\ref{chiral-dilatation-weights-LINEAR}.
\begin{table}[hbt!]
\begin{center}
\begin{tabular}{ |c||c| c| c| c| c| c| c|} 
 \hline
& $G_{ij}$ 
& $\varphi_{\hal i}$
& $F$
& ${H}_{\ha}$
&$b_{\hm \hn \hp}$
\\ 
\hline
 \hline
$\mathbb{D}$
&$3$
&$7/2$
&$4$
&$4$
&$0$
\\
\hline
\end{tabular}
\caption{\footnotesize{Dilatation weights of the off-shell $\cO (2)$ multiplet.}
\label{chiral-dilatation-weights-LINEAR}}
\end{center}
\end{table}

Given that $q^{i\underline{i}}$ and $\rho_{\hal}^{\underline{i}}$ describe an on-shell hypermultiplet in a standard Weyl multiplet background with transformation rules \eqref{hyper-susy}, it can be verified that the following composite fields define a triplet of $\cO(2)$ multiplets
\bsubeq\label{composite-linear}
\bea
G_{ij}{}^{\underline{i}\underline{j}}
&=&
q_{(i}{}^{\underline{i}}q_{j)}{}^{\underline{j}}
=q_{i}{}^{(\underline{i}}q_{j}{}^{\underline{j})}
~,\\
\varphi_{\hal i}{}^{\underline{i}\underline{j}}
&=& 
-\frac{1}{2} q_{i}{}^{(\underline{i}}\rho_{\hal}^{\underline{j})}
~,\\
F^{\underline{i}\underline{j}} 
&=& 
\frac{\ri}{8}\rho^{(\underline{i}}\rho^{\underline{j})}
~,
\label{composite-F}
\\
{H}^{\ha \, \underline{i}\underline{j}}
&=& 2q^{i(\underline{i}} \nabla^{\ha}{q_i}^{\underline{j})}
-\frac{\ri}{8}
\rho^{(\underline{i}}\G^{\ha} \rho^{\underline{j})}
~.
\label{linear-H}
\eea
\esubeq
These fields all transform according to 
\eqref{linear-multiplet} and
 each of the previous fields is symmetric in $\underline{i}$ and $\underline{j}$.
The field 
${H}^{ \ha \,\underline{i}\underline{j}}$ 
can be used to express the 
${\rm SU}(2)_R$ connection 
$\phi_{\hm}{}^{ij}$ as a composite field.
To see this, we introduce a new covariant derivative 
\bea
\mathbf{D}_{\ha} 
&=& 
{e_{\ha}}^{\hm}\left( \partial_{\hm} - \frac{1}{2}{\omega_{\hm}}^{\hc \hd}M_{\hc \hd} 
- b_{\hm }\mathbb{D}\right) 
= \cD_{\ha} + {e_{\ha}}^{\hm}{\phi_{\hm}}^{ij}J_{ij}~,
\eea
which then allows us to rearrange eq.~\eqref{linear-H} for the SU(2)$_R$ gauge connection:
\bea
 \phi_{\ha}{}^{ij}  
 &=& 
 4q^{-4}q^{(i}{}_{\underline{i}}q^{j)}{}_{\underline{j}}
 \Bigg[\, 
 q^{k\underline{i}} \mathbf{D}_{\ha} q_k{}^{\underline{j}}
 - \frac{1}{4}q^{k\underline{i}} 
 ({\psi_{\ha}}{}_k \rho^{\underline{j}}) 
  - \frac{\ri}{16}
 \rho^{\underline{i}}
 \G_{\ha}
 {\rho}^{\underline{j}}
 -\hf H_{\ha}{}^{\underline{i}\underline{j}}\,\Bigg] 
 ~.~~~~~~~~~
 \label{compositeSU2}
\eea

Our analysis demonstrates that the hyper-dilaton Weyl multiplet defines a new representation of the off-shell local 5D $\cN=1$ superconformal algebra. The multiplet comprises of the following independent fields: 
${e_{\hm}}^{\ha}$, $b_{\hm}$, $w_{\ha \hb}$, $q^{i\underline{i}}$, $b_{\hm \hn \hp}{}^{\underline{i}\underline{j}}$,
$\psi_{\hm\, i}$, and $\rho^{\underline{i}}$. It also possesses the same number of off-shell
degrees of freedom as the standard Weyl multiplet,
$32+32$. 
Table \ref{dof2} summarises the counting of degrees of freedom, 
underlining the symmetries acting on the fields. 
%%%%%%%%%%%%
\begin{table}[hbt!]
\begin{center}
\begin{tabular}{ |c c c c c c c |c c c c|} 
 \hline
${e_{\hm}}^{\ha}$ & $\omega_{\hm}{}^{\ha \hb}$ & $b_{\hm}$ & ${\mathfrak{f}_{\hm}}{}^{\ha}$ & $\phi_{\hm}{}^{ij}$ & $\psi_{\hm}{}_i$ & $\phi_{\hm}{}^i$ & $w_{\ha \hb}$ & $\rho^{\underline{i}}$ & $q^{i\underline{i}}$ & $b_{\hm \hn \hp}{}^{\underline{i}\underline{j}}$\\ 
$25B$ & $0$ & $5B$ & $0$ & $0$ & $40F$ & $0$ & $10B$ & $8F$ & $4B$ & $30B$\\
\hline
$P_{\ha}$ & $M_{\ha \hb}$ & $\mathbb D$ & $K_{\ha}$ & $J^{ij}$  & $Q$ & $S$ & {} & ${}$ & ${}$ & $\lambda_{\hm \hn}{}^{\underline{i}\underline{j}}$-sym\\
$-5B$ & $-10B$ & $-1B$ & $-5B$ & $-3B$ & $-8F$ & $-8F$ & {} & {} & {} & $-18B$\\
\hline
\multicolumn{11}{|c|}{Result: $32+32$ degrees of freedom}\\
\hline
\end{tabular}
\caption{\footnotesize{Degrees of freedom and symmetries of the hyper-dilaton Weyl multiplet. Row one gives all the component fields. Row two gives the number of independent components of these fields --- composite connections are counted with zero degrees of freedom. 
Row three gives the gauge symmetries. Note that $\lambda_{\hm \hn}{}^{\underline{i}\underline{j}}=-\lambda_{\hn \hm}{}^{\underline{i}\underline{j}}$ 
corresponds to the symmetry associated with the gauge three-forms
$b_{\hm \hn \hp}{}^{\underline{i}\underline{j}}$ 
with field strength
four-forms $h_{\hm \hn \hp \hq}{}^{\underline{i}\underline{j}}$ 
and ${H}^{\ha}{}^{\underline{i}\underline{j}}$.
Row four gives the number of gauge degrees of freedom to be subtracted when counting the total degrees of freedom.}\label{dof2}}
\end{center}
\end{table}
%%%%%%%%%%%%

Note that with the ingredients provided so far, it is 
a straightforward
exercise to obtain
the locally superconformal 
transformations of the fields
of the hyper-dilaton Weyl multiplet
written only in terms of the independent fields and they are given as follows:
\bsubeq\label{transf-hyper-Dilaton-Weyl5D}
\bea
%%%%%%
\delta e_{\hm}{}^{\ha} 
&=& 
\ri\, (\xi_i\G^{\ha}{\psi}_{\hm}{}^i)
- \lambda_{\mathbb{D}}{e}_{\hm}{}^{\ha}
+ \lambda^{\ha}{}_{\hb}e_{\hm}{}^{\hb}
~,
\label{d-vielbein5D}\\
\delta \psi_\hm{}_\hal^i &=&
	2 \mathbf{D}_\hm \xi_\hal^i + 8 q^{-4}q^{(i}{}_{\underline{i}}q^{j)}{}_{\underline{j}}
 \Bigg[\, 
 q^{k\underline{i}} \mathbf{D}_{m} q_k{}^{\underline{j}}
 - \frac{1}{4}q^{k\underline{i}} 
 ({\psi_{m}}{}_k \rho^{\underline{j}}) 
  - \frac{\ri}{16}
 \rho^{\underline{i}}
 \G_{m}
 {\rho}^{\underline{j}}
 -\hf H_{m}{}^{\underline{i}\underline{j}}\,\Bigg]  \xi_{\a j} \non \\ &&
	- \frac{1}{4} w_{\hc\hd} \Big(
		(\G_\hm \S^{\hc\hd})_\hal{}^\hbe 
		- 3 (\Sigma^{\hc\hd} \G_\hm)_\hal{}^\hbe \Big) \xi_\hbe^i
	+ 2 \ri \,(\G_\hm \eta^i)_\hal \non\\
&&+\frac{1}{2}\lambda^{\ha \hb}(\S_{\ha \hb} \psi_{\hm}{}^{i})_{\hal}
+ \lambda^{i}{}_{j} \,\psi_{\hm}{}_{\hal}^{j}
- \frac{1}{2}\lambda_{\mathbb{D}}\,{\psi}_{\hm}{}^{i}_{\hal}
~,
\label{d-gravitino5D}
\\
\delta b_{\hm}
&=& \partial_{\hm} \lambda_{\mathbb{D}} 
+\frac{1}{3}\,q^{-2}q^{i\underline{i}} (\xi_i \G_{\hm})^{\a} [
 (\nabla_{\ha}
\rho_{\underline{i}}\,
\G^{\ha}\e)_{\hal}
+\frac{3}{4} w_{\hc \hd} 
(\rho_{\underline{i}}\S^{\hc \hd} \e)_{\hal}
\,] \non \\ &&
-\xi_i   \phi_m{}^i
-\psi_{\hm}{}^i\eta_i  
- 2\lambda_{\hm}
~,
\label{d-dilatation5D}\\
\delta w_{\ha \hb} &=& 
2 \ri \, \xi_{i} R(Q)_{\ha \hb}{}^{i} 
+ \frac{4}{3} q^{-2}q^{i\underline{i}} (\xi_{i} \S_{\ha \hb})^{\a} [
 (\nabla_{c}
\rho_{\underline{i}}\,
\G^{c}\e)_{\hal}
+\frac{3}{4} w_{\hc \hd} 
(\rho_{\underline{i}}\S^{\hc \hd} \e)_{\hal}
\,] \non \\ &&
- 2 \l_{[\ha}{}^{\hc} w_{\hb] \hc} 
+ \l_{\mathbb{D}} w_{\ha \hb}~,
  \label{d-W5D}\\
  \delta q^{i\underline{i}} 
&=& 
\frac{1}{2}\xi^{ i}\rho^{\underline{i}}
+\lambda^{i}{}_{k}q^{k\underline{i}} 
+ \frac{3}{2}\lambda_{\mathbb{D}}q^{i\underline{i}}
~,
\label{d-qii5D}
\\
\delta \rho_{\hal}^{\underline{i}} 
&=&
-4\ri(\G^{\ha} {\xi}_i)_{\hal} 
\nabla_{\ha} q^{i\underline{i}}
 +\hf \lambda_{\ha \hb} (\S^{\ha \hb} \rho^{\underline{i}})_{\hal}    
+2 \lambda_{\mathbb{D}}\rho_{\hal}^{\underline{i}} 
-12 \eta_{\hal}{}^i q_i{}^{\underline{i}}~,
\label{d-rho5D}\\
\d b_{\hm \hn \hp}
&=&
2 \, \ve_{\ha \hb \hc \hd \he} e_{\hm} {}^{\ha} e_{\hn}{}^{\hb} e_{\hp}{}^{\hc} (\xi_i\S^{\hd \he} \varphi^i)
- 12 \ri (\psi_{[\hm}{}^{i} \S_{\hn \hp]} \xi^j) G_{ij}
+3 \pa_{[\hm}l_{\hn \hp]}~,
\eea
\esubeq
%%%%%%%%%%%%%%%%%%%%%%%%%%%%%%%%%
It would be useful to have \eqref{vect_der_hyper5D} expressions in terms of the derivative $\mathbf{D}_a$ instead of $\cD_a$, which has an implicit dependence on this new composite field $\phi_a{}^{ij}$. It holds that
\bsubeq
\bea
\nabla_{\ha} q^{i \underline{i}} &=& \frac{1}{2} \mathbf{D}_a q^{i \underline{i}} -\frac{1}{8} \psi_a^i \rho^{\underline{i}} - q^{-2} q^i{}_{\underline{j}} q^{k \underline{i}} \mathbf{D}_a q_k{}^{\underline{j}} +\frac{1}{4} q^{-2} q^i{}_{\underline{j}} q^{k\underline{i}} (\psi_{a k} \rho^{\underline{j}}) \non \\ &&+ \frac{\ri}{8} q^{-2} q^i{}_{\underline{j}} (\rho^{(\underline{i}} \Gamma_a \rho^{\underline{j})}) + q^{-2} q^i{}_{\underline{j}} H_a{}^{\underline{i}\underline{j}}\\
\nabla_{\ha} \rho^{\hal\underline{i}}
&=& 
\mathbf{D}_{a}\rho^{\hal \underline{i}} 
+2\ri (\psi_{\ha k}\G^{\hb})^{\hal} 
\nabla_{b} q^{k\underline{i}} 
+6\phi_{\ha}{}^{\hal k}q_{k}{}^{\underline{i}}
~,
%%%%%%
%%%%%%
\\
\Box q^{i\underline{i}} 
&=& \mathbf{D}^a \nabla_a q^{i \underline{i}} 
- 3 {\mathfrak{f}_{\ha}}^{\ha}  q^{i \underline{i}}
\non\\
&&
+(\nabla_a q_k{}^{\underline{i}})\Big{[}\,
  2q^{-2}q^{(i}{}_{\underline{j}}
  \mathbf{D}^{\ha} q^{k)\underline{j}}
  -2q^{-4}q^{(i}{}_{\underline{i}}q^{k)}{}_{\underline{j}} H^{\ha\underline{i}\underline{j}}
   - \frac{1}{2}q^{-2}q^{(i}{}_{\underline{j}}
 ({\psi^{\ha}}{}^{k)} \rho^{\underline{j}}) 
  \non\\
  &&~~~~~~~~~~~~~
  - \frac{\ri}{4}q^{-4}q^{(i}{}_{\underline{i}}q^{k)}{}_{\underline{j}}
 \rho^{\underline{i}}
 \G^{\ha}
 {\rho}^{\underline{j}}
 \,
\Big{]} 
  \nonumber\\
&&
-\frac{1}{8} w^{\hc \hd} (\psi_{\hc}{}^i  \G_{\hd} \rho^{\underline{i}})
+4\ri ({\psi_{\ha}}^{ i} \G^{\ha} \chi_{k}) q^{k \underline{i}} 
- \frac{1}{4}{\psi^{\ha i}}\de_{\ha}\rho^{\underline{i}}
+ \frac{\ri}{4}\phi_{\ha}{}^i \G^{\ha} \rho^{\underline{i}}
 ~.
\eea
\esubeq
where the composite connection $\phi_m{}^i$ and $\mathfrak{f}_{a}{}^{b}$ are now given in terms of  $\mathbf{D}_a$ by:
\begin{subequations} 
\bea 
\ri \,\phi_{\hm}{}^i
	&=& \frac{2}{3} (\G^{[\hp} \delta_\hm{}^{\hq]} + \frac{1}{4} \G_\hm \S^{\hp\hq})
		\bigg{[}
		\mathbf{D}_{[\hp} \psi_{\hq]}{}^i  
			+ \frac{1}{8} w_{\hc\hd}
			\big(3 \S^{\hc\hd} \G_{[\hp} \psi_{\hq]}{}^i
			- \G_{[\hp} \S^{\hc\hd} \psi_{\hq]}{}^i
\big)
\non\\
&&~~~
+4q^{-4}q^{(i}{}_{\underline{i}}q^{j)}{}_{\underline{j}}
 \Big\{\, 
 q^{k\underline{i}} \mathbf{D}_{[p} q_k{}^{\underline{j}}
 - \frac{1}{4}q^{k\underline{i}} 
 ({\psi_{[p}}{}_k \rho^{\underline{j}}) 
  - \frac{\ri}{16}
 \rho^{\underline{i}}
 \G_{[p}
 {\rho}^{\underline{j}}
 -\hf H_{[p}{}^{\underline{i}\underline{j}}\,\Big\} \psi_{\hq] j}
	\bigg{]}~, ~~~~~~
	\label{composite_phi_5D_dilaton} \\
{\mathfrak f}_\ha{}^\hb &=&
	- \frac{1}{6}\cR(\omega)_{\ha \hc}{}^{\hb\hc}
	+ \frac{1}{48} \delta_\ha{}^\hb \cR(\omega)_{\hc\hd}{}^{\hc\hd}
	- \frac{\ri}{6} \psi_{\hc j} \G^{[\hb} R(Q)_\ha{}^{\hc]j}
	- \frac{\ri}{12} \psi_{\hc j} \G_\ha R(Q)^{\hb\hc j}
		\eol &&
	+ \frac{1}{12}  q^{-2}q^{i\underline{i}} (\psi_{\ha  i} \G^\hb)^\a \Big{[}
 (\nabla_{\ha}
\rho_{\underline{i}}\,
\G^{\ha} \e)_{\a}
+\frac{3}{4} w_{\hc \hd} 
(\rho_{\underline{i}}\S^{\hc \hd} \e)_{\a}
\,\Big{] }
	\eol &&
	+ \frac{1}{3} \psi_{[\ha  j} \S^{\hb\hd} \phi_{\hd]}{}^j
	- \frac{1}{24} \delta_\ha{}^\hb (\psi_{\hc j} \S^{\hc\hd}  \phi_\hd{}^j)
		- \frac{\ri}{12} \psi_{\ha j} \psi_\hc{}^j w^{\hb\hc}
	+ \frac{\ri}{24} (\psi_{\ha j} \G_\he \psi_\hd{}^j) \tilde w^{\hb\hd\he}
	\eol &&
	+ \frac{\ri}{192} \delta_\ha{}^\hb
		\Big{[}2 (\psi_{\hc j} \psi_\hd{}^j) w^{\hc\hd} -
		(\psi_{\hc j} \G_\he \psi_\hd{}^j) \tilde w^{\hc\hd \he}\Big{]}~.
\eea
\end{subequations}
Note that the expression of ${\mathfrak f}_\ha{}^\hb$ has explicit as well as implicit dependence on the composite connection $\phi_{a}{}^i$ via $\de_a\rho_{\underline{i}}$, which can now be substituted from \eqref{composite_phi_5D_dilaton}.  For later use, it is convenient to have the bosonic expression of $\Box q^{i\underline{i}}$, which, by also using that $\mathbf{D}^a H_{a}{}^{\underline{i}\underline{j}}=0$ up to fermions, is given by:
\bea
\Box q^{i\underline{i}} 
 &=&  \frac{1}{2} \mathbf{D}^a\mathbf{D}_a q^{i \underline{i}} +\frac{3}{4} q^{-2} q^{i \underline{i}} (\mathbf{D}^a q^{k \underline{k}}) \mathbf{D}_a q_{k \underline{k}}  - q^{-2} q^i{}_{\underline{j}} q^{k \underline{i}} \mathbf{D}^a\mathbf{D}_a q_k{}^{\underline{j}} + \hf
 q^{-2}q^{k}{}_{\underline{j}}
 ( \mathbf{D}^{\ha} q^{i \underline{j}})\mathbf{D}_a q_k{}^{ \underline{i}}  \non \\ &&  
  +    q^{-4} q_{l \underline{l}} q^i{}_{\underline{j}} q^{k \underline{i}} (\mathbf{D}^a q^{l \underline{l}})  \mathbf{D}_a q_k{}^{\underline{j}} - \hf q^{-4} q^{i \underline{i}} H^{a \underline{j}\underline{k}}H_{a \underline{j}\underline{k}}
+ \frac{3}{16} \mathcal{R}  q^{i \underline{i}} + \text{fermions}~.
\eea

\section{Recovering Poincar\'e supergravity in 5D}
\label{section-3}

The goal of this section is to explicitly show that our hyper-dilaton Weyl multiplet constructed in the previous section can be used to derive a $40+40$ off-shell multiplet of 5D $\cN=1$ 
Poincar\'e supergravity, by making use of superconformal approaches. We first elaborate on the structure of the multiplet and then explain how to construct a Poincar\'e supergravity
action, pointing out some peculiarities which do not hold in the 4D $\cN=2$ supergravity case \cite{Gold:2022bdk}. 
As an extension of the results of \cite{Gold:2022bdk}, we describe a new type of $BF$-coupling which induces 
a scalar potential for the dilaton without a standard 
$R$-symmetry gauging that admits AdS$_5$ vacua.

\subsection{Hyper-dilaton multiplet of Poincar\'{e} supergravity}

To recover a multiplet of Poincar\'{e} supergravity, compensating multiplets must be coupled to an 
off-shell conformal supergravity multiplet to gauge fix some of the local superconformal symmetries.
As will be discussed below, from the symmetry point of view, it suffices to use the components of the new hyper-dilaton Weyl multiplet alone to appropriately gauge fix and eliminate all symmetries except local supersymmetry, Lorentz, and the gauge symmetry of the gauge three-forms
$b_{\hm \hn \hp}{}^{\underline{i}\underline{j}}$.  This peculiar feature is different from the construction of the 4D $\cN=2$ hyper-dilaton Poincar\'e multiplet \cite{Gold:2022bdk}. In the latter case, we are required to gauge fix the scalar field of the compensating vector multiplet in order to fix the extra U(1)$_R$ symmetry.  
In the 5D case, the only purpose to couple to a compensator is simply to obtain the Einstein-Hilbert kinetic term in a Poincar\'{e} supergravity action. 
The simplest choice is to couple our hyper-dilaton Weyl multiplet to a single off-shell, Abelian vector multiplet compensator. The scalar field in the compensator is assumed to be nowhere vanishing.

Let us first define an off-shell 5D $\cN=1$ Abelian vector multiplet in a standard Weyl multiplet background. 
Its component structure follows directly from the superfield definitions \eqref{components-vect}. The multiplet contains a real scalar field $\phi := W \loco$,
gaugini $\lambda_{\hal}^i:= \lambda_{\hal}^i \loco$, a triplet of auxiliary fields $X^{ij} := X^{ij} \loco$,
and a real Abelian gauge connection $v_{\hm} := \cV_m \loco$ or, equivalently,
its real field strength $f_{mn} :=  \cF_{mn}\loco = 2 \pa_{[m} v_{n]}$.  The  field strength $f_{mn}$ may be expressed in terms of the bar-projected, covariant field strength $F_{ab}:=\cF_{ab} \loco$ via the relation
\bea
F_{ab} = f_{ab} + \ri (\G_{[a})_{\a}{}^{\b} \psi_{b]}{}^{\a}_{k} \l_{\b}^k  
+ \frac{\ri}{2} \psi_{[a}{}^{\g}_{k} \psi_{b]}{}^{k}_{\g} \phi~, \qquad f_{ab}:= e_{a}{}^{m} e_{b}{}^{n} f_{mn}~.
\eea
The dilatation weights of the vector multiplet fields are summarised in Table
\ref{weights-vector}.
\begin{table}[hbt!]
\begin{center}
\begin{tabular}{ |c| c| c| c| c| c|} 
 \hline
& $\phi$ 
& $\lambda_{\hal}^i$
& $X^{ij}$
& $F_{\ha \hb}$
& $v_{\hm}$
\\ 
\hline
 \hline
$\mathbb{D}$
&$1$
&$3/2$
&$2$
&$2$
&$0$
\\
\hline
\end{tabular}
\caption{\footnotesize{Dilatation weights of the
Abelian vector multiplet.}\label{weights-vector}}
\end{center}
\end{table}

The transformation rules of the vector multiplet fields in a standard Weyl multiplet background can be obtained from the corresponding superfields. They read
\bsubeq
\bea
\d \phi
&=&
\ri \xi_i \l^i + \l_\mathbb{D} \phi~,
\\ 
\d\lambda_{\hal}^i
&=&
- (\S^{\ha \hb} \xi^i)_{\hal} F_{\ha \hb} - (\S^{\ha \hb} \xi^i)_{\hal} w_{\ha \hb} \phi
+ \xi_{\hal}{}_j X^{ij} + (\G^{\ha} {\xi}^i)_{\hal} \de_{\ha} \phi \nonumber \\
&&+ \frac{1}{2}\l^{\ha \hb}(\S_{\ha \hb} \l^i)_{\hal}  + \l^i{}_j \l^j_{\hal} + \frac{3}{2} \l_{\mathbb{D}} \l^i_{\hal} + 2 \ri \eta_{\hal}{}^i \phi~,
\label{transf-lambda}
\\
\d X^{ij}
&=& -2 \ri \xi^{(i} \G^{\ha} \de_{\ha} {\l}^{j)} 
+ \frac{3 \ri}{2} \xi^{(i} \S^{\ha \hb} \l^{j)} w_{\ha \hb} 
-16 \ri  (\xi^{(i} \chi^{j)}) \phi
\non\\
&&+ 2 \l^{(i}{}_k X^{j)k} 
-2 \eta^{(i} \l^{j)}
+ 2 \l_{\mathbb{D}} X^{ij}~,
\\
\d v_{\hm}
&=& \ri (\xi^i \psi_{\hm i}) \phi
- \ri (\xi^{i} \G_{\hm} \l_{i}  )
+\partial_{\hm} \l_V
~,
\label{dvm}
\eea
\esubeq
where
\bsubeq
\bea
\de_{\ha} \phi
&=&
\cD_{\ha} \phi - \frac{\ri}{2}\psi_{\ha}{}_i \l^i
~,
\\
\de_{\ha} \l^i_{\hal}
&=&
\cD_{\ha} \l^i_{\hal} 
+ \hf (\S^{\hb \hc}\psi_{\ha}{}^{i})_{\hal} \Big( F_{\hb \hc}
+ w_{\hb \hc} \phi \Big)
-\hf \psi_{\ha}{}_{\hal}{}_j X^{ij} 
\non\\
&&
-\hf (\G^{\hb} {\psi}_{\ha}{}^i)_{\hal} \de_{\hb} \phi 
- \ri \phi_{\ha}{}_{\hal}{}^i \phi
~.
\eea
\esubeq
Note that we have also included in \eqref{dvm}
the gauge field transformation
parametrised by the local real parameter $\l_V$.
The transformations of the vector multiplet
in a hyper-dilaton Weyl multiplet background are precisely the same as above. The only subtlety is
that one has to interpret several 
standard Weyl multiplet fields as composites of 
$q^{i\underline{i}}$, $\rho_{\hal}^{{\underline{i}}}$
and $b_{\hm \hn \hp}{}^{\underline{i}\underline{j}}$. It should be emphasised that, being a compensator, one may require that the lowest component of the vector multiplet is non-zero and positive, $\phi > 0$. 

\subsubsection{Gauge fixing in a string frame}

We now describe the structure of 
the supergravity multiplet by imposing several gauge fixing constraints. In a string frame, we choose the gauge fixing condition
\bsubeq \label{gauge-conditions}
\bea
&q^{i\underline{i}} =- \ve^{i\underline{i}}
~~~\Longleftrightarrow~~~
q^{i}{}_{\underline{i}} = \delta^i_{\underline{i}}
~~~\Longleftrightarrow~~~
q_{i}{}^{\underline{i}} = -\delta_i^{\underline{i}}
~~~\Longleftrightarrow~~~
q_{i\underline{i}} = \ve_{i\underline{i}}
~.
\label{SU2-gauge-fixing}
\eea 
This condition fixes dilatation and SU(2)$_R$ symmetries. By imposing 
\bea
b_m = 0~,
\eea
the special conformal $K^{\ha}$ symmetry is now fixed. 
In order to fix $S$-supersymmetry, we impose the constraint
\bea
\rho_{\hal}^{\underline{i}}=0~.
\label{S-susy-gauge}
\eea
\esubeq
The compensating vector multiplet contains $8+8$ off-shell 
degrees of freedom. Once added to the remaining fundamental fields in the hyper-dilaton Weyl multiplet, we obtain $40+40$ off-shell degrees of freedom
of a Poincar\'e supergravity multiplet, as shown in Table \ref{dof3}.
\begin{table}[hbt!]
\begin{center}
\begin{tabular}{ |c c c c c c c c c |} 
 \hline
${e_{\hm}}^{\ha}$ & ${\omega_{\hm}}{}^{\ha \hb}$  & $\psi_{\hm}{}^{\hal}_i$ & $w_{\ha \hb}$ 
& $b_{\hm \hn \hp}{}^{\underline{i}\underline{j}}$
& $\phi$ 
& $\l_{\hal}^{\underline{i}}$ & $X^{ij}$ & $v_{\hm}$\\ 
$25B$ & $0$ & $40F$ & $10B$ & $30B$ & $1B$ & $8F$  & $3B$ & $5B$\\
\hline
$P_{\ha}$ & $M_{\ha \hb}$  & $Q$ & ${}$ &($\lambda_{\hm \hn}{}^{\underline{i}\underline{j}}$)& ${}$ & ${}$ & 
 {} & 
($\l_V$)\\
$-5B$ & $-10B$ & $-8F$ & {} & $-18B$ & {} & {}  & {} & $-1B$\\
\hline
\multicolumn{9}{|c|}{Result: $40+40$ degrees of freedom}\\
\hline
\end{tabular}
\caption{\footnotesize{A Poincar\'e supergravity multiplet. Row one gives all fields in the multiplet. Row two gives the number of independent components of these fields. Row three gives the surviving gauge symmetries. Row four gives the number of gauge degrees of freedom to be subtracted when counting the total degrees of freedom. The parameter $\lambda_{\hm \hn}{}^{\underline{i}\underline{j}}$ describes the symmetry associated with the 
triplet of gauge three-form $b_{\hm \hn \hp}{}^{\underline{i}\underline{j}}$. The gauge parameter $\l_V$ describes the scalar symmetry of $v_{\hm}$.} \label{dof3}}
\end{center}
\end{table}
The fundamental fields are the vielbein ${e_{\hm}}^{\ha}$,
the gravitino $\psi_{\hm}{}^{\hal}_i$,
a real antisymmetric tensor $w_{\ha \hb}$,
a real scalar field that plays the role of a dilaton $\phi$,
a real triplet of scalar fields $X^{ij}$,
a triplet of gauge three forms $b_{\hm \hn \hp}{}^{\underline{i}\underline{j}}$,
a gauge field $v_{\hm}$ that plays the role of the graviphoton,
and a spinor field $\l_{\hal}^{\underline{i}}$.
Note that we kept the distinction of SU(2)$_R$ and SU(2) flavour indices. 
However, the gauge condition \eqref{SU2-gauge-fixing} implies that the two indices can be identified, after gauge fixing.

The transformation rules of the resulting Poincar\'e supergravity multiplet are those that preserve the previous set of gauge conditions \eqref{gauge-conditions}.
Since we fix $\rho_{\a}^{\underline{i}}=0$, to preserve \eqref{SU2-gauge-fixing}, we require 
$\l_{\mathbb D}= \l^{ij}=0$.
To preserve \eqref{S-susy-gauge}, it can be shown that any $Q$-supersymmetry transformation
must be accompanied by a compensating  $S$-supersymmetry transformation with the following parameter
\bea
\eta_{\hal}{}^i (\xi)
&=& 
-\frac{\ri}{3}(\G^{\ha}\xi_{i})_{\hal}\,\phi_{a}{}^{ij}
~,
\eea
A similar analysis shows that to preserve the condition $b_m=0$ one needs to enforce nontrivial compensating 
special conformal $K$-transformations with a parameter $\l^{\ha}(\xi)$. However, since all the other supergravity fields
are conformal (not necessarily superconformal) primaries, not transforming under special conformal boosts, 
in practice we will never have to worry about inserting the compensating $\l^{\ha}(\xi)$ parameter (whose expression is quite involved)
in any Poincar\'e supergravity transformations.

\subsubsection{Gauge fixing in the Einstein frame}
It is possible to choose a different gauge fixing to Poincar\'{e} supergravity, where we also impose constraints on some of the fields of the compensating vector multiplet.  
This gauge fixing choice, which is analogous to that in the 4D $\cN=2$ case \cite{Gold:2022bdk}, corresponds to the Einstein frame and leads to a different Poincar\'{e} supergravity multiplet. 

We now adopt the gauge where
\bsubeq\label{gauge-conditions-E}
\bea
&\phi=1
~,
\label{fixing-phi}
\\
&
b_{\hm}=0
~.
\label{fixing-K}
\eea 
Condition \eqref{fixing-phi} fixes dilatation symmetry, while \eqref{fixing-K} 
fixes special conformal $K^{\ha}$ symmetry.
In order to fix $S$-supersymmetry, we impose
\bea
\lambda_{\hal}^i=0~.
\label{S-susy-gauge-E}
\eea
A characterising feature of the hyper-dilaton Weyl multiplet is that it contains an SU(2)$_R$ compensator, 
the $q^{i\underline{i}}$ fields. 
By imposing
\be
q^{i\underline{i}} =- \ve^{i\underline{i}} \re^{-U}
~~~\Longleftrightarrow~~~
q^{i}{}_{\underline{i}} = \delta^i_{\underline{i}} \re^{-U}
~~~\Longleftrightarrow~~~
q_{i}{}^{\underline{i}} = -\delta_i^{\underline{i}} \re^{-U}
~~~\Longleftrightarrow~~~
q_{i\underline{i}} = \ve_{i\underline{i}} \re^{-U}
~,
\label{SU2-gauge-fixing-E}
\ee
\esubeq
we break the SU(2)$_R$ symmetry.
The resulting Poincar\'{e} supergravity multiplet is shown in Table \ref{dof3-E}.
\begin{table}[hbt!]
\begin{center}
\begin{tabular}{ |c c c c c c c c c |} 
 \hline
${e_{\hm}}^{\ha}$ & ${\omega_{\hm}}{}^{\ha \hb}$  & $\psi_{\hm}{}^{\hal}_i$ & $w_{\ha \hb}$ 
& $\rho_{\hal}^{\underline{i}}$ & $U$ & $b_{\hm \hn \hp}{}^{\underline{i}\underline{j}}$ & $X^{ij}$ & $v_{\hm}$\\ 
$25B$ & $0$ & $40F$ & $10B$ & $8F$ & $1B$ & $30B$ & $3B$ & $5B$\\
\hline
$P_{\ha}$ & $M_{\ha \hb}$  & $Q$ & ${}$ & ${}$ & ${}$ & 
($\lambda_{\hm \hn}{}^{\underline{i}\underline{j}}$) & {} & 
($\l_V$)\\
$-5B$ & $-10B$ & $-8F$ & {} & {} & {} & $-18B$ & {} & $-1B$\\
\hline
\multicolumn{9}{|c|}{Result: $40+40$ degrees of freedom}\\
\hline
\end{tabular}
\caption{\footnotesize{A variant Poincar\'e supergravity multiplet. Row one gives all the fields in the multiplet. Row two gives the number of independent components. Row three gives the surviving gauge symmetries. Row four gives the number of gauge degrees of freedom to be subtracted when counting the total degrees of freedom. The parameter $\lambda_{\hm \hn}{}^{\underline{i}\underline{j}}$ corresponds to the symmetry associated with the 
triplet of gauge three-form $b_{\hm \hn \hp}{}^{\underline{i}\underline{j}}$. The gauge parameter $\l_V$ describes the scalar symmetry of $v_{\hm}$.} \label{dof3-E}}
\end{center}
\end{table}
The fundamental fields are the vielbein ${e_{\hm}}^{\ha}$,
the gravitini $\psi_{\hm}{}^{\hal}_i$,
a real antisymmetric tensor $w_{\ha \hb}$,
a real scalar field that plays the role of a dilaton $U$,
a real triplet of scalar fields $X^{ij}$,
a triplet of gauge three forms $b_{\hm \hn \hp}{}^{\underline{i}\underline{j}}$,
a gauge field $v_{\hm}$ that plays the role of the graviphoton,
and a spinor field $\rho_{\hal}^{\underline{i}}$.

The transformation rules of the resulting Poincar\'e supergravity multiplet are those preserving \eqref{gauge-conditions-E}.
To preserve \eqref{fixing-phi}, we require 
$\l_{\mathbb D}\equiv0$.
Since $Q$-supersymmetry does not preserve the gauge fixing conditions, it is necessary 
to accompany these transformations with appropriate $S$-supersymmetry, special conformal, and SU(2)$_R$
compensating transformations.
To preserve \eqref{S-susy-gauge-E}, it can be shown that any $Q$-supersymmetry transformation
must be accompanied by a compensating  $S$-supersymmetry transformation with the following parameter
\bea
\eta_{\hal}{}^i (\xi)
&=& 
-\frac{\ri}{2}(\S^{\ha \hb}\xi^{i})_{\hal}\Big(F_{\ha \hb}+ w_{\ha \hb}\Big)
+\frac{\ri}{2}\xi_{\hal j}X^{ij} 
~.
\eea
A similar analysis shows that to preserve the condition $b_m=0$, one needs to enforce nontrivial compensating 
special conformal $K$-transformations with a parameter $\l^{\ha}(\xi)$. However, since all the other supergravity fields
are conformal (not necessarily superconformal) primaries, not transforming under special conformal boosts, 
in practice we will never have to worry about inserting the compensating $\l^{\ha}(\xi)$ parameter (whose expression is quite involved)
in any Poincar\'e supergravity transformations.
Finally, we can easily check that the requirement
$\d q^{(i\underline{i})}=0$ is satisfied by implementing in \eqref{d-qii} a 
compensating SU(2)$_R$ transformation with the parameter
\bea
\lambda^{ij}(\xi)
= 
-\hf\,\re^U
\xi^{(i}\rho^{j)} 
~,
\eea
where 
$\rho^{i}=\d^i_{\underline{i}}\rho^{\underline{i}}$.

\subsection{Hyper-dilaton Poincar\'e supergravity  action and dilaton potential}
\label{chiral analysis}
We turn to deriving a Poincar\'e supergravity action by considering the two-derivative action of the vector multiplet compensator \cite{BKNT-M14} in a hyper-dilaton Weyl multiplet
background and then imposing 
appropriate gauge fixing conditions leading to the two frames described above.
As shown in \cite{BKNT-M14}, the component form of such a vector multiplet action may be derived from the bosonic part of the $BF$ Lagrangian 
\bea
e^{-1}\cL_{BF}|_{bosonic}  
&=&
-\frac{1}{4} \Big( F \phi 
+ G_{i j} X^{i j} 
- \frac{1}{12} \ve^{{a} {b} {c} {d} {e}} f_{{a} {b}} b_{{c} {d} {e}}  \Big)
\non\\
&=& -\frac{1}{4} \Big( 
F \phi 
+ G_{i j} X^{i j} 
+ v^{a} h_{a} \Big)
~, \label{BF}
\eea
with the fields of the $\cO(2)$ multiplet being composite. More precisely, this amounts to taking the bosonic sector of the bar projection of eqs.~\eqref{O2composite-N}:
\bsubeq \label{composite-O2-BF}
\bea 
G^{ij}|_{bosonic}  
&=& 2 \phi X^{ij}~, \\
%%%%%%%%%%%%%%%%%%%%%%%
F|_{bosonic} &=& X^{ij} X_{ij} - f^{\ha\hb} f_{\ha\hb} 
+ 4\phi \nabla^\ha \nabla_\ha \phi + 2 (\nabla^\ha \phi )\nabla_\ha \phi  \non\\
&&- 6 \phi \,w^{\ha\hb} f_{\ha\hb} - \frac{39}{8} \phi^2 w^{\ha\hb} w_{\ha\hb}
-16 \phi^2 D~, \\
%%%%%%%%%%%%%%%%%%%%%%%%%
h_\ha|_{bosonic} &=& - \frac{1}{2} \eps_{\ha\hb\hc\hd\he} f^{\hb\hc} f^{\hd\he}
+ 4 \nabla^\hb (\phi f_{\hb\ha} + \frac{3}{2} \phi^2 w_{\hb\ha})~,
\eea
\esubeq
and plugging \eqref{composite-O2-BF} back into \eqref{BF}.
This procedure results in 
\bea
e^{-1}\, \cL_{BF}|_{bosonic}
&=&
- \frac{1}{2} \phi ({\de}^{\ha} \phi) {\de}_{\ha} \phi
- \phi^2 \de^a \de_a \phi
- \frac{3}{4} \phi X^{ij} X_{ij} 
+\frac{1}{8} \ve_{\ha \hb \hc \hd \he}v^{\ha} f^{\hb \hc} f^{\hd \he} \non\\
&&+ \frac{3}{4} \phi f^{\ha \hb} f_{\ha \hb}
+ \frac{9}{4} \phi^2 w^{\ha \hb} f_{\ha \hb}
+ \frac{39}{32}\phi^3 w^{ab} w_{ab}
+ 4 \phi^3 D~. 
\label{BF-2}
\eea
The expression \eqref{BF-2} can be written in terms of the degauged covariant derivative $\cD_a$, where we note the following relation
\bea
\de^a \de_a \phi = \cD^a \cD_a \phi 
+\frac{1}{8} \phi\, \cR~. 
\eea
After performing integration by parts, one then arrives at the following bosonic Lagrangian
\bea
e^{-1}\, \cL_{BF}|_{bosonic}
&=&
- \frac{1}{8} \phi^3 {\cR}
+ \frac{3}{2} \phi ({\cD}^{\ha} \phi) {\cD}_{\ha} \phi
- \frac{3}{4} \phi X^{ij} X_{ij}
+\frac{1}{8} \ve_{\ha \hb \hc \hd \he}v^{\ha} f^{\hb \hc} f^{\hd \he} 
\non\\
&&+ \frac{3}{4} \phi f^{\ha \hb} f_{\ha \hb}
+ \frac{9}{4} \phi^2 w^{\ha \hb} f_{\ha \hb}
+ \frac{39}{32} \phi^3 w^{\ha \hb} w_{\ha \hb} + 4 \phi^3 D~.
\label{bosonic-swm}
\eea

The action \eqref{bosonic-swm} is given in the standard Weyl multiplet background.
When working with a hyper-dilaton Weyl multiplet, we need to take into account that the auxiliary field $D$ is composite (and that \eqref{D_def} has to be used). 
The algebraic expression for $D$ takes the following form
\bea
D = -\frac{3}{32} {\cR} -\frac{3}{128} w^{\ha \hb} w_{\ha \hb}-\frac{1}{2 q^2} q_{i \underline{i}} {\cD}^{\ha} {\cD}_{\ha} q^{i \underline{i}} + \text{fermionic terms}~.
\eea
Upon substituting this, one obtains
\bea
e^{-1}\,\cL|_{bosonic} 
&=&
- \frac{1}{2} \phi^3 {\cR} -\frac{2}{q^2}\phi^3 q_{i \underline{i}} {\cD}^{\ha} {\cD}_{\ha} q^{i \underline{i}}
+ \frac{3}{2} \phi ({\cD}^{\ha} \phi) {\cD}_{\ha} \phi
- \frac{3}{4} \phi X^{ij} X_{ij}
\non\\
&&+\frac{1}{8} \ve_{\ha \hb \hc \hd \he}v^{\ha} f^{\hb \hc} f^{\hd \he} 
+ \frac{3}{4} \phi f^{\ha \hb} f_{\ha \hb}
+ \frac{9}{4} \phi^2 w^{\ha \hb} f_{\ha \hb}
+ \frac{9}{8} \phi^3 w^{\ha \hb} w_{\ha \hb}
~. \label{action-before-fixing}
\eea
In eq.~\eqref{action-before-fixing}, there is still a dependence upon 
the triplet of gauge three-forms 
$b_{\hm \hn \hp}{}^{\underline{i}\underline{j}}$, 
which is hidden in the SU(2)$_R$ connection inside the 
${\cD}_{\ha}$ derivatives. It is straightforward to obtain the analogue expressions in terms of $\mathbf{D}_a$.

As a direct generalisation, coupling to $n$ vector multiplets leads to the following action
\bea
e^{-1}\,\cL|_{bosonic} 
&=& C_{IJK} \bigg(
- \frac{1}{2} \phi^I \phi^J \phi^K {\cR} 
-\frac{2}{q^2}\phi^I \phi^J \phi^K q_{i \underline{i}} {\cD}^{\ha} {\cD}_{\ha} q^{i \underline{i}}
+ \frac{3}{2} \phi^I ({\cD}^{\ha} \phi^J) {\cD}_{\ha} \phi^K \non\\
&&- \frac{3}{4} \phi^I X^{ij\, J} X^{K}_{ij}
+\frac{1}{8} \ve_{\ha \hb \hc \hd \he}v^{\ha\, I} f^{\hb \hc\, J} f^{\hd \he\, K} 
+ \frac{3}{4} \phi^I f^{\ha \hb\, J} f^K_{\ha \hb}\non\\
&&+ \frac{9}{4}\phi^I \phi^J w^{\ha \hb} f^K_{\ha \hb}
+ \frac{9}{8} \phi^I \phi^J \phi^K w^{\ha \hb} w_{\ha \hb}
\bigg)~. 
\eea

The final step to obtain the bosonic sector of the Poincar\'e supergravity action is to impose the set of gauge fixing conditions on \eqref{action-before-fixing} or appropriate generalisations when physical matter multiplets are included. Here we give the gauge-fixed supergravity action in both the string and Einstein frames. 

\subsubsection{String frame}
Upon implementing the constraints \eqref{gauge-conditions}, the resulting BF action
turns out to be
\bea
e^{-1}\,\cL|_{bosonic} 
&=&
- \frac{1}{2} \phi^3 {\cR}
+\frac{1}{8} \ve_{\ha \hb \hc \hd \he}v^{\ha} f^{\hb \hc} f^{\hd \he} 
+ \frac{3}{4} \phi f^{\ha \hb} f_{\ha \hb}
+ \frac{9}{4}  \phi^2 w^{\ha \hb} f_{\ha \hb}
+ \frac{9}{8} \phi^3 w^{\ha \hb} w_{\ha \hb}
\non\\
&&+ \frac{3}{2} \phi (\pa^{\hm} \phi) \pa_{\hm} \phi
+ \frac{1}{4} h^{\ha}{}_{ij}h_{\ha}{}^{ij}
- \frac{3}{4} \phi X^{ij} X_{ij}~.
\label{sugra-action}
\eea
Here
$h_{\ha}{}^{ij}
=
\d^i_{\underline{i}}\d^j_{\underline{j}}
h_{\ha}{}^{\underline{i}\underline{j}}$ 
since we have stopped distinguishing between underlined and 
non-underlined SU(2) indices after gauge fixing. We also stress that $\phi > 0$ as the compensator is nowhere vanishing.

We can further analyse the on-shell structure of 
\eqref{sugra-action}.
It is clear that $w_{\ha \hb}$ and $X^{ij}$ are auxiliary 
fields that can be algebraically integrated out by using the 
equations of motion
\bea
f_{\ha \hb}+ \phi \, w_{\ha \hb}= 0~, \qquad 
X^{ij}=0
~.
\eea
The on-shell Lagrangian then reads
\bea
e^{-1}\,\cL|_{bosonic} 
&=&
- \frac{1}{2} \phi^3 {\cR}
+\frac{1}{8} \ve_{\ha \hb \hc \hd \he}v^{\ha} f^{\hb \hc} f^{\hd \he} 
- \frac{3}{8} \phi f^{\ha \hb} f_{\ha \hb}
\non\\
&&+ \frac{3}{2} \phi (\pa^{\hm} \phi) \pa_{\hm} \phi
+ \frac{1}{4} h^{\ha}{}_{ij}h_{\ha}{}^{ij}
~.
\eea
The first three terms are kinetic terms
for minimal on-shell $\cN=1$ Poincar\'e supergravity
with a dynamical graviton and graviphoton, described in a string frame.
The last two terms describe a dilaton and a triplet of dynamical
gauge three-forms which are not part of the minimal
on-shell $\cN=1$ Poincar\'e supergravity multiplet.

By construction, the supersymmetric $BF$-action  \eqref{BF} is also well defined as an invariant in a
hyper-dilaton Weyl background. We can readily construct an invariant of this form by
considering the off-shell vector multiplet compensator used in this section and an off-shell
linear multiplet given by
\bsubeq\label{composite-linear-xi}
\bea
e^{-1}\cL_{\xi}|_{bosonic}  
= -\frac{1}{4}
\Big(
\phi\, F_{\xi} + G_\xi{}_{ij} X^{ij} + v_a h_\xi{}^{\ha}
\Big)~,
\eea
where we have defined
\bea
&G_\xi{}_{ij}
:=
\xi_{\underline{i}\underline{j}}\,
G_{ij}{}^{\underline{i}\underline{j}}
~,~~~
\varphi_\xi{}_{\hal i}
:=\xi_{\underline{i}\underline{j}}\,
\varphi_{\hal i}{}^{\underline{i}\underline{j}}
~,~~~
\non\\
&
F_\xi
:=
\xi_{\underline{i}\underline{j}}\,
F^{\underline{i}\underline{j}}
~,~~~
b_\xi{}_{\hm \hn \hp}
:=\xi_{\underline{i}\underline{j}}\,
b_{\hm \hn \hp}{}^{\underline{i}\underline{j}}
~,~~~
h_\xi{}^{\ha}
=\xi_{\underline{i}\underline{j}}\,
h^{\ha}{}^{\underline{i}\underline{j}}
~.
\eea
\esubeq
Here 
$G_{ij}{}^{\underline{i}\underline{j}}$,
$\varphi_{\hal i}{}^{\underline{i}\underline{j}}$,
$F^{\underline{i}\underline{j}}$,
$b_{\hm \hn \hp}{}^{\underline{i}\underline{j}}$, 
and
$h^{\ha}{}^{\underline{i}\underline{j}}$
are  fields of the composite triplet of linear multiplets
\eqref{composite-linear} constructed in terms of 
fundamental fields of the hyper-dilaton Weyl multiplet,
while 
$\xi_{\underline{i}\underline{j}}
=
\xi_{\underline{j}\underline{i}}$
is a real triplet of
(structure group invariant) 
constants.
The bosonic part of the resulting Lagrangian is given by
\bea
e^{-1}\cL_{\xi}|_{bosonic}  
&=& -\frac{1}{4}
\xi_{\underline{i}\underline{j}}
\Big(
q_{i}{}^{\underline{i}}q_{j}{}^{\underline{j}}X^{i j} 
- \frac{1}{12} \ve^{mnpqr} b_{mnp}{}^{\underline{i}\underline{j}} f_{{q} {r}} 
\Big) \non\\
&=& -\frac{1}{4}
\xi_{\underline{i}\underline{j}}
\Big(
q_{i}{}^{\underline{i}}q_{j}{}^{\underline{j}}X^{i j} 
+ h^{\hm}{}^{\underline{i}\underline{j}} v_{\hm}
\Big)
~.~~~~~~~~~
\label{Lxi}
\eea
Upon imposing the gauge fixing
\eqref{gauge-conditions}
and adding \eqref{Lxi} into
\eqref{sugra-action},
we get
\bea
e^{-1}\,\cL|_{bosonic}  
&=& 
- \frac{1}{2} \phi^3 {\cR}
+\frac{1}{8} \ve_{\ha \hb \hc \hd \he}v^{\ha} f^{\hb \hc} f^{\hd \he} 
+ \frac{3}{4} \phi f^{\ha \hb} f_{\ha \hb}
+ \frac{9}{4}  \phi^2 w^{\ha \hb} f_{\ha \hb}
+ \frac{9}{8}  \phi^3 w^{\ha \hb} w_{\ha \hb}
\non\\
&&+ \frac{3}{2} \phi (\pa^{\hm} \phi) \pa_{\hm} \phi
+ \frac{1}{4} h^{\ha}{}_{ij}h_{\ha}{}^{ij}
- \frac{3}{4} \phi X^{ij} X_{ij}
\non\\
&&+\frac{1}{48} \xi_{ij} \ve^{\hm \hn \hp {q} {r}} b_{\hm \hn \hp}{}^{ij} f_{{q} {r}} -\frac{1}{4}\xi_{ij}  X^{ij}~,
\label{deformed-L}
\eea
where, after gauge fixing, we have used
$\xi_{ij}=\d_i^{\underline{i}}\d_j^{\underline{j}}
\xi_{\underline{i}\underline{j}}$
and $b_{\hm \hn \hp}{}^{ij}
=
\d^i_{\underline{i}}\d^j_{\underline{j}}
b_{\hm \hn \hp}{}^{\underline{i}\underline{j}}$.
We can then integrate $w_{\ha \hb}$ and $X^{ij}$ out as they are auxiliary fields. 
With the $\xi$-deformation turned on,
the equations of motion obtained from \eqref{deformed-L} become
\bea
f_{\ha \hb}+ \phi \, w_{\ha \hb}=0~, \qquad
\phi \, X^{ij} + \frac{1}{6} \xi^{i j} = 0~.
\eea
This leads to the on-shell Lagrangian
\bea
e^{-1}\,\cL|_{bosonic}  
&=& 
 - \frac{1}{2} \phi^3 {\cR}
+\frac{1}{8} \ve_{\hm \hn \hp {q} {r}}v^{\hm} f^{\hn \hp} f^{{q} {r}} 
- \frac{3}{8} \phi f^{\hm \hn} f_{\hm \hn}
+ \frac{3}{2} \phi (\pa^{\hm} \phi) \pa_{\hm} \phi
+ \frac{1}{4} h^{\hm}{}_{kl}h_{\hm}{}^{kl} 
\non\\
&&+ \frac{1}{24 \phi } \xi^2 +\frac{1}{48} \xi_{ij} \ve^{\hm \hn \hp {q} {r}} b_{\hm \hn \hp}{}^{ij} f_{{q} {r}} 
~,
\eea
where
\be
\xi^2:=\hf \xi^{ij}\xi_{ij}
\geq0
~.
\ee
As a result, we have obtained a non-trivial, negatively defined, potential for the dilaton. The previous Lagrangian admits a constant dilaton, AdS$_5$ vacua.

\subsubsection{Einstein frame}
If we instead adopt the gauge fixing conditions \eqref{gauge-conditions-E} to \eqref{action-before-fixing}, we obtain the following Poincar\'{e} supergravity action
\bea
e^{-1}\,\cL|_{bosonic} 
&=&
- \frac{1}{2} {\cR}
+\frac{1}{8} \ve_{\ha \hb \hc \hd \he}v^{\ha} f^{\hb \hc} f^{\hd \he} 
+ \frac{3}{4} f^{\ha \hb} f_{\ha \hb}
+ \frac{9}{4}  w^{\ha \hb} f_{\ha \hb}
+ \frac{9}{8}  w^{\ha \hb} w_{\ha \hb}
\non\\
&&-2 (\pa^{\hm} U) \pa_{\hm} U
+ \frac{1}{4} \re^{4U} h^{\ha}{}_{ij}h_{\ha}{}^{ij}
- \frac{3}{4} X^{ij} X_{ij}~.
\label{sugra-action-E}
\eea

We can further analyse the on-shell structure of 
\eqref{sugra-action-E}.
It is clear that $w_{\ha \hb}$ and $X^{ij}$ are auxiliary 
fields that can be algebraically integrated out by using the 
equations of motion
\bea
w_{\ha \hb}= - f_{\ha \hb}
~,~~~
X^{ij}=0
~.
\eea
The on-shell Lagrangian then reads
\bea
e^{-1}\,\cL|_{bosonic} 
&=&
- \frac{1}{2} {\cR}
+\frac{1}{8} \ve_{\ha \hb \hc \hd \he}v^{\ha} f^{\hb \hc} f^{\hd \he} 
- \frac{3}{8} f^{\ha \hb} f_{\ha \hb}
\non\\
&&-2 (\pa^{\hm} U) \pa_{\hm} U
+ \frac{1}{4} \re^{4U} h^{\ha}{}_{ij}h_{\ha}{}^{ij}
~.
\eea
The first three terms describe the standard kinetic terms
for minimal on-shell $\cN=1$ Poincar\'e supergravity
with a dynamical graviton and graviphoton.
The last two terms describe a dilaton and a triplet of dynamical
gauge three-forms which are not part of the minimal
on-shell $\cN=1$ Poincar\'e supergravity multiplet.
This is a standard feature of dilaton multiplets, where on-shell a physical dilaton multiplet adds to the degrees of freedom of the multiplet. In the case of hyper-dilaton Poincar\'e supergravity, the extra multiplet is a hypermultiplet where three of the scalars have been dualised to a triplet of  gauge three forms in complete analogy to the 4D $\cN=2$ case of \cite{Muller_hyper:1986ts,Gold:2022bdk}.

Let us now add the second supersymmetric invariant \eqref{Lxi} to the action \eqref{action-before-fixing}. Upon imposing the gauge fixing conditions
\eqref{gauge-conditions-E}, we arrive at
\bea
e^{-1}\,\cL|_{bosonic}  
&=& 
- \frac{1}{2} {\cR}
+\frac{1}{8} \ve_{\ha \hb \hc \hd \he}v^{\ha} f^{\hb \hc} f^{\hd \he} 
+ \frac{3}{4} f^{\ha \hb} f_{\ha \hb}
+ \frac{9}{4}  w^{\ha \hb} f_{\ha \hb}
+ \frac{9}{8}  w^{\ha \hb} w_{\ha \hb}
\non\\
&&-2 (\pa^{\hm} U) \pa_{\hm} U
+ \frac{1}{4} \re^{4U} h^{\ha}{}_{ij}h_{\ha}{}^{ij}
- \frac{3}{4} X^{ij} X_{ij}
\non\\
&&+ \frac{1}{48} \xi_{ij} \ve^{mnpqr} b_{mnp}{}^{ij} f_{qr} - \frac{1}{4}\xi_{ij} \re^{-2U} X^{ij}~.
\label{deformed-L-E}
\eea
We can then integrate $w_{\ha \hb}$ and $X^{ij}$ out as they are auxiliary fields. 
With the $\xi$-deformation turned on,
the equations of motion obtained from \eqref{deformed-L-E} become
\bea
w_{\ha \hb}= -f_{\ha \hb}
~,~~~
X^{ij}= -\frac{1}{6} \xi^{i j} \re^{-2U}
~.
\eea
This leads to the on-shell Lagrangian
\bea
e^{-1}\,\cL|_{bosonic}  
&=& 
 - \frac{1}{2} {\cR}
+\frac{1}{8} \ve_{mnpqr}v^{\hm} f^{\hn \hp} f^{qr} 
- \frac{3}{8} f^{\hm \hn} f_{\hm \hn}
-2 (\pa^{\hm} U) \pa_{\hm} U
+ \frac{1}{4} \re^{4U} h^{\hm}{}_{kl}h_{\hm}{}^{kl} 
\non\\
&&+ \frac{1}{24} \re^{-4U} \xi^2 +\frac{1}{48} \xi_{ij} \ve^{\hm \hn \hp qr} b_{\hm \hn \hp}{}^{ij} f_{qr} 
~.
\eea

\section{Superconformal multiplets in 6D $\cN=(1,0)$ superspace}
\label{6DMultipletSuperspace}

The following section reviews the relevant details of various superconformal multiplets required in this work. We first describe the 6D $\cN=(1,0)$ standard Weyl multiplet of conformal supergravity in superspace before moving on to the discussion of matter multiplets: the on-shell hypermultiplet and linear  multiplets. Here we make use of the conformal superspace formulation in the traceless frame \cite{BNT-M17} and results from \cite{Butter:2018wss}. We also refer the reader to the following list of papers for other work
on flat superspace and multiplets in six dimensions \cite{Koller:1982cs,Howe:1983fr,Kugo:1982bn,oweAR,DragonNV,GrundbergLindstrom_6D,GatesPenatiTartaglino_6D} while 
see also
\cite{BreitenlohnerRQ,GatesQV,GatesJV,SmithWAA,AwadaER,BergshoeffSU,BergshoeffRB,Sokatchev:1988aa,Linch:2012zh}
for alternative curved superspace approaches to describe supergravity multiplets in six dimensions.

\subsection{The standard Weyl multiplet}
\label{6DSWM-superspace}
The standard Weyl multiplet of 6D $\cN=(1,0)$ conformal supergravity \cite{Bergshoeff:1985mz} contains $40+40$ physical components, and is associated with the gauging of the superconformal algebra ${\rm OSp}(6,2|1)$. Associated respectively with local translations, $Q$-supersymmetry, ${\rm SU(2)}_R$ and dilatations are the vielbein $e_\hm{}^\ha$,
the gravitino $\psi_\hm{}_\hal^i$,
the ${\rm SU(2)_R}$ gauge field $\phi_\hm{}^{ij}$, and a dilatation gauge field $b_\hm$. There are three composite connections which are associated with the remaining gauge symmetries: these are
the spin connection  $\omega_\hm{}^{\ha\hb}$, the $S$-supersymmetry connection
$\phi_\hm{}_\hal^i$, and the special conformal connection
$\mathfrak{f}_\hm{}^\ha$ --- they are algebraically determined in terms of the other fields by imposing constraints on some of the
curvature tensors. To achieve an off-shell representation, one needs to introduce three covariant auxiliary fields: an anti-self-dual tensor $T^-_{abc}$; a real scalar field $D$; and a chiral fermion $\chi^{\a i}$. In this subsection, we review how to embed this in conformal superspace \cite{BKNT16,BNT-M17}. The component structure of the multiplet is given in section \ref{6DCSWM}. 

The 6D $\cN=(1,0)$ conformal superspace is parametrised by
local bosonic $(x^{m})$ and fermionic $(\theta^\mu_i)$ coordinates 
$z^{M} = (x^{m},\q^{{\mu}}_i)$, 
where $m = 0, 1,2,3,4, 5$, $\mu = 1,2,3, 4$ and $i = 1, 2$.
By gauging the
full 6D $\cN=(1,0)$ superconformal algebra in superspace, we introduce
covariant derivatives $ {\nabla}_{A} = (\nabla_{a} , \nabla_{\a}^i)$ which take the form
\bea
\nabla_A
= E_{A} - \o_{A}{}^{\underline{b}}X_{\underline{b}} &=& E_A - \hf  {\Omega}_A{}^{ab} M_{ab} - \Phi_A{}^{ij} J_{ij} - B_A \mathbb D
	-  {\mathfrak{F}}_A{}_B K^B \ , \\ 
	&=& E_A - \hf  {\Omega}_A{}^{ab} M_{ab} - \Phi_A{}^{ij} J_{ij} - B_A \mathbb D
	-  {\mathfrak{F}}_A{}_\a^i S^\a_i - {\mathfrak{F}}_A{}^a K_a \ .
\eea
Here $E_A = E_A{}^M\partial_M$ is the inverse super-vielbein (which plays a role of a connection for local super-translations), $M_{a b}$ are the Lorentz generators, $J_{ij}$ are generators of the
${\rm SU(2)}_R$ $R$-symmetry group,
$\mathbb D$ is the dilatation generator and $K^{A} = (K^{a}, S^{\a}_{i})$ are the special superconformal
generators.\footnote{Note here the change in the SU$(2)_R$ index structure of the 6D $S$-supersymmetry generator, $S^\a_i$, relative to the $5D$ case where we it was originally introduced as $S_{\a i}$. Though this difference might seem unnatural, and introduce minus signs in similar expressions in 5D and 6D, we decided to keep adhering to the notations used in \cite{BKNT-M14,Butter:2018wss}.} The super-vielbein one-form is given by  $E^{A} =\rd z^{M} E_{M}{}^{A}$ and satisfies $E_{M}{}^{A} E_{A}{}^{N} =\d_{M}^{N}$,
     $E_{A}{}^{M} E_{M}{}^{B}=\d_{A}^{B}$. Associated with each structure group generator $X_{\underline{a}} = (M_{ a b},J_{ij},\bbD, S^{\a}_{i}, K_a)$ there is a connection superfield one-form given by
$\omega^{\underline{a}} = (\O^{a b},\F^{ij},B,\mathfrak{F}_{\a}^{i},\mathfrak{F}^{a})= \rd z^M \omega_M{}^{\underline{a}} = E^{A} \o_{A}{}^{\underline{a}}$.

To describe the standard 6D $\cN=(1,0)$ Weyl multiplet in conformal superspace, the algebra of covariant derivatives 
\begin{align}
[ \nabla_\hA , \nabla_\hB \}
	&= -\mathscr{T}_{\hA\hB}{}^\hC \nabla_\hC
	- \frac{1}{2} {\mathscr{R}(M)}_{\hA\hB}{}^{\hc\hd} M_{\hc\hd}
	- {\mathscr{R}(J)}_{\hA\hB}{}^{kl} J_{kl}
	\non \\ & \quad
	- {\mathscr{R}(\mathbb{D})}_{\hA\hB} \mathbb D
	- {\mathscr{R}(S)}_{\hA\hB}{}_{\hga}^{ k} S^{\hga}_{ k}
	- {\mathscr{R}(K)}_{\hA\hB}{}^\hc K_\hc~,
	\label{6Dnablanabla}
\end{align}
is constrained to be completely determined in terms of the symmetric super-Weyl tensor superfield $W^{\a\b}$, which is a superconformal primary with conformal dimension one
\be
W^{\a \b} = W^{\b\a} \ , \quad
K^AW^{\a\b} = 0 \ ,
 \quad \mathbb D W^{\a\b} = W^{\a\b} \ ,
\ee
obeying the Bianchi identities
\bsubeq\label{6DWBI}
\bea
\nabla_\a^{(i} \nabla_{\b}^{j)} W^{\g\d} &=& - \d^{(\g}_{[\a} \nabla_{\b]}^{(i} \nabla_{\r}^{j)} W^{\d) \r} \ , \\
\nabla_\a^k \nabla_{\g k} W^{\b\g} - \frac{1}{4} \d^\b_\a \nabla_\g^k \nabla_{\d k} W^{\g\d}
 &=& 8 \ri \nabla_{\a \g} W^{\g \b} \ .
\eea
\esubeq
The relation $W^{\a\b}=1/6(\tilde{\g}^{abc})^{\a\b}W_{abc}$ means that the super-Weyl tensor $W^{\a\b}$ is equivalent to an anti-self-dual rank-3 tensor superfield $W_{abc}$. In  \eqref{6Dnablanabla}
$ \mathscr{T}_{\hA \hB}{}^C$ is the torsion curvature, and $ \mathscr{R}(M)_{\hA \hB}{}^{\hc \hd}$,
$ \mathscr{R}(J)_{\hA \hB}{}^{kl}$, $ \mathscr{R}(\mathbb D)_{\hA \hB}$, $ \mathscr{R}(S)_{\hA \hB}{}_{\g}^{k}$, and $ \mathscr{R}(K)_{\hA \hB}{}^{\hc}$
are the curvatures associated with Lorentz, ${\rm SU(2)}_R$,
dilatation, $S$-supersymmetry, and special conformal
boosts, respectively. Their expressions in terms of the super-Weyl tensor $W^{\a\b}$ and its descendant superfields of
dimension 3/2
\begin{align}
X^{\a i} := - \frac{\ri}{10} \nabla_{\b}^i W^{\a\b}
~,\quad
X_\g^k{}^{\a\b} :=
- \frac{\ri}{4} \nabla_\g^k W^{\a\b} - \d^{(\a}_{\g} X^{\b) k}
~,
\label{6DXfields}
\end{align}
and of dimension 2
\bsubeq \label{6DYfielDs}
\begin{align}
Y_\a{}^\b{}^{ij} &:= - \frac{5}{2} \Big( \nabla_\a^{(i} X^{\b j)} - \frac{1}{4} \d^\b_\a \nabla_\g^{(i} X^{\g j)} \Big)
= - \frac{5}{2} \nabla_\a^{(i} X^{\b j)} \ , \\
Y &:= \frac{1}{4} \nabla_\g^k X^\g_k \ , \\
Y_{\a\b}{}^{\g\d} &:=
\nabla_{(\a}^k X_{\b) k}{}^{\g\d}
- \frac{1}{6} \d_\b^{(\g} \nabla_\r^k X_{\a k}{}^{\d) \r}
- \frac{1}{6} \d_\a^{(\g} \nabla_\r^k X_{\b k}{}^{\d) \r} \ ,
\end{align}
\esubeq
are given in appendix \ref{6Dconformal-identities}. Just like the 5D case, we consider the superspace and component structures for 6D corresponding to the ``traceless" choice of conventional constraints, which was first considered in \cite{BNT-M17}. The component and superspace structures are summarised in section \ref{6DCSWM} and appendix \ref{6Dconformal-identities}.

The superfields
$X^{\a i}$,
$X_\g^k{}^{\a\b}$,
$Y_\a{}^\b{}^{ij}$,
$Y$, and $Y_{\a\b}{}^{\g\d}$ are the only independent descendants of $W^{\a\b}$. All the other higher dimension descendants obtained by the action of spinor derivatives on $W^{\a\b}$ are  vector derivatives of these independent fields as a result of the non-trivial Bianchi identities \eqref{6DWBI}. Eq.~\eqref{6DS-on-X_Y-a} gives the action of the $S$-generators on these independent descendants that prove to  all be annihilated by $K^{a}$.

The conformal supergravity gauge group $\cG$ is generated by
{\it covariant general coordinate transformations},
$\delta_{\rm cgct}$, associated with a local superdiffeomorphism parameter $\xi^{\hA}$ and
{\it standard superconformal transformations},
$\delta_{\cH}$, associated with the following local superfield parameters:
the dilatation $\s$, Lorentz $\L^{\ha \hb}=-\L^{\hb \ha}$,  ${\rm SU(2)}_R$ $\L^{ij}=\L^{ji}$,
, and special conformal transformations $\L_A=(\eta_\a^i, \L_a)$.
The covariant derivatives transform as
\bea
\d_\cG \nabla_A &=& [\cK , \nabla_A] \ ,
\label{6DTransCD}
\eea
where $\cK$ denotes the first-order differential operator
\bea\label{transformation_6D_generator}
\cK = \xi^C  {\nabla}_C + \hf  {\L}^{ab} M_{ab} +  {\L}^{ij} J_{ij} +  \s \mathbb D +  {\L}_A K^A ~.
\eea
A covariant (or tensor) superfield $U$ transforms as
\be
\d_{\cG} U =
(\d_{\rm cgct}
+\d_{\cH}) U =
 \cK U
 ~.
\ee
The superfield $U$ is said to
be \emph{superconformal primary} and of dimension $\D$ if $K_A U = 0$ and $\mathbb D U = \D U$.

\subsection{The on-shell hypermultiplet}
Analogously to the 5D case, our starting point is the on-shell realisation for the 6D $\cN=(1,0)$ hypermultiplet with $4+4$ degrees of freedom. In conformal superspace, it is described by a Lorentz scalar superfield $q^{i\underline{i}}$ subject to the constraint
\bea
\de_{\hal}^{(i} q^{j) \underline{j}} = 0~.
\label{6Donshell-sspace}
\eea
Here, the index $\underline{i}=\underline{1},\underline{2}$ denotes an SU(2) flavour 
index. The superfield $q^{i \underline{i}}$ is a Lorentz scalar superconformal primary,
\bea
M_{\ha \hb} q^{i \underline{i}} =0~, \qquad S^\a
_{ j} q^{i \underline{i}} = K_a q^{i \underline{i}} =0~, \qquad J^{kl} q^{i \underline{i}} = \epsilon^{i(k} q^{l) \underline{i}} ~. \label{6Dhyper-primary}
\eea
Eqs.~\eqref{6Donshell-sspace}, \eqref{6Dhyper-primary}, and the relation \eqref{6Dalg-S-spinor} tell us that $ \mathbb{D} q^{i \underline{i}} = 2 q^{i \underline{i}}$. The only independent descent superfield of $q^{i \underline{i}}$ is a dimension 5/2 spinor superfield
\bea
\rho_{\hal}^{\underline{k}}:= \de_{\hal}^{j}q_{j}{}^{\underline{k}}~.
\eea 
Equation \eqref{6Donshell-sspace} can now equivalently be written in terms of this spinor superfield 
\bea
\de_{\hal}^{i}q^{k \underline{k}} = -\hf \ve^{ik} \rho_{\hal}^{\underline{k}}~. \label{6Donshell-sspace-1}
\eea
Applying spinor derivatives on these independent fields leads to several implications of the (anti-)commutation relations \eqref{6Dnew-frame_spinor-spinor}, \eqref{6Dnew-vectspinor}, along with the constraints \eqref{6Dhyper-primary}, and \eqref{6Donshell-sspace-1}: 
\bea\label{6Dtwospinors-on-q}
\de_{\hal}^i \de_{\hbe}^j q_{j}{}^{\underline{i}} &=& \de_{\hal}^i \rho_{\hbe}^{\underline{i}} = -4 \ri \de_{\hal \hbe} q^{i \underline{i}}~,
\eea
with $\de_{\hal \hbe}:= (\g^{\ha})_{\hal \hbe} \de_{\ha}$. 
Next, we shall consider 
\bea
\e^{\g\d\a\b}\{ \de_{\a}^k, \de_{\b k} \} \rho_{\d}^{\underline{i}}
&=&  8 \ri \tilde{\de}^{\g\d} \rho_{\d}^{\underline{i}} 
-24 \ri W^{\g\d} \rho_{\d}^{\underline{i}} 
+ 288 X^{\g k} q_{k}{}^{\underline{i}}~,
\label{6Ddd-rho-right}
\eea
here $\tilde{\de}^{\hal \hbe}:= (\tilde{\g}^{\ha})^{\hal \hbe} \de_{\ha}$ and we have made use of \eqref{6Dnew-frame_spinor-spinor} and the $S$-supersymmetry transformation
\bea\label{6DS-rho}
S^{\g}_{ i} \rho_{\hbe}^{\underline{i}} = 16 \d_{\b}^{ \g} q_{i}{}^{\underline{i}} \qquad \Longrightarrow \qquad K_{\ha}\, \rho_{\hbe}^{\underline{i}} = 0~.
\eea
On the other hand, by virtue of \eqref{6Dtwospinors-on-q}, we also have that
\bea
\e^{\g\d\a\b}\{ \de_{\a}^k, \de_{\b k} \} \rho_{\d}^{\underline{i}} &=& 
16 \ri (\tilde{\g}^{\a})^{\a \g} \de_\a^i \de_a q_i{}^{\underline{i}}
\non\\
&=& 16 \ri (\tilde{\g}^{\a})^{\a \g} [\de_\a^i, \de_a] q_i{}^{\underline{i}} +16 \ri (\tilde{\g}^{\a})^{\a \g} \de_a \rho_\a{}^{\underline{i}}~. \label{6Ddd-rho-left}
\eea
Applying the commutation relation \eqref{6Dnew-vectspinor}, we can then equate \eqref{6Ddd-rho-right} and \eqref{6Ddd-rho-left} to obtain
\bea
(\tilde{\g}^{a}\nabla_{a}
\rho^{\underline{i}}\,
)^{\a}
&=&
-W^{\a\b} \rho_\b^{\underline{i}}
-4 \ri  \,  {X}^{\a k}q_{k}{}^{\underline{i}}~. \label{6Dvect-rho-sspace}
\eea
We can then hit both sides of \eqref{6Dvect-rho-sspace} with $\de_{\hal}^i$ and make use of \eqref{6Dtwospinors-on-q}, \eqref{6Dnew-vectspinor}, and the identity \eqref{6Deq:Wdervs}. This results in the equation 
\bea\label{6DBox-q1}
\Box q^{i \underline{i}} 
&=&
\frac{1}{2}X^{i}\rho^{\underline{i}} 
-\frac{1}{2} Y \,q^{i\underline{i}}~, \qquad \Box := \nabla^{a}\nabla_{a}~.
\eea

The local superconformal $\d= \d_{Q} + \d_{\cH}$ transformations (except translation, i.e. $\xi^a = 0$) of the covariant superfields $q^{i \underline{i}}$ and $\r_{\a}^{\underline{i}}$ can be derived using \eqref{transformation_6D_generator} and the relations \eqref{6Donshell-sspace-1}, \eqref{6Dtwospinors-on-q}, and \eqref{6DS-rho}. This leads to
\bsubeq \label{6Dd-hyper-sspace}
\bea
\delta q^{i\underline{i}} 
&=& 
\frac{1}{2}\xi^{ i}\rho^{\underline{i}}
+\L^{i}{}_{k}q^{k\underline{i}} 
+ 2 \s q^{i\underline{i}}
~,
\\
\delta \rho_{\hal}^{\underline{i}} 
&=&
-4\ri( {\xi}_i \g^{\ha})_{\hal} 
\nabla_{\ha} q^{i\underline{i}}
 -\frac{1}{4} \L_{\ha \hb} (\g^{\ha \hb} \rho^{\underline{i}})_{\hal}    
+\frac{5}{2} \s \rho_{\hal}^{\underline{i}} 
+16 \eta_{\hal}^i q_i{}^{\underline{i}}~.
\eea
\esubeq

\subsection{The $\cO(2)$ multiplet}

The $6D$ linear multiplet, or $\cO(2)$ multiplet can be described in terms of an SU$(2)_R$ triplet of Lorentz scalar superfields $L^{ij}$, with $(L^{ij})^{*}= \ve_{ik} \ve_{jl} L^{kl}$ and satisfies the defining constraint
\bea
\de_{\a}^{(i} L^{jk)} = 0~.
\eea
Here $L^{ij}$ is a superconformal primary dimension-4 superfield,
\bea
S^{\g}_{k} L^{ij}= K^a L^{ij}=0~, \qquad \mathbb{D} L^{ij} = 4 L^{ij}~, \qquad J^{ij} L^{kl} = \e^{k(i} L^{j)l} + \e^{l(i} L^{j)k}~.
\eea
The tower of component fields of the superfield $L^{ij}$ is given by the following set of useful identities:
\bsubeq
\begin{align}
\nabla_\hal^i L^{jk} &= - 2 \eps^{i(j} \varphi_\a^{k)} \ , \\
\nabla_\hal^i \varphi_\hbe^j &= -  \frac{\ri}{2} \eps^{ij} H_{\hal\hbe} - \ri \nabla_{\hal\hbe} L^{ij} \ ,  \\
\nabla_\g^k H_{\a\b} &= -8 \de_{\g[\a} \varphi_{\b]}^k - 2 \de_{\a\b} \varphi_\g^k + 2 \e_{\a\b\g\d} W^{\d\r} \varphi_\r^k \ ,
\end{align}
\esubeq
where we have defined the independent descendant superfields
\bsubeq \label{6DO2superfieldComps}
\begin{align}
\varphi_\hal^i &:= - \frac{1}{3} \nabla_{\hal j} L^{ij} \ , \\
H_a &:= -\frac{\ri}{4} (\tilde{\g}_a)^{\a\b} \nabla_{\a}^k \varphi_{\b k} \ , \qquad
H_{\a\b} := (\g^a)_{\a\b} H_a \ .
\end{align}
\esubeq
We will be using these results later when analysing the component structure of the multiplet. Further, it can be checked that $H^{a}$ obeys the differential condition 
\be 
\nabla_\ha H^\ha = 0~, \qquad {H}^{\ha}:= \frac{1}{5!}\ve^{a b c d e f} H_{b c d e f}~.
\ee
The descendants \eqref{6DO2superfieldComps} prove to be annihilated by $K_a$ and to satisfy
\bsubeq
\begin{align} 
S^\a_i \varphi_\b^j &= 8 \d^\a_\b L^{ij} \ , \\
S^\g_k H_{\a\b} &= -40 \ri \d^\g_{[\a} \varphi_{\b]k} \ .
\end{align}
\esubeq
We refer the reader to \cite{Butter:2018wss} for a superform description of the $\cO(2)$ multiplet.

\section{The standard Weyl multiplet in components}
\label{6DCSWM}
Similar to the 5D case, we begin by identifying the various component fields of the 6D $\cN=(1,0)$ standard Weyl multiplet \cite{Bergshoeff:1985mz} within the geometry of conformal superspace. The vielbein ($e_m{}^a$) and gravitino ($\psi_m{}^\a_i$) are identified with the coefficients of $\rd x^\hm$ of the super-vielbein $E^\hA = (E^\ha, E^\hal_i) = \rd z^\hM\, E_\hM{}^\hA$,
\bea
e_{\hm}{}^{\ha} (x):= E_{\hm}{}^{\ha}(z)|~, \qquad \psi_{\hm}{}^{i}_{ \hal}(x) := 2 E_{\hm}{}^{i}_{\hal} (z)|~.
\eea  
In a local coordinate independent way they are given by
\begin{align}
e{}^a = \rd x^m e_m{}^a = E^a\doubar~,~~~~~~
\psi^\a_i = \rd x^m \psi_m{}^\a_i =  2 E^\a_i \doubar ~.
\end{align}
Similar to the 5D case, look at \eqref{singlebar} and \eqref{doublebar}, the single bar denotes setting $\q=0$ and the double-bar denotes setting $\q = \rd \q = 0$. Analogously, the remaining fundamental and composite one-forms correspond to
double-bar projections of superspace one-forms,
\begin{align}
	\phi^{kl} := \Phi{}^{kl} \doubar~, \quad
	b := B\doubar ~, \quad
\omega^{cd} := \Omega{}^{cd} \doubar ~, \quad  
\phi_\g^k := 2 \,\mathfrak F{}_\g^k\doubar~, \quad
\mathfrak{f}{}_c := \mathfrak{F}{}_c\doubar~. 
\end{align}
The covariant matter fields are contained within the super-Weyl tensor $W_{abc}$
and its independent descendants,
\bsubeq
\bea
T^-_{abc} &:=& -2 W_{abc}\loco \ ,
\\
\chi^{\a i} &:=& \frac{15}{2}X^{\a i} \loco=-\frac{3\ri}{4}\de^i_\b W^{\a\b}\loco~,
\\
D&:=&
\frac{15}{2}Y \loco=
-\frac{3\ri}{16}\de_\a^k\de_{\b k}W^{\a\b}\loco
~.
\eea
\esubeq
The lowest components of the
other nontrivial descendants of $W^{\a\b}$, specifically $X_{\a}^i{}^{\b\g}|$, $Y_{\a}{}^{\b}{}^{kl}|$
and $Y_{\a\b}{}^{\g\d}|$,
prove to be directly related to component curvatures and hence
are composite fields.

The component gauge connections can now be used to define the locally superconformal covariant derivative $\de_a$, which coincide with the bar projection of the conformal superspace covariant derivative $\de_a|$   
\bea
e_m{}^a{\de}_{a} = \pa_{m}
- \frac{1}{2} \psi_{m}{}^{\a}_i \de_{\a}^i| 
- \frac{1}{2}  {\o}_{m}{}^{c d} M_{c d}
-  \phi_{m}{}^{ij} J_{ij}
-  {b}_{m} \mathbb{D}
- \frac{1}{2}  {\phi}_{m}{}^{ i}_{\a} S^{\a}_{i}
-  {{\mathfrak f}}_{m}{}_{a} K^{a}~.
\eea
This satisfies the algebra
\bea 
[\de_a,\de_b] &=& - R(P)_{ab}{}^{c} \de_c - R(Q)_{ab}{}^\a_i Q_\a^i - \frac{1}{2} R(M)_{ab}{}^{cd} M_{cd} - R(J)_{ab}{}^{kl} J_{kl}\non \\ && - R(\mathbb{D})_{ab} \mathbb{D} - R(S)_{ab}{}_\g^i S^\g_i -R(K)_{ab}{}^c K_c ~.
\eea 
where we identified $R(P)_{\ha \hb}{}^{\hc} = \scT_{\ha \hb}{}^\hc \loco$, 
$R(Q)_{\ha \hb}{}_\hal^i = \scT_{\ha\hb}{}_\hal^i \loco$,
while
$ {R}(M)_{\ha \hb}{}^{\hc \hd}$, $ {R}(J)_{\ha \hb}{}^{ij}$, $ {R}(\mathbb D)_{\ha \hb}$, $ {R}(S)_{\ha \hb}{}_{\hga}^{k}$, and $ {R}(K)_{\ha \hb}{}^{\hc}$ are coinciding with the lowest components of the corresponding superspace curvature tensors given in appendix \ref{6Dconformal-identities}.

The constraints on the superspace curvatures determine the  supercovariantised component curvature by taking the double-bar projection of the superspace two forms. This leads to (see \cite{BNT-M17} for details)
\bsubeq\label{6Dcomp-curv_2-1}
\bea
 {R}(P)_{ab}{}^c
&=&
2 \,e_\ha{}^\hm e_\hb{}^\hn \cD_{[\hm} e_{\hn]}{}^\hc
	+ \frac{\ri}{2} \psi_{[\ha  j} \g^\hc \psi_{\hb]}{}^j
~,
\label{6Dcomp-curv_2-1-a}
\\
%%%%%%
 {R}(Q)_{ab}{}_k
&=&
\frac{1}{2} {\Psi}_{ab}{}_k
+\ri \tilde{\g}_{[a} {\phi}_{b]}{}_{k}
+\frac{1}{24}T^-_{cde}\tilde{\g}^{cde}\g_{[a}\psi_{b]}{}_k
~,
\label{6Dcomp-curv_2-1-b}
\\
%%%%%%
 {R}(\mathbb D)_{ab}
&=&
2e_a{}^me_b{}^n\pa_{[m}b_{n]}
 +4  {\mathfrak{f}}_{[a}{}_{b]}
+\psi_{[a}{}^{ i}  {\phi}_{b]}{}_{ i}
+\frac{\ri}{15} \psi_{[a}{}^j\g_{b]} \chi_{ j}
~,
\label{6Dcomp-curv_2-1-c}
\\
%%%%%%
 {R}(M)_{ab}{}^{cd}
&=&
\cR_{ab}{}^{cd}( {\o})
+8 \d_{[a}^{[c}  {\mathfrak{f}}_{b]}{}^{d]}
+\ri\psi_{[a}{}_j\g_{b]} {R}(Q)^{cd}{}^j
+2\ri\psi_{[a}{}_j\g^{[c} {R}(Q)_{b]}{}^{d]}{}^{ j}
\non\\
&&
-\psi_{[a}{}_j  \g^{cd} {\f}_{b]}{}^j
-\frac{2\ri}{15} \d_{[a}^{[c}\psi_{b]}{}_j \g^{d]}\chi^{j}
+\frac{\ri}{2} \psi_{[a}{}^{ j}\g^e\psi_{b]}{}_j\,T^-_{e}{}^{cd}
~,
\label{6Dcomp-curv_2-1-d}
\\
%%%%%%
 {R}(J)_{ab}{}^{kl}
&=&
\cR_{ab}{}^{kl}(\phi)
+4\psi_{[a}{}^{ (k}  {\f}_{b]}{}^{l)}
+\frac{4\ri}{15} \psi_{[a}{}^{ (k}\g_{b]} \chi^{ l)}
~,
\label{6Dcomp-curv_2-1-e}
\eea
\esubeq
where ${\cD}_m $ is the spin, dilatation, and ${\rm SU}(2)_R$ covariant derivative 
\be \label{6DhatDder}
 {\cD}_m = \partial_m
- \hf  \omega_m{}^{bc} M_{bc}
- b_m \bbD
- \phi_m{}^{ij} J_{ij}
\ , \quad
 {\cD}_a = e_a{}^m  {\cD}_m \ ,
\ee
along with the field strengths and curvature
\bsubeq
\bea
 {\Psi}_{ab}{}^\g_k
&:=&
2e_a{}^me_b{}^n {\cD}_{[m}\psi_{n]}{}^\g_k
~,
\\
 {\cR}_{ab}{}^{cd}
&:=&
 {\cR}_{ab}{}^{cd}(\o)
=
e_a{}^me_b{}^n\Big(
2\pa_{[m}\o_{n]}{}^{cd}
-2\o_{[m}{}^{ce} \o_{n]}{}_e{}^d
\Big)
~,
\label{6DRomega}
\\
 {\cR}_{ab}{}^{kl}
&:=&
 {\cR}_{ab}{}^{kl}(\phi)
=
e_a{}^me_b{}^n\Big(
2\pa_{[m}\phi_{n]}{}^{kl}
+2\phi_{[m}{}^{p(k} \phi_{n]}{}_p{}^{l)}
\Big)
~.
\eea
\esubeq
The component curvatures turn out to obey ``traceless'' conventional constraints \cite{BNT-M17}
\begin{align} \label{6Dconventional constraints}
R(P)_{ab}{}^{c} =& 0~, \qquad
\g^b R(Q)_{ab i} = 0~, \qquad
R(M)_{ab}{}^{cb} = 0~.
\end{align}
The constraints \eqref{6Dconventional constraints} can be solved for the composite connections as follows:
\bsubeq\label{6Dcomposite connection}
\bea
 {\o}_{a}{}_{bc}
&=&
\omega(e)_{a}{}_{bc}
-2\eta_{a[b}b_{c]}
-\frac{\ri}{4}\psi_b{}^k\g_a\psi_c{}_k
-\frac{\ri}{2}\psi_a{}^k\g_{[b}\psi_{c]}{}_k
~,
\\
%%%
 {\phi}_m{}^{k}
&=&
\frac{\ri}{16}\Big(
\g^{bc}\g_m-\frac{3}{5}\g_m\tilde{\g}^{bc}
\Big)
\Big( 
 {\Psi}_{bc}{}^{k}
+\frac{1}{12}T^-_{def}\tilde{\g}^{def}\g_{[b}\psi_{c]}{}^{k}
\Big)
~,~~~~~~
\label{6Deq:SConn-new}
\\
%%%
 {\mathfrak{f}}_{a}{}^{b}
&=&
-\frac{1}{8}\cR_{a}{}^{b}( {\o})
+\frac{1}{80}\d_{a}^{b}\cR( {\o})
+\frac{1}{8}\psi_{[a}{}_j  \g^{bc} {\f}_{c]}{}^j
-\frac{1}{80}\d_{a}^{b}\psi_{c}{}_j  \g^{cd} {\f}_{d}{}^j
 \non\\
&&
+\frac{\ri}{16}\psi_{c}{}_j\g_{a} {R}(Q)^{bc}{}^j+\frac{\ri}{8}\psi_{c}{}_j\g^{[b} {R}(Q)_{a}{}^{c]}{}^{ j}
+\frac{\ri}{60} \psi_{a}{}_j\g^b\chi^j
\non\\
&&
+\frac{\ri}{16}\psi_{a}{}^{ j}\g_c\psi_{d}{}_j\,T^-{}^{bcd}
-\frac{\ri}{160}\d_{a}^{b} \psi_{c}{}^{ j}\g_d\psi_{e}{}_j\,T^-{}^{cde}
~,
\label{6Deq:KConn-new}
\eea
\esubeq
where $\omega(e)_{a}{}_{bc} = -\frac{1}{2} (\cC_{abc} + \cC_{cab} - \cC_{bca})$
is the torsion-free spin connection given
in terms of the anholonomy coefficient
$\cC_{mn}{}^a := 2 \,\partial_{[m} e_{n]}{}^a$,  $\cR(\o)={\cR}_{ab}{}^{ab}(\o)$ is  the scalar curvature and $\cR_a{}^b(\o)={\cR}_{ac}{}^{bc}(\o)$ is the Ricci curvature. It is important to emphasise that in the traceless frame the composite connection $\phi_m{}^k$ does neither depend on $\chi$ and nor on $D$, however, the composite connection $ {\mathfrak{f}}_{a}{}^{b}$ has a dependence on $\chi$.

The supersymmetry transformations of the fundamental gauge connections of the Weyl
multiplet can be derived directly from the transformations of their corresponding
superspace one-forms, see \cite{BNT-M17} for details. In components, the local superconformal transformations, except covariant general coordinate transformations, are identified by the following operator $\delta$
\bea
\delta
= 
\xi^{\a}_iQ_{\a}^i
+\hf \l^{a b}M_{a b} 
+ \l^{ij}J_{ij} 
+ \l_{\mathbb{D}}\mathbb{D} 
+ \l_{a} K^{a}
+ \eta_{\a}^{i} S^{\a}_{i}
~.
\eea
Here, the local component parameters are respectively defined as the $\q=0$ components of the corresponding superfield parameters, $\xi^{\hal}_i := \xi^{\hal}_i \loco$, $\l^{\ha \hb} := \L^{\ha \hb} \loco$, $\l^{ij} := \L^{ij} \loco$, $\l_{\mathbb{D}} := \sigma \loco$, $\l^{\ha} := \L^{\ha} \loco$, and $\eta_{\hal}^{i} := \eta_{\hal}^{i} \loco$.
The local superconformal transformations of 
the independent fields of the standard Weyl multiplet
are given by
\bsubeq\label{6Dtransf-standard-Weyl}
\bea
\delta e_{m}{}^{a} 
&=& 
-\ri\,\xi_i\g^{a}{\psi}_{m}{}^{ i}
+ \lambda^{a}{}_{b}e_{m}{}^{b}
- \lambda_{\mathbb{D}}{e}_{m}{}^{a}
~,
\label{6Dd-vielbein}
\\
%%%%%%
\delta \psi_{m}{}^{\a}_i  
&=& 2\mathcal{D}_{m} \xi^{\a}_i  
-\frac{1}{12} (\xi_i \g_m \tilde{\g}^{abc})^{\a}  T^{-}_{abc} 
+ \frac{1}{4} \lambda^{ab}(\psi_m{}_i\g_{ab})^\a
\non \\  && - \lambda_{i}{}^{j} \,\psi_{m}{}^{\a}_{j}
- \frac{1}{2}\lambda_{\mathbb{D}}\,{\psi}_{m}{}^{\a}_i
-2\ri (\eta_{ i}\tilde{\g}_{m})^{\a }
~,
\label{6Dd-gravitino}
 \\
  \delta  \phi_{m}{}^{k l} &=& 
 -4 \xi^{(k} \phi_{m}{}^{l)} 
 -\frac{4\ri}{15}  \xi^{(k} \g_{m} \chi^{l)}
 +\pa_{m}\l^{k l}
 -2 \phi_{m}{}^{(k}{}_{i} \l^{l) i}
  -4 \eta^{(k} \psi_{m}{}^{l)}
  ~,
 \label{6Dd-SU2}
\\
\delta b_{m}
&=& 
\frac{\ri}{15}\, \xi_i \g_{m} \chi^{i} 
+\xi_i   \phi_m{}^i
+\partial_{m} \lambda_{\mathbb{D}} 
-\psi_{m}{}^{ i}\eta_{ i}  
- 2\lambda_{m}
~,
\label{6Dd-dilatation}
%%%%%
\\
  \delta T^{-}_{a b c} &=& 
  -\frac{\ri}{8} \xi^{ k} \g^{e f} \g_{abc} R(Q)_{e f}{}_{k} 
  -\frac{2 \ri}{15} \xi_i \g_{abc} \chi^{i}
 - 3 \l^e{}_{[a} T^{-}_{b c] e}
 + \l_{\mathbb{D}} T^{-}_{a b c}~,
   \label{6Dd-W} 
 %%%%%%%
 \\
 \delta  \chi^{\a i}
 &=& \hf \xi^{\a i} D
 +\frac{3}{4} R(J)_{ab}{}^{ij} (\xi_j {\g}^{ab})^\a
 -\frac{1}{4} (\xi^{i} \g^a  \tilde{\g}^{bcd})^{\a }  \de_a  T^{-}_{ b c d}  
 +\frac{1}{4} \l^{cd} (\chi^{ i} \g_{cd})^\a \non\\
 &&
  + \l^i{}_j \chi^{\a j}
 +\frac{3}{2} \l_{\mathbb{D}} \chi^{\a i}
 + \ri (\eta^{i}\tilde{\g}^{abc})^{\a}  T^{-}_{a b c} ~,
 \label{6Dd-chi}
 %%%%%%%%%
 \\
 \delta D
 &=& -2 \ri \, \xi_{i}\g^{a}\de_{a} \chi^{i}
 +2 \l_{\mathbb{D} }D
 -4  \chi^{ i} \eta_{ i} ~,
 \label{6Dd-D}
\eea
\esubeq
where 
\bsubeq\label{6Dnabla-on-W-and-Chi}
\bea
\de_d  T^{-}_{ a b c } &=& 
\mathcal{D}_{a}T^{-}_{ a b c }
+\frac{\ri}{15}  \psi_{d k} \g_{abc} \chi^{ k}
+4 \ri \psi_d{}_k {X^{k}}^{(-)}_{abc} ~, \\
%%%%%%%%
\de_{a} \chi^{\b j}
&=& 
\cD_{a}\chi^{\b j}
-\frac{3}{8} (\psi_{a i} \g^{bc})^\b  R(J)_{bc}{}^{ij}
-\frac{1}{8} (\psi_a{}^{j}  \g^e \tilde{\g}^{bcd})^{\b}  \de_e T^-_{bcd} \non
\\&&+\frac{1}{4} \psi_a{}^{\b j} D+\frac{\ri}{2} T^{-}_{b c d}   (\phi_a{}^{j}\tilde{\g}^{bcd})^{\b}~,
\eea
\esubeq
and we have defined ${X_\g^{k}}^{(-)}_{abc} := \frac{1}{8} (\g_{abc})_{\a \b} {X_\g^{k \a \b}}$.

Note that the transformations 
\eqref{6Dtransf-standard-Weyl} 
form an algebra that
closes off-shell on a local extension of ${\rm OSp}(6,2|1)$. 
To conclude this subsection, 
for convenience, we include Table \ref{6Dchiral-dilatation-weights} 
which summarises the non-trivial dilatation
weights of the fields and local gauge parameters 
of the standard Weyl multiplet. 
\begin{table}[hbt!]
\begin{center}
\begin{tabular}{ |c||c| c| c| c| c| c|c|} 
\hline
& $e_m{}^a$
&$\psi_m{}_i$, $\xi_i$
& $\phi_m{}^i$, $\eta^i$ 
&$\mathfrak{f}_{m}{}_c$
& $T^-_{abc}$
&$\chi^{i}$ 
&$D$
\\ 
\hline
\hline
$\mathbb{D}$
&$-1$
&$-1/2$
&$1/2$
&$1$
&$1$
&$3/2$
&$2$
\\
\hline
\end{tabular}
\caption{\footnotesize{Summary of the non-trivial 
dilatation weights in the standard Weyl 
multiplet.}\label{6Dchiral-dilatation-weights}}
\end{center}
\end{table}

\section{The hyper-dilaton Weyl multiplet in 6D}
\label{6Dhyper+HDWM}
The aim of this section is to construct the $40+40$
hyper-dilaton Weyl multiplet of off-shell 
$\cN=(1,0)$ conformal supergravity in six dimensions. The construction mimics the 5D $\cN=1$ case elaborated earlier in our paper and the 4D $\cN=2$ case of \cite{Gold:2022bdk}.

We begin with the component structure of the on-shell hypermultiplet. This can be readily extracted from the previous superspace realisation via the bar projection. The independent components of the on-shell hypermultiplet are simply: the Lorentz scalar field $q^{i \underline{i}} \loco$ which is a superconformal primary,  and the spinor field $\rho_{\hal}^{\underline{i}} \loco$, which account for $4+4$ on-shell degrees of freedom.  All other descendants are derivatives of these two fields.

In what follows, we will associate the same symbol for the covariant component fields and the corresponding superfields, when the interpretation is clear from the context. The local superconformal transformations of the component fields follow directly from the projections of \eqref{6Dd-hyper-sspace}, which give
\bsubeq\label{6Dhyper-susy}
\bea
\delta q^{i\underline{i}} 
&=& 
\frac{1}{2}\xi^{ i}\rho^{\underline{i}}
+\l^{i}{}_{k}q^{k\underline{i}} 
+ 2 \lambda_{\mathbb{D}} q^{i\underline{i}}
~,
\\
\delta \rho_{\hal}^{\underline{i}} 
&=&
-4\ri( {\xi}_i \g^{\ha})_{\hal} 
\nabla_{\ha} q^{i\underline{i}}
 -\frac{1}{4} \l_{\ha \hb} (\g^{\ha \hb} \rho^{\underline{i}})_{\hal}    
+\frac{5}{2} \lambda_{\mathbb{D}} \rho_{\hal}^{\underline{i}} 
+16 \eta_{\hal}^i q_i{}^{\underline{i}}~,
\eea
\esubeq
where 
\bea
\nabla_{a} q^{i\underline{i}}
=
\cD_{a} q^{i\underline{i}}
 -\frac{1}{4}\psi_{a}{}^{ i}
 \rho^{\underline{i}}~.
 \label{6DDD'q}
\eea
Unlike the standard Weyl multiplet, the algebra of the local transformations \eqref{6Dhyper-susy} closes only when the equations of motion for the fields $q^{i\underline{i}}$ and $\rho_{\a}^{\underline{i}}$ are imposed. These equations of motion can be obtained by taking the bar projection of \eqref{6Dvect-rho-sspace} and \eqref{6DBox-q1}. Specifically, in the traceless frame, the lowest component of the descendants $\nabla_{a} \nabla_\a^iq_i{}^{\underline{i}}|$ and $\Box q^{i \underline{i}}|$ leads to the following two constraints:
\bsubeq \label{6Don-shell-hyper}
\bea
(\nabla_{a}
\rho^{\underline{i}}\,
\tilde{\g}^{a})^{\a}
&=&
-\frac{1}{12}
({\rho}^{\underline{i}}\,
\tilde{\g}^{a b c})^{\a} T^-_{a b c}
+\frac{8\ri}{15}  \,  {\chi}^{\a k}q_{k}{}^{\underline{i}}~, \\
\Box q^{i \underline{i}} 
&=&
\frac{1}{15}\chi^{i}\rho^{\underline{i}} 
-\frac{1}{15} D \,q^{i\underline{i}}~, \qquad \Box := \nabla^{a}\nabla_{a}~.
\eea
\esubeq
The expressions for 
$\nabla_{a}\rho^{ \underline{i}}_\a$ 
and $\Box q^{i\underline{i}}$ in terms of the derivatives 
$\cD_{a}$ are given by 
\bsubeq
\bea
\nabla_{a} \rho_\a^{\underline{i}}
&=& 
\cD_{a}\rho_\a^{ \underline{i}} 
+2\ri (\psi_{a}{}_k \g^c)_{\a} \left( 
\cD_{c} q^{k\underline{i}}
 -\frac{1}{4}\psi_{c}{}^{k}
 \rho^{\underline{i}}
\right) 
-8\phi_{a}{}_\a^{ k}q_{k}{}^{\underline{i}}
~,\label{6Dnabla-rho}
%%%%%%
%%%%%%
\\
\Box q^{i\underline{i}} 
&=& 
\cD^a\cD_a q^{i\underline{i}}
- 4 \mathfrak{f}_a{}^a q^{i \underline{i}}
-\frac{1}{4} \cD_a (\psi^{a i} \rho^i)
-\frac{\ri}{4} \phi_a{}^i \tilde{\g}^a \rho^{\underline{i}}
\non\\
&&
+ \frac{1}{4} \psi_a{}^{\a i}\Big{[}
-\cD^a \rho_{\a}^{\underline{i}} 
- 2 \ri  (\psi^{a}{}_{j} \gamma^c)_{\a}\cD_c q^{j \underline{i}}
+\frac{1}{24} (\rho^{\underline{i}}\tilde{\g}^{bcd} \gamma^a)_{\a} T^-_{bcd} 
\non\\&& ~~~~~~~~~~~~
- \frac{8 \ri}{15}(\chi^{k}\g^a)_{\a}q_{k}^{\underline{i}}
+\frac{\ri}{2}  (\psi^{a}{}_{j}\gamma^c)_{\a} \psi_c{}^{ j} \rho^{\underline{i}}
+ 8 \phi^{a}{}_{\a}^{j} q_{j}{}^{\underline{i}}\Big{]}
\label{6Dbox-q}~.
\eea
\esubeq
Equations \eqref{6Don-shell-hyper} can then be interpreted as algebraic equations for the 
standard Weyl multiplet that determine
the auxiliary fields $\chi^{\a i}$ and $D$ in terms of 
$q^{i\underline{i}}$ and $\rho_{\a}^{\underline{i}}$, 
together with the other independent fields of the standard Weyl multiplet. It can be noted that in the traceless frame equations \eqref{6Dnabla-rho} and \eqref{6Dbox-q} do not depend on the fields $\chi^{\a i}$ and $D$ respectively making it trivial to find these auxiliary fields. If we assume that $q^{i\underline{i}}$ is an invertible matrix, which is equivalent to imposing
\bea
q^2:=q^{i\underline{i}}q_{i\underline{i}}=\ve_{ij}\ve_{\underline{i}\underline{j}}q^{i\underline{i}}q^{j\underline{j}}
=2\det{q^{i\underline{i}}}\ne0
~,
\eea
then the following relations hold
\bsubeq\label{6DcompositeSSbD}
\bea
{\chi}^{\alpha}_{i} & = & \frac{15 \ri}{4} q^{-2}q_{i\underline{i}}\bigg[
(\nabla_{a}
\rho^{\underline{i}}\tilde{\g}^{}
)^{\a}
+\frac{1}{12}
({\rho}^{\underline{i}}\,
\tilde{\g}^{a b c})^{\a} T^-_{a b c}
\,\bigg] 
~,
\label{6Dsigma}
\\
%%%%%%%%%%%%
D & = & 
- 15 q^{-2} q^{i\underline{i}}\Box q_{i \underline{i}}
+\frac{15 \ri}{8} q^{-2}\bigg[
\nabla_{a}
\rho^{\underline{j}} \tilde{\g}^{a} \rho_{ \underline{j}}
+\frac{1}{12}
({\rho}^{\underline{j}}\,
\tilde{\g}^{a b c} \rho_{ \underline{j}}) T^-_{a b c}\,
\,\bigg]  
~. 
\label{6DD_def}
\eea
\esubeq
Once more, we stress that the right-hand side of equation \eqref{6Dsigma} does not have any dependence on field $\chi$, thus making it a composite field. Similarly, the right-hand side of equation \eqref{6DD_def} does not depend on $D$, however it has an implicit dependence on $\chi$ through the special conformal connection $\mathfrak{f}_a{}_b$, eq.~\eqref{6Deq:KConn-new} and see \eqref{6Dbox-q}, that is hidden in the expression of $\Box q^{i\underline{i}}$. 
It is straightforward to pull out the explicit dependence upon $\chi$ and then use \eqref{6Dsigma}.
As a result, both
 $\chi$ and $D$ are composite.

As a next step in the construction of the hyper-dilaton Weyl multiplet, we note that
associated to an on-shell hypermultiplet one can construct a triplet of linear multiplets, exactly as in the 4D $\cN=2$ and 5D $\cN=1$ cases. The component fields of the $\cN=1$ off-shell linear (or $\cO(2)$) multiplet are defined in terms of the bar projections of \eqref{6DO2superfieldComps}:
an SU$(2)_R$ triplet of Lorentz scalar fields $L^{ij} = L^{ij}\loco$\,; 
a spinor field $\varphi_{\a}^i =\varphi_{\a}^i\loco$\,;
and a closed anti-symmetric five-form field strength $h_{m n p q r} := H_{ m n p q r}\loco$\,, which is equivalent to a conserved dual vector 
${h}^{a}:= \frac{1}{5!}\ve^{a m n p q r}h_{m n p q r}$. Defining $H^a = H^a\loco$, at the component level it holds that
\bea
H_a = h_a - \psi_b{}_i \g^{ab}\varphi^i
- \frac{\ri}{2} \psi_b{}_i\g^{abc} \psi_c{}_j  L^{ij}
\ ~.
\eea
The covariant conservation equation of $H_a$ is
\bea
\de^{\ha} {H}_{a} 
&=& 
0~.
\label{6Dcovariant-current}
\eea
The constraint implies the existence of a gauge four-form potential, $b_{m n p q}$, 
and its exterior derivative $h_{m n p q r}:= 5 \pa_{[m}b_{n p q r]}$.
The local superconformal transformations of the linear multiplet in a standard Weyl multiplet background
are given by
\bsubeq
\label{6Dlinear-multiplet}
\bea
\d L^{ij}
&=&
2 \xi^{\a(i} \varphi^{j)}_\a
+ 2\lambda^{(i}{}_{k}  L^{j) k} 
+4 \lambda_{\mathbb{D}} L^{ i j}
~,
%%%%%%
\\
\d \varphi_{\a }^i
&=&
\frac{\ri}{2}\xi^{\b i} H_{\b\a}
-\ri \xi^\b_j \de_{\b\a} L^{ij}
+\frac{1}{4} \l^{ab}  (\varphi^i \tilde{\g}_{ab})_\a
-\l^{i j} \varphi_{\a j}
+\frac{9}{2}\lambda_{\mathbb{D}}\varphi_{\a}^{i}
+8 \eta_{\a}^{j}   L_{j}{}^i ~,
\\
\d H_a 
&=&
2 (\xi_i \g_{ab})^\b \de^b \varphi_\b^i
+\frac{1}{12} (\xi^{ i} \g_a \tilde{\g}^{bcd}  \varphi_{ i}) T^{-}_{bcd}
+\l_a{}^d H_d 
+ 5 \l_{\mathbb{D}} H_a
\non\\&&- 10 \ri \eta^i \tilde{\g}_a\varphi_{ i}
~,
\eea
\esubeq
where
\bsubeq
\bea
{\nabla}_a L^{ij} &=& {\cD}_aL^{ij} - \psi_a{}^{(i} \varphi^{j)}
\label{6DdeL}
\ , \\
{\nabla}_a \varphi^i
&=&
{\cD}_a \varphi^i
- \frac{\ri}{4}\psi_a{}^{i}\g_b{ H}^b
-\frac{\ri}{2}\psi_a{}_j{\slashed{\nabla}}L^{ij}
+ 4 {\phi}_{a j} L^{ij}
\ .
\eea
\esubeq
The locally superconformal transformations of $b_{m n p q}$ are 
\bea
\d b_{m n p q}
&=&
- \, \ve_{m n p q e f} (\xi_i \g^{e f} \varphi^i)
+8 \ri (\psi_{[mi} \g_{npq]} \xi_j) L^{i j}
+4 \pa_{[m}l_{n p q]}~,
\label{6Dd-bmn}
\eea
where we have also included 
the gauge transformation
$\d_l b_{m n p q}= 4 \pa_{[m}l_{n p q]}$
leaving 
$h_{m n p q r}$
and
${H}^{a}$ invariant.
For convenience, we have summarised the dilatation weights
of the fields of the $\cO(2)$ multiplet in Table 
\ref{6Dchiral-dilatation-weights-LINEAR}.
\begin{table}[hbt!]
\begin{center}
\begin{tabular}{ |c||c| c| c| c| c| c| c|} 
 \hline
& $L_{ij}$ 
& $\varphi_{\a i}$
& ${H}_{a}$
&$b_{m n p q}$
\\ 
\hline
 \hline
$\mathbb{D}$
&$4$
&$9/2$
&$5$
&$0$
\\
\hline
\end{tabular}
\caption{\footnotesize{Summary of the 
dilatation weights in the off-shell $\cO (2)$ multiplet.}
\label{6Dchiral-dilatation-weights-LINEAR}}
\end{center}
\end{table}

Now that we have reviewed the structure of a locally superconformal $\cO(2)$ multiplet, we return to constructing a triplet of linear multiplets from an on-shell hypermultiplet. Given that $q^{i\underline{i}}$ and $\rho_{\a}^{\underline{i}}$ describe an on-shell hypermultiplet in a standard Weyl multiplet background with transformation rules \eqref{6Dhyper-susy}, it can be shown that the following composite fields define a triplet of $\cO(2)$ multiplets
\bsubeq\label{6Dcomposite-linear}
\bea
L_{ij}{}^{\underline{i}\underline{j}}
&=&
q_{(i}{}^{\underline{i}}q_{j)}{}^{\underline{j}}
=q_{i}{}^{(\underline{i}}q_{j}{}^{\underline{j})}
~,\\
\varphi_{\a i}{}^{\underline{i}\underline{j}}
&=& 
\frac{1}{2} q_{i}{}^{(\underline{i}}\rho_{\a}{}^{\underline{j})}
~,
\\
{H}^{\ha \, \underline{i}\underline{j}}
&=& 2q^{i(\underline{i}} \nabla^{\ha}{q_i}^{\underline{j})}
-\frac{\ri}{8}
\rho^{(\underline{i}}\tilde{\g}^{a} \rho^{\underline{j})}
~.
\label{6Dlinear-H}
\eea
\esubeq
These fields all transform according to 
\eqref{6Dlinear-multiplet} and
 each of the previous fields is symmetric in $\underline{i}$ and $\underline{j}$.
The field 
${H}^{ \ha \,\underline{i}\underline{j}}$, in particular, is interesting as it 
can be used to express the 
${\rm SU}(2)_R$ connection 
$\phi_{\hm}{}^{ij}$ as a composite field.
To see this, we introduce a new covariant derivative 
\bea
\mathbf{D}_{\ha} 
&=& 
{e_{\ha}}^{\hm}\left( \partial_{\hm} - \frac{1}{2}{\omega_{\hm}}^{\hc \hd}M_{\hc \hd} 
- b_{\hm }\mathbb{D}\right) 
= \cD_{\ha} + {e_{\ha}}^{\hm}{\phi_{\hm}}^{ij}J_{ij}~,
\eea
which then allows us to rearrange eq.~\eqref{6Dlinear-H} for the SU(2)$_R$ gauge connection:
\bea
 \phi_{\ha}{}^{ij}  
 &=&
 4q^{-4}q^{(i}{}_{\underline{i}}q^{j)}{}_{\underline{j}}
 \Bigg[\, 
 q^{k\underline{i}} \mathbf{D}_{\ha} q_k{}^{\underline{j}}
 - \frac{1}{4}q^{k\underline{i}} 
 ({\psi_{a}}{}_k \rho^{\underline{j}}) 
  - \frac{\ri}{16}
 \rho^{\underline{i}}
 \tilde{\g}_{a}
 {\rho}^{\underline{j}}
 -\hf H_{\ha}{}^{\underline{i}\underline{j}}\,\Bigg] 
 ~.~~~~~~~~~
 \label{6DcompositeSU2}
\eea

This concludes the definition of the hyper-dilaton Weyl multiplet. Our analysis demonstrates that the hyper-dilaton Weyl multiplet defines a new representation of the off-shell local 6D, $\cN=(1,0)$ superconformal algebra. The multiplet comprises the following independent fields: 
${e_{\hm}}^{\ha}$, $b_{\hm}$, $T^-_{a b c}$, $q^{i\underline{i}}$, $b_{m n p q}{}^{\underline{i}\underline{j}}$,
$\psi_{m\, i}$, and $\rho^{\underline{i}}$. It also possesses the same number of off-shell
degrees of freedom as the standard Weyl multiplet,
$40+40$. 
Table \ref{6Ddof2} summarises the counting of degrees of freedom, 
underlining the symmetries acting on the fields. 
%%%%%%%%%%%%
\begin{table}[hbt!]
\begin{center}
\begin{tabular}{ |c c c c c c c |c c c c|} 
 \hline
${e_{\hm}}^{\ha}$ & $\omega_{\hm}{}^{\ha \hb}$ & $b_{\hm}$ & ${\mathfrak{f}_{\hm}}{}^{\ha}$ & $\phi_{\hm}{}^{ij}$ & $\psi_{\hm}{}_i$ & $\phi_{\hm}{}^i$ & $T^-_{a b c}$ & $\rho^{\underline{i}}$ & $q^{i\underline{i}}$ & $b_{m n p q}{}^{\underline{i}\underline{j}}$\\ 
$36$ & $0$ & $6B$ & $0$ & $0$ & $48F$ & $0$ & $10B$ & $8F$ & $4B$ & $45B$\\
\hline
$P_{\ha}$ & $M_{\ha \hb}$ & $\mathbb D$ & $K_{\ha}$ & $J^{ij}$  & $Q$ & $S$ & {} & ${}$ & ${}$ & $\lambda_{m n p}{}^{\underline{i}\underline{j}}$-sym\\
$-6B$ & $-15B$ & $-1B$ & $-6B$ & $-3B$ & $-8F$ & $-8F$ & {} & {} & {} & $-30B$\\
\hline
\multicolumn{11}{|c|}{Result: $40+40$ degrees of freedom}\\
\hline
\end{tabular}
\caption{\footnotesize{Degrees of freedom and symmetries of the hyper-dilaton Weyl multiplet. Row one gives all the fields in the multiplet. Row two gives the number of independent components of these fields --- composite connections are counted with zero degrees of freedom. 
Row three gives the gauge symmetries. Note that the parameter $\lambda_{m n p}{}^{\underline{i}\underline{j}}$ 
describes the symmetry associated with the gauge four-forms
$b_{m n p q}{}^{\underline{i}\underline{j}}$ 
with field strength
five-forms $h_{m n p q r}{}^{\underline{i}\underline{j}}$ 
and ${E}^{a}{}^{\underline{i}\underline{j}}$.
Row four gives the number of gauge degrees of freedom to be subtracted when counting the total degrees of freedom.}\label{6Ddof2}}
\end{center}
\end{table}
Note that with the ingredients provided so far, it is 
a straightforward
exercise to obtain
the locally superconformal 
transformations of the fields
of the hyper-dilaton Weyl multiplet
written only in terms of the independent fields and they are given as follows:
\bsubeq\label{6Dtransf-hyper-Dilaton-Weyl}
\bea
\delta e_{m}{}^{a} 
&=& 
-\ri\,\xi_i\g^{a}{\psi}_{m}{}^{ i}
+ \lambda^{a}{}_{b}e_{m}{}^{b}
- \lambda_{\mathbb{D}}{e}_{m}{}^{a}
~,
\\
%%%%%%
\delta \psi_{m}{}^{\a}_i  
&=& 2\mathcal{D}_{m} \xi^{\a}_i  
-\frac{1}{12} T^{-}_{abc} (\xi_i \g_m \tilde{\g}^{abc})^{\a }
+ \frac{1}{4} \lambda^{ab}(\psi_m{}_i\g_{ab})^\a
 - \lambda_{i}{}^{j} \,\psi_{m}{}^{\a}_{j}
\non \\  &&- \frac{1}{2}\lambda_{\mathbb{D}}\,{\psi}_{m}{}^{\a}_i
-2\ri (\eta_{ i} \tilde{\g}_{m})^{\a}
~,
 \\
\delta b_{m}
&=& 
\frac{1}{4}\,q^{-2}q_{i\underline{i}} (\xi^{ i} \g_{m})_{\a} \bigg[
\frac{1}{12}
({\rho}^{\underline{i}}\,
\tilde{\g}^{a b c})^{\a} T^-_{a b c}-(\tilde{\g}^{a}\nabla_{a}
\rho^{\underline{i}}
)^{\a}\bigg]
 \non \\ &&
+\xi_i   \phi_m{}^i
+\partial_{m} \lambda_{\mathbb{D}} 
 -\psi_{m}{}^{ i}\eta_{ i} 
- 2\lambda_{m}
~,
%%%%%
\\
  \delta T^{-}_{a b c} &=& 
  -\frac{\ri}{8} \xi^{ k} \g^{e f} \g_{abc} R(Q)_{e f}{}_{k} 
\non\\
&&
-\frac{1}{2} q^{-2}  q_{i\underline{i}} (\xi^{ i} \g_{abc})_{ \a}  \bigg[
\frac{1}{12}
({\rho}^{\underline{i}}\,
\tilde{\g}^{a b c})^{\a} T^-_{a b c}-(\tilde{\g}^{a}\nabla_{a}
\rho^{\underline{i}}\,
)^{\a}
\,\bigg] \non \\ &&
 - 3 \l^e{}_{[a} T^{-}_{b c] e}
 + \l_{\mathbb{D}} T^{-}_{a b c}~,
 %%%%%%%
 \\
 \delta q^{i\underline{i}} 
&=& 
\frac{1}{2}\xi^{ i}\rho^{\underline{i}}
+\lambda^{i}{}_{k}q^{k\underline{i}} 
+ 2\lambda_{\mathbb{D}}q^{i\underline{i}}
~,
\label{6Dd-qii}
\\
\delta \rho_{\a}^{\underline{i}} 
&=&
-4\ri ({\xi}_i \g^c)_{\a} \de_c q^{i\underline{i}}
+\frac{1}{4} \lambda_{a b}  (\rho^{\underline{i}}\tilde{\g}^{a b})_\a   
+\frac{5}{2} \lambda_{\mathbb{D}}\rho_{\a}^{\underline{i}} 
+16 \eta_{\a}^k q_k{}^{\underline{i}}~,
\label{6Dd-rho}
\\
\d b_{m n p q}
&=&
- \, \ve_{m n p q e f} \xi_i \g^{e f} \varphi^i
+8 \ri (\psi_{[mi} \g_{npq]} \xi_j) L^{i j}
+4 \pa_{[m}l_{n p q]}~.
\eea
\esubeq
For completeness, here we present the expressions relevant to the transformation in terms of this new covariant derivative $\mathbf{D}_a$ instead of $\cD_a$, which has an implicit dependence on the composite ${\rm SU}(2)_R$ connection $\phi_a{}^{ij}$.
\bsubeq
\bea
\nabla_{a} q^{i\underline{i}}
&=&
\hf \mathbf{D}_a q^{i\underline{i}} -\frac{1}{8}\psi_{a}{}^{i}
 \rho^{\underline{i}}
 - q^{-2}q^{i}{}_{\underline{k}}  q^{k\underline{i}} \mathbf{D}_{\ha} q_k{}^{\underline{k}}
  +\frac{1}{4} q^{-2}q^{i}{}_{\underline{k}}q^{k\underline{i}} 
 ({\psi_{a}}{}_k \rho^{\underline{k}})\non \\ &&
  +\frac{\ri}{8}  q^{-2}q^{i}{}_{\underline{k}} 
 (\rho^{(\underline{k}}
 \tilde{\g}_{a}
 {\rho}^{\underline{i})})
 + q^{-2}q^{i}{}_{\underline{k}}H_{\ha}{}^{\underline{k}\underline{i}}
 ~,\\
\nabla_{a} \rho_\a^{\underline{i}}
&=& 
\mathbf{D}_a \rho_\a^{ \underline{i}} 
+2\ri (\psi_{a}{}_k\g^c)_{\a} 
\nabla_{c} q^{k\underline{i}}
-8\phi_{a}{}_\a^{ k}q_{k}{}^{\underline{i}}
~,\label{6Dnabla-rho2}
%%%%%%
%%%%%%
\\
\Box q^{i\underline{i}} 
&=& 
\mathbf{D}_a\de^a q^{i\underline{i}} - 4 \mathfrak{f}_a{}^a q^{i \underline{i}} \non \\ &&+  4q^{-4}q^{(i}{}_{\underline{k}}q^{j)}{}_{\underline{l}} (\de^a q_j{}^{\underline{i}})
 \bigg[\, 
 q^{k\underline{k}} \mathbf{D}_{\ha} q_k{}^{\underline{l}}
 -\hf H_{\ha}{}^{\underline{k}\underline{l}}
 - \frac{1}{4}q^{k\underline{k}} 
 ({\psi_{a}}{}_k \rho^{\underline{l}}) 
 \frac{\ri}{16}
 \rho^{\underline{k}}
 \tilde{\g}_{a}
 {\rho}^{\underline{l}}
 \,\bigg]  
\non \\ &&
-\frac{1}{96} (\psi_a{}^{ i} \gamma^a\tilde{\g}^{bcd}\rho^{\underline{i}}) T^-_{bcd}  + \frac{2 \ri}{15} (\psi_a{}^{ i}\g^a \chi^{ k}) q_{k}^{\underline{i}} -\frac{1}{4} \psi_a{}^{\a i}\nabla_{a} \rho_{\a}^{\underline{i}} 
-\frac{\ri}{4} \phi_a{}^i\tilde{\g}^a \rho^{\underline{i}} ~, ~~~
\label{6Dbox-q2}
\eea
\esubeq
where the composite connection $\phi_m{}^i$ and $\mathfrak{f}_{a}{}^{b}$ are now given in terms of  $\mathbf{D}_a$ by:
\bsubeq
\bea
 {\phi}_m{}^{k}
&=&
\frac{\ri}{16}\Big(
\g^{bc}\g_m-\frac{3}{5}\g_m\tilde{\g}^{bc}
\Big)\bigg[2 \mathbf{D}_{[b} \psi_{c]}{}^k  +\frac{1}{12}T^-_{def}\tilde{\g}^{def}\g_{[b}\psi_{c]}{}^{k} \non \\ && ~~~
+8q^{-4}q^{(k}{}_{\underline{i}}q^{j)}{}_{\underline{j}}
 \Big\{\, 
 q^{k\underline{i}} \mathbf{D}_{[b} q_k{}^{\underline{j}}
 -\hf H_{[b}{}^{\underline{i}\underline{j}} 
 -\frac{1}{4}q^{k\underline{i}} 
 ({\psi_{[b}}{}_k \rho^{\underline{j}}) 
 - \frac{\ri}{16}
 (\rho^{\underline{i}}
 \tilde{\g}_{[b}
 {\rho}^{\underline{j}})
 \,\Big\} \psi_{c]j}
\bigg]
~,~~~~~~~~~
\label{6Deq:SConn-dilaton}
\\
%%%
 {\mathfrak{f}}_{a}{}^{b}
&=&
-\frac{1}{8}\cR_{a}{}^{b}( {\o})
+\frac{1}{80}\d_{a}^{b}\cR( {\o})
+\frac{\ri}{16}\psi_{c}{}_j\g_{a} {R}(Q)^{bc}{}^j
+\frac{\ri}{8}\psi_{c}{}_j\g^{[b} {R}(Q)_{a}{}^{c]}{}^{ j}
\non\\
&&-\frac{1}{16} q^{-2}q^{j}{}_{\underline{i}} (\psi_{a}{}_j\g^b)_{\a} \bigg[
\frac{1}{12}
({\rho}^{\underline{i}}\,
\tilde{\g}^{c d e})^{\a} T^-_{c d e}-(\tilde{\g}^{c}\nabla_{c}
\rho^{\underline{i}}\,
)^{\a}
\,\bigg]
+\frac{1}{8}\psi_{[a}{}_j  \g^{bc} {\f}_{c]}{}^j
\non \\ &&
-\frac{1}{80}\d_{a}^{b}\psi_{c}{}_j  \g^{cd} {\f}_{d}{}^j
+\frac{\ri}{16}\psi_{a}{}^{ j}\g_c\psi_{d}{}_j\,T^-{}^{bcd}
-\frac{\ri}{160}\d_{a}^{b} \psi_{c}{}^{ j}\g_d\psi_{e}{}_j\,T^-{}^{cde}
~.
\eea
\esubeq
 Note that the expression of ${\mathfrak f}_\ha{}^\hb$ has an explicit as well as implicit dependence on the composite connection $\phi_{a}{}^i$ via $\de_a\rho^{\underline{i}}$, which can now be substituted from \eqref{6Deq:SConn-dilaton}.
It is also convenient to provide the bosonic part of $\Box q^{i\underline{i}}$. By using that $\mathbf{D}^a H_{a}{}^{\underline{i}\underline{j}}=0$ up to fermions, it holds:
\bea
\Box q^{i\underline{i}} 
 &=&  \frac{1}{2} \mathbf{D}^a\mathbf{D}_a q^{i \underline{i}} +\frac{3}{4} q^{-2} q^{i \underline{i}} (\mathbf{D}^a q^{k \underline{k}}) \mathbf{D}_a q_{k \underline{k}}  - q^{-2} q^i{}_{\underline{j}} q^{k \underline{i}} \mathbf{D}^a\mathbf{D}_a q_k{}^{\underline{j}} + \hf
 q^{-2}q^{k}{}_{\underline{j}}
  \mathbf{D}^{\ha} q^{i \underline{j}}\mathbf{D}_a q_k{}^{ \underline{i}}  \non \\ &&  
  +    q^{-4} q_{l \underline{l}} q^i{}_{\underline{j}} q^{k \underline{i}} (\mathbf{D}^a q^{l \underline{l}})  \mathbf{D}_a q_k{}^{\underline{j}}  - \hf q^{-4} q^{i \underline{i}} H^{a \underline{j}\underline{k}}H_{a \underline{j}\underline{k}} 
+ \frac{1}{5} \mathcal{R}  q^{i \underline{i}} + \text{fermions}~.
\eea

In analogy to the 5D $\cN=1$ hyper-dilaton Weyl multiplet, we will end this subsection by underlining the following two remarks about the 6D $\cN=(1,0)$ hyper-dilaton Weyl multiplet:
\begin{enumerate}
    \item From a symmetry point of view, the hyper-dilaton Weyl multiplet contains all the fields that are required to gauge fix the extra symmetries of the superconformal group, i.e., it contains a triplet of scalar field $q^{i \underline{i}}$, which can be used to gauge fix dilatation and ${\rm SU}(2)_R$ symmetry; the spinor field $\rho_\a^{\underline{i}}$ and the dilatation connection $b_m$ can be used to fix $S$-supersymmetry and special conformal symmetry, respectively. An example of such a gauge choice is as follows:
    \bsubeq\label{6Dgauge-conditions}
\bea
&q^{i \underline{i}}=-\epsilon^{i\underline{i}}
~,
\label{6Dfixing-phi}
\\ &
\rho_{\a}^{\underline{i}}=0~,
\label{6DS-susy-gauge}
\\&
b_{\hm}=0
~.
\label{6Dfixing-K}
\eea
\esubeq
This indicates that in the gauge fixed version we would obtain an off-shell irreducible multiplet of Poincar\'e supergravity.

\item The second point would be to obtain a supersymmetric completion of the Einstein-Hilbert term by using an appropriate compensating multiplet. In 4D $\cN=2$ and 5D $\cN=1$ this was achieved by using an off-shell vector multiplet compensator. In 6D the $\cN=(1,0)$ vector multiplet has no scalar fields \cite{Koller:1982cs,Howe:1983fr} that can be used for this purpose. The natural choice would be a tensor multiplet. However, known versions of an off-shell tensor multiplet involve infinite number of auxiliary fields \cite{Sokatchev:1988aa,Linch:2012zh}. One might wonder whether it suffices to use an off-shell linear multiplet. The action for an improved linear multiplet in a standard Weyl multiplet background take the form of the following  $BF$ Lagrangian
\cite{Bergshoeff:1985mz,Butter:2018wss}
\bea
e^{-1}\,\cL|_{bosonic} 
&=&
-\frac{2}{15}\big(3\cR+D\big) L
-\frac{1}{8L}H^a H_a
-\frac{1}{2L} H^a \phi_a^{ij} L_{ij}
-\cD^a\cD_a L
\non\\&&
+\frac{1}{4L}(\cD^a L^{ij}) \cD_a L_{ij}
-\frac{1}{8L^3} \tilde{b}^{mn}L_i{}^j (\partial_m L^{ki}) \partial_n L_{jk}
~.
\label{6Dbosonic-swm}
\eea
When working with a hyper-dilaton Weyl multiplet, we need to take into account that the auxiliary field $D$ and the ${\rm SU}(2)_R$ connection are composite fields (and that \eqref{6DD_def} and \eqref{6DcompositeSU2} have to be used). The combination $(3\cR+D)$ turns out to not depend on the scalar curvature $\cR$,
\bea
3\cR+D &=&
 -15{q^{-2}} \Bigg\{{q_{i \underline{i}}} \mathbf{D}_a \mathbf{D}^a q^{i \underline{i}} +( \mathbf{D}^a q^{i \underline{i}})\mathbf{D}_a q_{i \underline{i}} 
 - q^{-2}  q^{i \underline{i}} q^{j \underline{j}}(\mathbf{D}_a q_{i \underline{i}}) \mathbf{D}^a q_{j \underline{j}} \non\\
 &&~~~~~~~~~~~~
- \hf q^{-2} H_{a \underline{i}\underline{j}} H^{a \underline{i}\underline{j}}\Bigg\} 
~.
\eea
Clearly, by plugging this into \eqref{6Dbosonic-swm}, the result is independent of $\cR$ and fails to be a good starting point to engineer a supersymmetric extension of the Einstein-Hilbert term to obtain a two-derivative Poincar\'e' supergravity Lagrangian.

It is worth mentioning that coupling the hyper-Dilaton Weyl multiplet to any number of linear multiplet will encounter the same problem. This is not too surprising since the linear multiplet is on-shell equivalent to the on-shell hypermultiplet up to trading a scalar field with a gauge four-form. We expect that the same would be true by using other variant off-shell hypermultiplets, such as the off-shell charged hypermultiplet, coupled to conformal supergravity \cite{Sokatchev:1988aa,Linch:2012zh}. We will come back in the future to engineer 6D $\cN=(1,0)$ off-shell Poincar\'e supergravity theories, and their matter couplings, by using our new hyper-dilaton Weyl multiplet in a superconformal setting.
\end{enumerate}

%%%%%%%%%%%%%%%%%%%%%%%%%%%%%%%%%%
%%%%%%%%%%%%%%%%%%%%%%%%%%%%%%%%%%
\vspace{0.3cm}
\noindent
{\bf Acknowledgements:}\\
We are grateful to Gregory Gold and William Kitchin
for discussions and collaborations related to this work.
This work is supported by the Australian Research Council (ARC)
Future Fellowship FT180100353, and by the Capacity Building Package of the University
of Queensland.
S.K. is supported 
by the postgraduate scholarships 
at the University of Queensland.
J.W., during part of this work, thanks the support of The Australian Mathematical Sciences Institute (AMSI) Vacation Research Scholarship 2021/22 for providing financial support, and the European Organization for Nuclear Research (CERN) for recently providing financial support, hospitality, and further research experience during June-August 2022.
J.H., S.K., and G.T.-M.
thank the MATRIX Institute in Creswick for hospitality and support during part of this
project. 
J.H. thanks the Simons Foundation for support through a MATRIX-Simons travel grant.

%%%%%%%%%%%%%%%%%%%%%%%%%%%%%%
%%%%%%%%%%%%%%%%%%%%%%%%%%%%%%

%%%%%%%%%%%%%%%%%%%%%%%%%%%%%%%%%%%%%%%%%%%%%%%%%%%%%%
%%%%%%%%%%%%%%%%%%%%%%%%%%%%%%%%%%%%%%%%%%%%%%%%%%%%%%

\appendix

%%%%%%%%%%%%%%%%%%%%%%%%%%%%%%%%%%%%%%%%%%%%%%%%%%%%%%
%%%%%%%%%%%%%%%%%%%%%%%%%%%%%%%%%%%%%%%%%%%%%%%%%%%%%%

\section{5D $\cN=1$ conformal superspace identities}
\label{conformal-identities}

In this appendix we collect results about conformal superspace in the traceless frame of \cite{BKNT-M14} focusing on the ingredients relevant to our discussion in the paper. We adhere with the notations and conventions of \cite{BKNT-M14}.

The Lorentz generators act on the superspace covariant derivatives $\de_{\hA} = (\de_{\ha}, \de_{\hal}^{i})$ in the following way
\bsubeq \label{SCA}
\begin{align} 
[M_{\ha \hb} , M_{\hc \hd}] &= 2 \eta_{\hc [\ha} M_{\hb] \hd} - 2 \eta_{{d} [ \ha} M_{\hb ] \hc}~, \\
[M_{\ha \hb}, \de_\hc] &= 2 \eta_{\hc [\ha} \de_{\hb]}~, \\ 
[M_{\hal \hbe} , \de_\hga^i ] &= \eps_{\hga (\hal} \de_{\hbe)}^i~, 
\end{align}
where $M_{\hal \hbe} = 1/2 (\S^{\ha \hb})_{\a\b} M_{\ha \hb}$. 
The $\rm SU(2)_{R}$ and dilatation generators satisfy
\begin{align}
[J^{ij} , J^{kl}] &= \eps^{k(i} J^{j) l} + \eps^{l (i} J^{j) k}~,
\quad 
[J^{ij} , \de_\hal^k ] = \eps^{k (i} \de_\hal^{j)}~\, \\
[\mathbb{D} , \de_\ha] &= \de_\ha~, 
\quad 
[\gD , \de_\hal^i ] = \hf \de_\hal^i~.
\end{align}
The Lorentz and $\rm SU(2)_R$ generators act on the special conformal generators $K_\hA = (K_\ha , S_{\hal i})$ according to the rules
\begin{align} 
[M_{\ha \hb} , K_\hc] &= 2 \eta_{\hc [\ha} K_{\hb]} \ , 
\quad
[M_{\hal \hbe} , S_\hga^i ] = \eps_{\hga (\hal} S_{\hbe)}^i \,
\quad
[J^{ij} , S_\hal^k ] = \eps^{k (i} S_\hal^{j)}~, 
\end{align}
while the dilatation generator acts on $K_{\hA}$ as
\begin{align} 
[\mathbb{D} , K_\ha] = - K_{\ha}~, 
\quad 
[\gD , S_{\hal i} ] = - \hf S_{\hal i}~.
\end{align}
Among themselves, the generators $K_{\hA}$ obey the only nontrivial anti-commutation relation
\begin{align} 
\{ S_\hal^i , S_\hbe^j \} &= - 2 \ri \,\eps^{ij} (\G^{\hc})_{\hal\hbe} K_{\hc}~.
\end{align}
The algebra of $K_{\hA}$ with $\de_{\hA}$ is given by
\begin{align}
[K_\ha , \de_\hb] &= 2 \eta_{\ha \hb} \gD + 2 M_{\ha\hb}~, \\
{[} K_\ha , \de_\hal^i {]} &=  \ri (\G_\ha)_\hal{}^\hbe S_{\hbe}^i~, \\
\{ S_{\hal i} , \de_\hbe^j \} &= 2 \eps_{\hal \hbe} \d_i^j \gD - 4 \d_i^j M_{\hal\hbe} + 6 \eps_{\hal\hbe} J_i{}^j~, \label{alg-S-spinor} \\
[S_{\hbe i}, \nabla_\ha] &= 
	\ri (\Gamma_\ha)_\hbe{}^\hal \nabla_{\hal i}
	- \frac{1}{4} W_\ha{}^\hb (\Gamma_\hb)_\hbe{}^\hal S_{\hal i}
	+ \frac{\ri}{8} (\G_\ha \G^\hb)_\hbe{}^\hga X_{\hga i} K_\hb
	- \frac{\ri}{4} W_{\ha\hb\hbe i} K^\hb~.
\end{align}
\esubeq

The anticommutator of two spinor derivatives,
$\{{\de}_{\hal}^i,{\de}_{\hbe}^j\}$, has the following non-zero
 torsion and curvatures
\begin{subequations}  \label{new-frame_spinor-spinor}
\begin{align}
\scT_{\hal}^i{}_\hbe^j{}^\hc &= 2 \ri \ve^{ij} (\Gamma^\hc)_{\hal\hbe}~, \\
{\mathscr{R}}(M)_{\hal}^i{}_\hbe^j{}^{\hc\hd}
	&= 2 \ri \ve^{ij} \ve_{\hal\hbe} W^{\hc\hd}
		+ \ri \ve^{ij} (\Gamma_\hb)_{\hal\hbe} \tilde W^{\hb\hc\hd} ~, \\
{\mathscr{R}}(S)_{\hal}^i{}_\hbe^j{}^{\hga k}
	&= \frac{3\ri}{4} \ve^{ij} \ve_{\hal\hbe} X^{\hga k}
		+ \ri \ve^{ij} \delta^\hga_{[\hal} X_{\hbe]}^k ~, \\
{\mathscr{R}}(K)_{\hal}^i{}_\hbe^j{}^{\hc}
	&= - \frac{\ri}{2} \eps^{ij} \eps_{\hal\hbe} \nabla^\hb W_\hb{}^\hc
	+ \frac{\ri}{2} \eps^{ij} (\Gamma^\ha)_{\hal\hbe} \nabla^\hd \tilde W_{\hd\ha}{}^\hc
	- \frac{\ri}{32}  \eps^{ij} (\Gamma^\hc)_{\hal\hbe} Y
	\non\\ & \quad
	+ \frac{\ri}{4} \eps^{ij} \eps_{\hal\hbe} \tilde W_\hc{}^{\hd\he} W_{\hd\he}
	+ \frac{\ri}{2} \eps^{ij} (\Gamma^\ha)_{\hal\hbe}
	\Big(W_{\ha\hd} W^{\hc\hd} 
	- \frac{3}{16} W^{\hb\hd}W_{\hb\hd} \delta_\ha{}^\hc	
	\Big)~.
\end{align}
\end{subequations}
The non-vanishing torsion and curvatures in the spinor-vector commutator $[\de_{\hb}, \de_{\hal}^i]$ are:
\begin{subequations} \label{new-vectspinor}
\begin{align}
\scT_{\hb}{}_{\hal}^i{}^{{\g}}_k
	&= \frac{1}{4} \delta^i_k \Big(
		3(\Gamma_\hb)_\hal{}^\hbe W_\hbe{}^{{\g}}
		- W_\hal{}^\hbe (\Gamma_\hb)_\hbe{}^{{\g}}
	\Big)~, \\
{\mathscr{R}}(D)_{\hb}{}_{\hal}^i &= -\frac{1}{4} (\Gamma_\hb)_\hal{}^\hga X_\hga^i~, \\
{\mathscr{R}}(J)_{\hb}{}_\hal^i{}^{jk} &= -\frac{3}{4}(\Gamma_\hb)_\hal{}^\hga \eps^{i(j} X_\hga^{k)}~, \\
{\mathscr{R}}(M)_\hb{}_\hal^i{}^{\hc\hd} 
	&= -(\Gamma_\hb)_\hal{}^\hga W^{\hc\hd}{}_\hga^i 
	- \frac{1}{4} \ve_\hb{}^{\hc\hd\he{f}} W_{\he{f}}{}_\hal^i 
	+ \hf \delta_\hb^{[\hc} (\Gamma^{\hd]})_\hal{}^\hga X_\hga^i ~,\\
{\mathscr{R}}(S)_{\hb}{}_{ \hal}^i{}^{\hga j}
	&=
	\frac{1}{16} X_{\hc\hd}{}^{ij} (\S^{\hc\hd} \G_\hb - 2 \G_\hb \S^{\hc\hd})_\hal{}^\hga
	\non\\ & \quad
	- \frac{3\ri}{8} \ve^{ij}\, \nabla_{[\hb} W_{\hc\hd]} (\S^{\hc\hd})_\hal{}^\hga 
	- \frac{\ri}{8} \ve^{ij}\,\nabla_\hd W^{\hd\hc} (\S_{\hc\hb})_\hal{}^\hga 
	\non\\ & \quad
	+ \frac{3\ri}{16} \ve^{ij} \, \nabla^\hd W_{\hd\hb} \,\delta_\hal^\hga
	- \frac{\ri}{8} \ve^{ij}\, \nabla^\hc \tilde W_{\hc\hb}{}^\hd (\G_\hd)_\hal{}^\hga
	\non\\ & \quad
	+ \frac{\ri}{16} \ve^{ij} \tilde W^{\hc\hd\he} W_{\hd\he} (\S_{\hc\hb})_\hal{}^\hga
	- \frac{3\ri}{32} \ve^{ij} \tilde W_{\hb\hd\he} W^{\hd\he} \delta_\hal{}^\hga
	\non\\ & \quad
	+ \frac{\ri}{4} \ve^{ij}\, W_{\hb\hd} W^{\hc\hd} (\G_\hc)_\hal{}^\hga 
	- \frac{3\ri}{64} \ve^{ij}\,W^{\hc\hd}W_{\hc\hd} (\G_\hb)_\hal{}^\hga ~, \\
{\mathscr{R}}(K)_{\hb}{}_{ \hal}^i{}^\hc &=
	\frac{1}{6} (\G^\hc)_\hal{}^\hbe \nabla^\hd W_{\hd\hb}{}_\hbe^i
	+ \frac{1}{12} (\G_\hb)_\hal{}^\hbe \nabla^\hd W{}_\hd{}^{\hc}{}_\hbe^i
	+ \frac{1}{6} {\nabla}_\hal{}^\hbe W_\hb{}^\hc{}_\hbe^i
	- \frac{1}{24} \ve_\hb{}^{\hc\hd\he f} \nabla_\hd W_{\he\hat{f}}{}_\hal^i
	\non\\ & \quad
	+ \frac{1}{8} (\G^\hc)_\hal{}^\hbe \nabla_\hb X^i_\hbe
	+ \frac{1}{64} W^{\hd\he} (3 \G_\hb \S_{\hd\he} \G^\hc - \S_{\hd\he} \G_\hb \G^\hc)_\hal{}^\hbe X_\hbe^i
	\non\\ & \quad
	- \frac{1}{48} \tilde W_{\hb\hd\he} (\G^\hc)_\hal{}^\hbe W^{\hd\he}{}_\hbe^i
	+ \frac{1}{8} \delta_\hb{}^\hc W^{\hd\he} W_{\hd\he}{}_\hal^i
	\non\\ & \quad
	+ \frac{1}{12} (\S_\hb{}^\hc)_\hal{}^\hbe W_{\hd\he}{}_\hbe^i W^{\hd\he}
	- \frac{1}{12} W^{\hd\he} (\S_{\hd\he})_\hal{}^\hbe W_{\hb}{}^{\hc}{}_\hbe{}^i
	\non\\ & \quad
	+ \frac{13}{48} W_{\hb\hd} W^{\hd\hc}{}_\hal^i
	+ \frac{11}{48} W_{\hb\hd}{}_\hal^i W^{\hd\hc}
	- \frac{13}{96} (\G_\hb)_\hal{}^\hbe W_{\hd\he}{}_\hbe^i \, \tilde W^{\hd\he\hc}
	~.
\end{align}
\end{subequations}
The commutator of two vector derivatives $[\de_{\ha}, \de_{\hb}]$ has the following non-zero torsion and curvatures:
\begin{subequations}
\begin{align}
\scT_{\ha\hb}{}^\hal_i &= -\frac{\ri}{2} W_{\ha\hb}{}^\hal_i~, \\
{\mathscr{R}}(J)_{\ha\hb}{}^{ij} &= -\frac{3\ri}{4} X_{\ha\hb}{}^{ij}~, \\
{\mathscr{R}}(M)_{\ha\hb}{}^{\hc\hd} &=
	-\frac{1}{4} (\Sigma_{\ha\hb})^{\hal\hbe}
	(\Sigma^{\hc\hd})^{\hga \hde}
	\Big( \ri W_{\hal\hbe\hga\hde}
		+ 3 W_{(\hal \hbe} W_{\hga \hde)}
	\Big)~, \label{eq:newRMab} \\
{\mathscr{R}}(S)_{\ha\hb}{}_\hal^i
	&=
	- \frac{1}{2} {\nabla}_\hal{}^\hbe W_{\ha\hb\hbe}{}^i
	- \frac{1}{2} (\G_{[\ha })_\hal{}^\hbe \nabla^\hc W_{\hb]\hc \hbe}{}^i
	\non\\ & \quad
	- \frac{1}{8} W_\hal{}^\hbe W_{\ha\hb\hbe}{}^i
	+ \frac{1}{16} (\S_{\ha\hb})_\hal{}^\hbe W^{\hc\hd} W_{\hc\hd\hbe}{}^i
	+ \frac{3}{8} W^\hc{}_{[\ha } W_{\hb]\hc\hal}{}^i~, \\
{\mathscr{R}}(K)_{\ha\hb}{}^\hc &=
	\frac{1}{4} \nabla_d  {\mathscr{R}}(M)_{\ha\hb}{}^{\hc\hd}
	- \frac{\ri}{16} W_{\ha b}{}^\hal_j (\G^\hc)_\hal{}^\hbe X_\hbe^j
	- \frac{\ri}{8} W_{\hd [\ha}{}^\hal_j (\G_{\hb]})_\hal{}^\hbe W^{\hc\hd}_\hbe{}^j
	\non\\ & \quad
	+ \frac{\ri}{8} W_{\ha \hd }{}^\hal_i (\G^\hc)_\hal{}^\hbe
		W_\hb{}^{\hd}{}_\hbe{}^i~.
\end{align}
\end{subequations}

%%%%%%%%%%%%%%%%%%%%%%%%%%%%%%%%%%%%%%%%%%%%%%%%%%%%%%
%%%%%%%%%%%%%%%%%%%%%%%%%%%%%%%%%%%%%%%%%%%%%%%%%%%%%%
%
Recall that the descendant superfields
$X_{\a}^{i}$, $W_{\a \b \g}{}^i$, $W_{\a \b \g \d}$, $X_{\a \b}{}^{ij}$, and $Y$, were defined in
\eqref{descendantsW-5d}. They transform under $S$-supersymmetry as
\begin{align}
S_{\hal i} W_{\hbe\hga\hde}{}^j &= 6 \d^j_i \eps_{\hal (\hbe} W_{\hga \hde)} \ , \qquad
S_{\hal i} X_\hbe^j = 4 \d_i^j W_{\hal\hbe}~, \eol
S_{\hal i} W_{\hbe\hga\hde\hrh} &= 24 \eps_{\hal (\hbe} W_{\hga\hde \hrh)}{}_i \ , \qquad
S_{\hal i} Y = 8 \ri X_{\hal i} ~, \eol
S_{\hal i} X_{\hbe\hga}{}^{jk} &= - 4 \d_i^{(j} W_{\hal\hbe\hga}{}^{k)} + 4 \d_i^{(j} \eps_{\hal (\hbe} X_{\hga)}^{k)} \ .\label{S-on-X_Y-a}
\end{align}
A complete list of the identities for spinor covariant derivative acting on these descendants can be found in subsection 2.3 of\cite{BKNT-M14}.\footnote{It should be noted that the identities given in \cite{BKNT-M14} were described in a non-traceless frame.} Here we only provide the relations which are useful for our analysis. The following relations are expressed in the traceless frame, which is our convention in this paper:
\bsubeq \label{eq:Wdervs}
\bea
\nabla_{\hga}^k W_{\hal\hbe} 
&=& W_{\hal\hbe\hga}{}^k + \eps_{\hga (\hal} X_{\hbe)}^k \ ,  \label{Wdervs-a}
\\
\nabla_\hal^i X_\hbe^j
&=&
 X_{\hal \hbe}{}^{ij} 
 + \frac{\ri}{8} \eps^{ij}\Big(
  \eps_{\hal\hbe} Y
+ 4 \eps^{\ha\hb\hc\hd\he} (\S_{\ha\hb})_{\hal\hbe} \nabla_\hc W_{\hd\he} \non\\
&& \hspace{1.5cm}- 4 (\G^\hb)_{\hal\hbe} \nabla^\ha W_{\ha \hb}
+ \ve^{abcde} (\G_{a})_{\a \b} W_{bc} W_{de}
\Big) \ . ~~~~~~~~~
\eea
\esubeq

%%%%%%%%%%%%%%%%%%%%%%%%%%%%%%%%%%%%%%%%%%%%%%%%%%%%%%
%%%%%%%%%%%%%%%%%%%%%%%%%%%%%%%%%%%%%%%%%%%%%%%%%%%%%%

\section{6D $\cN=(1,0)$ conformal superspace identities}
\label{6Dconformal-identities}
In this appendix we collect results about 6D $\cN=(1,0)$ conformal superspace \cite{BKNT16} in the traceless frame of \cite{BNT-M17,Butter:2018wss} focusing on the ingredients relevant to our discussion in the paper.

The Lorentz generators act on the superspace covariant derivatives $\de_A=(\de_a,\de_\a^i)$ as
\begin{align}\label{6DSCA}
[M_{ab} , M_{cd}] &= 2 \eta_{c[a} M_{b] d} - 2 \eta_{d [a} M_{b] c} \ , \\
[M_{ab} ,  {\nabla}_c ] &= 2 \eta_{c [a}  {\nabla}_{b]} \ , \\
 [M_\a{}^\b , \nabla_\g^k] &= - \d_\g^\b \nabla_\a^k + \frac{1}{4} \d^\b_\a \nabla_\g^k
~,
\end{align}
where $M_\a{}^\b=-\frac{1}{4}(\g^{ab})_\a{}^\b M_{ab}$.
The ${\rm SU(2)}_R$ and dilatation generators satisfy
\begin{align}
[J^{ij}, J^{kl}] &= \eps^{k(i} J^{j)l} + \eps^{l(i} J^{j)k} \ , \quad [J^{ij} , \nabla_\a^k] = \eps^{k(i} \nabla_\a^{j)} \ ,  \\
[\mathbb D ,  {\nabla}_a] &=  {\nabla}_a \ , \quad [\mathbb D , \nabla_\a^i] = \hf \nabla_\a^i \ .
\end{align}
The Lorentz and ${\rm SU(2)}_R$
generators
act
on the special conformal generators $K^A=(K^a,S^\a_i)$ as
\begin{align}
[M_{ab} , K^c] = 2 \d^c_{[a} K_{b]} \ , \quad
  [M_\a{}^\b , S^\g_k] = \d^\g_\a S^\b_k - \frac{1}{4} \d^\b_\a S^\g_k
\ , \quad
[J^{ij} , S^\g_k] = \d_k^{(i} S^{\g j)} \ ,
\end{align}
while the dilatation generator acts on  $K^A$ as
\begin{align}
[\mathbb D , K^a] = - K^a \ , \quad [\mathbb D, S^\a_i] &= - \hf S^\a_i \ .
\end{align}
Among themselves, the generators $K_A$ obey the only nontrivial anti-commutation relation
\begin{align}
\{ S^\a_i , S^\b_j \} = - 2 \ri \eps_{ij} (\tilde{\g}_c)^{\a\b} K^c \ .
\end{align}
The algebra of $K^A$ with ${\nabla}_A$ is given by
\bea
{[}K_a ,  {\nabla}_b{]} &=& 2 \eta_{ab} \mathbb D + 2 M_{ab} \ , \\
{[}K^a , \nabla_\a^i {]} &=& - \ri (\g^a)_{\a\b} S^{\b i} \ , \\
\{ S^\a_i , \nabla_\b^j \} &=& 2 \d^\a_\b \d^j_i \mathbb D - 4 \d^j_i M_\b{}^\a + 8 \d^\a_\b J_i{}^j \label{6Dalg-S-spinor} \ , \\
{[}S^\a_i ,  {\nabla}_b{]}
&=&
 -\ri (\tilde{\g}_b)^{\a\b}  {\de}_{\b i}
+\frac{1}{10}W_{bcd}(\tilde{\g}^{cd})^\a{}_\g
S^\g_i
-\frac{1}{4}X^{\a}_{ i}K_b
\non\\
&&
+\Big{[}
\frac{1}{4}(\tilde{\g}_{bc})^{\a}{}_{\b} X^{\b}_{ i}
+\frac{1}{2}({\g}_{bc})_\b{}^{\g}X_{\g i}{}^{\b\a} \Big{]}K^c
~.
\label{6DSnabla}
\eea

The anticommutator of two spinor derivatives,
$\{{\de}_\a^i,{\de}_\b^j\}$, has the following non-zero
 torsion and curvatures\bsubeq\label{6Dnew-frame_spinor-spinor}
\bea
 \mathscr{T}_{\a}^i{}_\b^j{}^c
&=&
2\ri\ve^{ij}(\g^c)_{\a\b}
~,
\\
 \mathscr{R}(M)_{\a}^i{}_\b^j{}^{cd}
&=&
4\ri\ve^{ij}(\g_a)_{\a\b} W^{acd}
~,
\\
%%%
 \mathscr{R}(S)_{\a}^i{}_\b^j{}_\g^k
&=&
-\frac{3}{2}\ve^{ij}\ve_{\a\b\g\d} X^{\d k}
~,
\\
 \mathscr{R}(K)_{\a}^i{}_\b^j{}_c
&=&
\ri\ve^{ij}(\g^a)_{\a\b}\left(
\frac{1}{4} \eta_{ac}Y
- {\de}^b W_{a b c}
+ W_{a}{}^{ef} W_{cef}
\right)
~.
\eea
\esubeq
The non-zero torsion and curvatures in the commutator ${[} {\de}_a,{\de}_\b^{j}{]}$ are:
\bsubeq\label{6Dnew-vectspinor}
\bea
 \mathscr{T}_{a}{}_\b^j{}^\g_k
&=&
-\frac{1}{2} (\g_{a})_{\b \d} W^{\d \g}\d^j_k
~,
\\
%%%
 \mathscr{R}(\mathbb D)_a{}_\b^j
&=&
- \frac{\ri}{2} (\gamma_{a})_{\b\g} X^{\g j}
~,
\\
%%%
 \mathscr{R}(M)_a{}_\b^j{}^{cd}
&=&
\ri \delta_{a}^{[c} (\g^{d]})_{\b\g} X^{\g j}
-\ri(\gamma_{a}{}^{ c d})_{\g \d} X_{\b}^j{}^{\g \d}
+2\ri(\gamma_{a})_{\b\g} (\gamma^{c d})_{\d}{}^{\r}X_{\r}^j{}^{\gamma \delta }
~,
\\
%%%
 \mathscr{R}(J)_a{}_\b^j{}^{kl}
&=&
2 \ri (\gamma_{a})_{\b\g} X^{\g (k}\ve^{l)j}
~,
\\
%%%
 \mathscr{R}(S)_a{}_\b^j{}_\g^k
&=&
- \frac{\ri}{4} (\gamma_{a})_{ \b \d} \, Y_{\g}{}^{\d}{}^{ jk}
+\frac{3\ri}{20}  (\gamma_{a})_{\g\d} Y_{\b}{}^{\d}{}^{ jk}
- \frac{\ri}{8} (\gamma_{a})_{\b \d}  {\de}_{\g \r}{W^{\d \r}}  \ve^{jk}
\non\\
&&
+ \frac{\ri}{40}(\gamma_{a})_{\g\d}  {\de}_{\b\r}{W^{\d \r}}\ve^{jk}
- \frac{\ri}{8} (\gamma_{a})_{\d \epsilon}\, \ve_{\b \r \t \g}\,W^{\d \r} W^{\epsilon \t} \ve^{jk}
~, \\
% % % %
 \mathscr{R}(K)_a{}_\b^j{}_c
&=&
\frac{\ri}{4} (\gamma_{c})_{\b \g}  {\de}_{a}{X^{\g j}}
- \frac{\ri}{4} (\gamma_{a c d})_{\g \d}  {\de}^{d}{X_{\b}^j{}^{\g \d }}
+ \frac{\ri}{3} (\gamma_{a})_{\b \d} (\gamma_{c d})_{\r}{}^{\g} {\de}^{d}{X_{\g}^j{}^{\d \r }}
\non\\
&&
- \frac{\ri}{8} (\gamma_{a})_{\b \g} (\gamma_{c})_{\d \r}W^{\g \d} X^{\r j}
+ \frac{5\ri}{12} (\gamma_{a})_{\b\r}(\gamma_{c})_{\g \epsilon} W^{\g \d} X_{\d}^j{}^{\r \epsilon }
\non\\
&&
+ \frac{\ri}{4} ( \gamma_{a})_{\g \r} (\gamma_{c})_{\b \epsilon} 	W^{\g \d} X_{\d}^j{}^{\r \epsilon }
- \frac{\ri}{2}(\gamma_{a})_{\g \r} (\gamma_{c})_{ \d \epsilon}W^{\g\d}X_{\b}^j{}^{\r\epsilon }
~.
%%%
\eea
\esubeq
The commutator of two vector derivatives,
$[ {\de}_a, {\de}_b]$,
has the following non-vanishing torsion and curvatures:
\bsubeq\label{6Dnew-frame_vector-vector}
\bea
 \mathscr{T}_{ab}{}^\gamma_k
&=&
(\g_{ab})_\b{}^\a X_{\a k}{}^{\b\g}
~,
\label{6Dnew-frame_ALL-T-c}
\\
%%%
 \mathscr{R}(M)_{a b}{}^{c d} &=&
Y_{a b}{}^{cd}
=\frac{1}{4}(\g_{ab})_\g{}^\a(\g^{cd})_\d{}^\b Y_{\a\b}{}^{\g\d}
~,
\label{6Dnew-frame_ALL-R-i}
\\
%%%
 \mathscr{R}(J)_{ab}{}^{kl}
&=&
\frac{1}{2}(\g_{ab})_\d{}^\g Y_\g{}^\d{}^{kl} = Y_{a b}{}^{kl}
~,
\label{6Dnew-frame_ALL-R-J}
\\
%%%
 \mathscr{R}(S)_{ab}{}_\gamma^k
&=&
- \frac{\ri}{3} (\gamma_{a b})_{\delta}{}^{\alpha}  {\de}_{\gamma \beta} X_{\alpha}^k{}^{\beta \delta }
- \frac{\ri}{6} (\gamma_{a b c})_{\alpha\beta}  {\de}^c X_{\gamma}^k{}^{\alpha \beta }
- \frac{\ri}{6} \ve_{\gamma \beta \epsilon \rho} (\gamma_{a b})_{\delta}{}^\rho
W^{\alpha \beta} X_{\alpha}^k{}^{\delta \epsilon}
~,~~~~~~~~~~~~
\label{6Dnew-frame_ALL-R-k}
\\
%%%
 \mathscr{R}(K)_{ab}{}_{c}
&=&
\frac{1}{4}  {\de}^d Y_{a b c d}
+\frac{\ri}{3} X_{\alpha}^k{}^{\beta \gamma } X_{\beta k}{}^{\alpha \delta}(\gamma_{a b c})_{\gamma\delta}
+\ri (\gamma_{a b})_{\epsilon}{}^\alpha
(\gamma_{c})_{\gamma\delta} X_{\alpha}^k{}^{\beta \gamma } X_{\beta k}{}^{\delta \epsilon}
\non\\
&&
+\frac{\ri}{4}X^{\alpha k} X_{\beta k}{}^{\gamma \delta}
(\gamma_{a b})_{\gamma}{}^{\beta} (\gamma_{c})_{\alpha\delta}
~.
\label{6Dnew-frame_ALL-R-l}
\eea
\esubeq
Recall the descendant superfields
$X^{\a i}$, $X_{\a i}{}^{\b\g}$, $Y$, $Y_\a{}^\b{}^{kl}$ ,
$Y_{\a\b}{}^{\g\d}$ (and equivalently $Y_{ab}{}^{cd}$), were defined in
\eqref{6DXfields} and \eqref{6DYfielDs}. They transform under $S$-supersymmetry as \cite{BKNT16}
\bsubeq\label{6DS-on-X_Y}
\bea
S^\a_i X^{\b j} &=& \frac{8\ri}{5} \d_i^j W^{\a\b} ~,
~~~~~~
S^\a_i X_\b^j{}{}^{\g\d}
=
-\ri\d^j_i \d^\a_\b W^{\g\d}
+\frac{2\ri}{5} \d^j_i \d^{(\g}_{\b} W^{\d) \a}
~,
\label{6DS-on-X_Y-a}
\\
%%%
S^\g_k Y_\a{}^{\b}{}^{ij}
&=&
-\d^{(i}_k \left(
16 X_\a^{j)}{}^{\g\b}
- 2 \d_\a^\b X^{\g j)}
+ 8 \d^\g_\a X^{\b j)} \right)
~,
\label{6DS-on-X_Y-b}
\\
%%%
S^\r_j Y_{\a\b}{}^{\g\d}
&=&
24 \left( \d^\r_{(\a} X_{\b) j}{}^{\g\d}
- \frac{1}{3} \d^{(\g}_{(\a} X_{\b) j}{}^{\d) \r} \right)
~ ,~~~~~~
S^\a_i Y = - 4 X^\a_i
~ .
\label{6DS-on-X_Y-c}
\eea
\esubeq

By using \eqref{6DWBI} and the previous definitions,
one can derive spinor covariant derivative acting on these descendants of the super-Weyl tensor and can be found in \cite{BNT-M17,Butter:2018wss} in the traceless frame. Here we only provide the relations which are useful for our analysis, which are
\bsubeq \label{6Deq:Wdervs}
\bea
\nabla_\a^i X^{\b j}
&=&
- \frac{2}{5} Y_\a{}^\b{}^{ij}
- \frac{2 }{5} \eps^{ij}  {\nabla}_{\a \g} W^{\g\b}
- \hf \eps^{ij} \d_\a^\b Y
\ , \\
%%%
\nabla_\a^i X_\b^j{}^{\g\d}
&=&
\hf \d^{(\g}_\a Y_\b{}^{\d)}{}^{ij} - \frac{1}{10} \d_\b^{(\g} Y_\a{}^{\d)}{}^{ij}
- \hf \eps^{ij} Y_{\a\b}{}^{\g\d}
- \frac{1}{4}\eps^{ij}  {\nabla}_{\a\b} W^{\g\d}
\non\\
&&
+ \frac{3}{20} \eps^{ij} \d_\b^{(\g}  {\nabla}_{\a \r} W^{\d) \r}
-\frac{1}{4}\eps^{ij} \d^{(\g}_\a  {\nabla}_{\b \r} W^{\d) \r}
\ ~.
\eea
\esubeq

%%%%%%%%%%%%%%%%%%%%%%%%%%%%%%
%%%%%%%%%%%%%%%%%%%%%%%%%%%%%%

\begin{footnotesize}

\end{footnotesize}


\begin{thebibliography}{66}




\bibitem{Deser} 
S.~Deser, 
``Scale invariance and gravitational coupling,''
Annals Phys.\  {\bf 59}, 248 (1970).

\bibitem{Zumino} B. Zumino, 
``Effective Lagrangians and broken symmetries," 
in {\it Lectures on Elementary Particles and Quantum Field Theory,
Vol. 2}, S. Deser, M. Grisaru and H. Pendleton (Eds.),
Cambridge, Mass. 1970, pp. 437-500.

\bibitem{freedman}
D.~Z.~Freedman and A.~Van~Proeyen,
\emph{Supergravity}, Cambridge University Press (2012).


\bibitem{Lauria:2020rhc}
E.~Lauria and A.~Van Proeyen,
``${\cal N}=2$ Supergravity in $D=4,5,6$ Dimensions,''
Lect. Notes Phys. \textbf{966} (2020), pp.
%doi:10.1007/978-3-030-33757-5
[arXiv:2004.11433 [hep-th]].



\bibitem{SUPERSPACE}
S.~J.~Gates, M.~T.~Grisaru, M.~Rocek and W.~Siegel,
``Superspace Or One Thousand and One Lessons in Supersymmetry,''
Front. Phys. \textbf{58}, 1-548 (1983)
[arXiv:hep-th/0108200 [hep-th]].


\bibitem{Buchbinder-Kuzenko}
I.~Buchbinder and S.~M.~Kuzenko.
\emph{Ideas and methods of supersymmetry and supergravity: 
Or a walk  through superspace},
IOP, Bristol (1998).


\bibitem{Bergshoeff:1985mz}
E.~Bergshoeff, E.~Sezgin and A.~Van Proeyen,
``Superconformal Tensor Calculus and Matter Couplings in Six-dimensions,''
Nucl. Phys. B \textbf{264}, 653 (1986)
[erratum: Nucl. Phys. B \textbf{598}, 667 (2001)].
%doi:10.1016/0550-3213(86)90503-1



\bibitem{CVanP}
  F.~Coomans and A.~Van Proeyen,
  ``Off-shell $N=(1,0)$, $D=6$ supergravity from superconformal methods,''
  JHEP {\bf 1102}, 049 (2011)
  Erratum: [JHEP {\bf 1201}, 119 (2012)]
  [arXiv:1101.2403 [hep-th]].


\bibitem{Bergshoeff:2012ax}
  E.~Bergshoeff, F.~Coomans, E.~Sezgin and A.~Van Proeyen,
  ``Higher Derivative Extension of 6D Chiral Gauged Supergravity,''
JHEP {\bf 1207}, 011 (2012).
[arXiv:1203.2975 [hep-th]].


\bibitem{OzkanThesis}
M.~Ozkan,
``Supersymmetric curvature squared invariants in five and six dimensions,''
\href{http://oaktrust.library.tamu.edu/bitstream/handle/1969.1/151223/OZKAN-DISSERTATION-2013.pdf?sequence=1}
{\it PhD Thesis}, Texas A\&M University, 2013.


\bibitem{BKNT16}
  D.~Butter, S.~M.~Kuzenko, J.~Novak and S.~Theisen,
  ``Invariants for minimal conformal supergravity in six dimensions,''
  JHEP {\bf 1612}, 072 (2016)
%  doi:10.1007/JHEP12(2016)072
  [arXiv:1606.02921 [hep-th]].



\bibitem{BNT-M17}
  D.~Butter, J.~Novak and G.~Tartaglino-Mazzucchelli,
  ``The component structure of conformal supergravity invariants in six dimensions,''
  JHEP {\bf 1705}, 133 (2017)
%  doi:10.1007/JHEP05(2017)133
  [arXiv:1701.08163 [hep-th]].


\bibitem{Butter:2018wss}
D.~Butter, J.~Novak, M.~Ozkan, Y.~Pang and G.~Tartaglino-Mazzucchelli,
``Curvature squared invariants in six-dimensional ${\cal N} = (1,0)$ supergravity,''
JHEP \textbf{04}, 013 (2019)
%doi:10.1007/JHEP04(2019)013
[arXiv:1808.00459 [hep-th]].


\bibitem{Sokatchev:1988aa}
E.~Sokatchev,
``Off-shell Six-dimensional Supergravity in Harmonic Superspace,''
Class. Quant. Grav. \textbf{5} (1988), 1459-1471.
%doi:10.1088/0264-9381/5/11/009


\bibitem{Linch:2012zh}
W.~D.~Linch, III and G.~Tartaglino-Mazzucchelli,
``Six-dimensional Supergravity and Projective Superfields,''
JHEP \textbf{08} (2012), 075
%doi:10.1007/JHEP08(2012)075
[arXiv:1204.4195 [hep-th]].


\bibitem{Bergshoeff:2001hc}
E.~Bergshoeff, T.~de Wit, R.~Halbersma, S.~Cucu, M.~Derix and A.~Van Proeyen,
``Weyl multiplets of N=2 conformal supergravity in five-dimensions,''
JHEP \textbf{06}, 051 (2001)
%doi:10.1088/1126-6708/2001/06/051
[arXiv:hep-th/0104113 [hep-th]].


\bibitem{Butter:2017pbp}
D.~Butter, S.~Hegde, I.~Lodato and B.~Sahoo,
``$N=2$ dilaton Weyl multiplet in 4D supergravity,''
JHEP \textbf{03}, 154 (2018)
%doi:10.1007/JHEP03(2018)154
[arXiv:1712.05365 [hep-th]].


\bibitem{Gold:2022bdk}
G.~Gold, S.~Khandelwal, W.~Kitchin and G.~Tartaglino-Mazzucchelli,
``Hyper-Dilaton Weyl Multiplet of 4D, ${\mathcal{N}}=2$ Conformal Supergravity,''
JHEP \textbf{09} (2022), 016,
[arXiv:2203.12203 [hep-th]].



 \bibitem{KT-M08}
S.~M.~Kuzenko and G.~Tartaglino-Mazzucchelli,
``Super-Weyl invariance in 5D supergravity,''
JHEP \textbf{04} (2008), 032
%doi:10.1088/1126-6708/2008/04/032
[arXiv:0802.3953 [hep-th]].



\bibitem{BKNT-M14}
D.~Butter, S.~M.~Kuzenko, J.~Novak and G.~Tartaglino-Mazzucchelli,
``Conformal supergravity in five dimensions: New approach and applications,''
JHEP \textbf{02}, 111 (2015)
%doi:10.1007/JHEP02(2015)111
[arXiv:1410.8682 [hep-th]].




\bibitem{Kugo:2000hn}
T.~Kugo and K.~Ohashi,
``Supergravity tensor calculus in 5-D from 6-D,''
Prog. Theor. Phys. \textbf{104}, 835-865 (2000)
%doi:10.1143/PTP.104.835
[arXiv:hep-ph/0006231 [hep-ph]].


\bibitem{Fujita:2001kv}
T.~Fujita and K.~Ohashi,
``Superconformal tensor calculus in five-dimensions,''
Prog. Theor. Phys. \textbf{106}, 221-247 (2001)
%doi:10.1143/PTP.106.221
[arXiv:hep-th/0104130 [hep-th]].


\bibitem{Kugo:2002vc}
T.~Kugo and K.~Ohashi,
``Gauge and nongauge tensor multiplets in 5-D conformal supergravity,''
Prog. Theor. Phys. \textbf{108}, 1143-1164 (2003)
%doi:10.1143/PTP.108.1143
[arXiv:hep-th/0208082 [hep-th]].



\bibitem{Bergshoeff:2002qk}
E.~Bergshoeff, S.~Cucu, T.~De Wit, J.~Gheerardyn, R.~Halbersma, S.~Vandoren and A.~Van Proeyen,
``Superconformal N=2, D = 5 matter with and without actions,''
JHEP \textbf{10}, 045 (2002)
%doi:10.1088/1126-6708/2002/10/045
[arXiv:hep-th/0205230 [hep-th]].


\bibitem{Bergshoeff:2004kh}
E.~Bergshoeff, S.~Cucu, T.~de Wit, J.~Gheerardyn, S.~Vandoren and A.~Van Proeyen,
``N = 2 supergravity in five-dimensions revisited,''
Class. Quant. Grav. \textbf{21}, 3015-3042 (2004)
%doi:10.1088/0264-9381/23/23/C01
[arXiv:hep-th/0403045 [hep-th]].







\bibitem{Galperin:1984av}
A.~Galperin, E.~Ivanov, S.~Kalitsyn, V.~Ogievetsky and E.~Sokatchev,
``Unconstrained N=2 Matter, Yang-Mills and Supergravity Theories in Harmonic Superspace,''
Class. Quant. Grav. \textbf{1} (1984), 469-498
[erratum: Class. Quant. Grav. \textbf{2} (1985), 127].
%doi:10.1088/0264-9381/1/5/004




\bibitem{Bagger:1987rc}
J.~A.~Bagger, A.~S.~Galperin, E.~A.~Ivanov and V.~I.~Ogievetsky,
``Gauging $N=2 \sigma$ Models in Harmonic Superspace,''
Nucl. Phys. B \textbf{303} (1988), 522-542.
%doi:10.1016/0550-3213(88)90392-6


\bibitem{Galperin:2001seg}
A.~S.~Galperin, E.~A.~Ivanov, V.~I.~Ogievetsky and E.~S.~Sokatchev,
``Harmonic superspace,''
Cambridge University Press, 2007,
% ISBN 978-0-511-53510-9, 978-0-521-02042-8, 978-0-521-80164-5, 978-0-511-03236-3
doi:10.1017/CBO9780511535109.


\bibitem{Karlhede:1984vr}
A.~Karlhede, U.~Lindstrom and M.~Rocek,
``Selfinteracting Tensor Multiplets in $N=2$ Superspace,''
Phys. Lett. B \textbf{147} (1984), 297-300.
%doi:10.1016/0370-2693(84)90120-5


\bibitem{Lindstrom:1987ks}
U.~Lindstrom and M.~Rocek,
``New Hyperkahler Metrics and New Supermultiplets,''
Commun. Math. Phys. \textbf{115} (1988), 21.
%doi:10.1007/BF01238851


\bibitem{Lindstrom:1989ne}
U.~Lindstrom and M.~Rocek,
``$N=2$ Superyang-mills Theory in Projective Superspace,''
Commun. Math. Phys. \textbf{128} (1990), 191.
%doi:10.1007/BF02097052

\bibitem{Lindstrom:2009afn}
U.~Lindstrom and M.~Rocek,
``Properties of hyperkahler manifolds and their twistor spaces,''
Commun. Math. Phys. \textbf{293} (2010), 257-278
%doi:10.1007/s00220-009-0923-0
[arXiv:0807.1366 [hep-th]].




\bibitem{K06} 
  S.~M.~Kuzenko,
  ``On compactified harmonic/projective superspace, 5D superconformal theories, 
  and all that,'' Nucl.\ Phys.\ B {\bf 745}, 176 (2006)
  [hep-th/0601177].
  
  
\bibitem{KT-M_5D2}
S.~M.~Kuzenko and G.~Tartaglino-Mazzucchelli,
  ``Five-dimensional superfield supergravity,''
 Phys.\ Lett.\  B {\bf 661}, 42 (2008)
  [arXiv:0710.3440 [hep-th]];


\bibitem{KT-M_5D3} 
S.~M.~Kuzenko and G.~Tartaglino-Mazzucchelli,
  ``5D supergravity and projective superspace,''
JHEP {\bf 0802}, 004 (2008)
  [arXiv:0712.3102 [hep-th]].

   





\bibitem{Kuzenko:2008ep}
S.~M. Kuzenko, U.~Lindstrom, M.~Rocek, and G.~Tartaglino-Mazzucchelli,
``4D N = 2 Supergravity and Projective Superspace,''
JHEP \textbf{09}, 051 (2008)
[arXiv:0805.4683 [hep-th]].



\bibitem{Kuzenko:2009zu}
S.~M.~Kuzenko, U.~Lindstrom, M.~Rocek and G.~Tartaglino-Mazzucchelli,
``On conformal supergravity and projective superspace,''
JHEP \textbf{08}, 023 (2009)
%doi:10.1088/1126-6708/2009/08/023
[arXiv:0905.0063 [hep-th]].


\bibitem{Butter:2014gha}
D.~Butter,
``New approach to curved projective superspace,''
Phys. Rev. D \textbf{92}, no.8, 085004 (2015)
%doi:10.1103/PhysRevD.92.085004
[arXiv:1406.6235 [hep-th]].


\bibitem{Butter:2014xua}
D.~Butter,
``Projective multiplets and hyperk\"ahler cones in conformal supergravity,''
JHEP \textbf{06}, 161 (2015)
%doi:10.1007/JHEP06(2015)161
[arXiv:1410.3604 [hep-th]].

\bibitem{Butter:2015nza}
D.~Butter,
``On conformal supergravity and harmonic superspace,''
JHEP \textbf{03}, 107 (2016)
%doi:10.1007/JHEP03(2016)107
[arXiv:1508.07718 [hep-th]].























% \bibitem{BNOPT-M18}
% D.~Butter, J.~Novak, M.~Ozkan, Yi~Pang and G.~Tartaglino-Mazzucchelli,
% ``Curvature squared invariants in six-dimensional
% $\cN = (1, 0)$ supergravity,''
% JHEP  13 (2019)
% %doi:10.1007/JHEP02(2015)111
% [arXiv::1808.00459 [hep-th]].




\bibitem{BSS1}
  E.~Bergshoeff, A.~Salam and E.~Sezgin,
  ``A supersymmetric $R^2$-action in six dimensions and torsion,''
  Phys.\ Lett.\ B {\bf 173}, 73 (1986).


\bibitem{LopesCardoso:1998tkj}
G.~Lopes Cardoso, B.~de Wit and T.~Mohaupt,
``Corrections to macroscopic supersymmetric black hole entropy,''
Phys. Lett. B \textbf{451}, 309-316 (1999)
%doi:10.1016/S0370-2693(99)00227-0
[arXiv:hep-th/9812082 [hep-th]].


\bibitem{Mohaupt:2000mj}
T.~Mohaupt,
``Black hole entropy, special geometry and strings,''
Fortsch. Phys. \textbf{49}, 3-161 (2001)
%doi:10.1002/1521-3978(200102)49:1/3\ensuremath{<}3::AID-PROP3\ensuremath{>}3.0.CO;2-\#
[arXiv:hep-th/0007195 [hep-th]].


\bibitem{Hanaki:2006pj}
K.~Hanaki, K.~Ohashi and Y.~Tachikawa,
``Supersymmetric Completion of an R**2 term in Five-dimensional Supergravity,''
Prog. Theor. Phys. \textbf{117}, 533 (2007)
%doi:10.1143/PTP.117.533
[arXiv:hep-th/0611329 [hep-th]].



\bibitem{Butter:2013lta}
 D.~Butter, B.~de Wit, S.~M.~Kuzenko and I.~Lodato,
  ``New higher-derivative invariants in N=2 supergravity and the Gauss-Bonnet term,''
  JHEP {\bf 1312} (2013) 062
  %doi:10.1007/JHEP12(2013)062
  [arXiv:1307.6546 [hep-th]].
  

\bibitem{Kuzenko:2013vha}
S.~M.~Kuzenko, J.~Novak and G.~Tartaglino-Mazzucchelli,
``N=6 superconformal gravity in three dimensions from superspace,''
JHEP \textbf{01}, 121 (2014)
%doi:10.1007/JHEP01(2014)121
[arXiv:1308.5552 [hep-th]].


\bibitem{OP131}
 M.~Ozkan and Y.~Pang,
``Supersymmetric Completion of Gauss-Bonnet Combination in Five Dimensions,''
  JHEP {\bf 1303} (2013) 158
   Erratum: [JHEP {\bf 1307} (2013) 152]
%  doi:10.1007/JHEP03(2013)158, 10.1007/JHEP07(2013)152
  [arXiv:1301.6622 [hep-th]].


\bibitem{OP132}
 M.~Ozkan and Y.~Pang,
  ``All off-shell $R^{2}$ invariants in five dimensional $\mathcal{N} =$ 2 supergravity,''
  JHEP {\bf 1308} (2013) 042
  %doi:10.1007/JHEP08(2013)042
  [arXiv:1306.1540, arXiv:1306.1540 [hep-th]].

 



\bibitem{Kuzenko:2015jxa}
S.~M.~Kuzenko and J.~Novak,
``On curvature squared terms in N=2 supergravity,''
Phys. Rev. D \textbf{92}, no.8, 085033 (2015)
%doi:10.1103/PhysRevD.92.085033
[arXiv:1507.04922 [hep-th]].




\bibitem{Butter:2016mtk}
D.~Butter, F.~Ciceri, B.~de Wit and B.~Sahoo,
``Construction of all N=4 conformal supergravities,''
Phys. Rev. Lett. \textbf{118}, no.8, 081602 (2017)
%doi:10.1103/PhysRevLett.118.081602
[arXiv:1609.09083 [hep-th]].



\bibitem{NOPT-M17}
  J.~Novak, M.~Ozkan, Y.~Pang and G.~Tartaglino-Mazzucchelli,
  ``Gauss-Bonnet supergravity in six dimensions,''
  Phys.\ Rev.\ Lett.\  {\bf 119}, no. 11, 111602 (2017).
%  doi:10.1103/PhysRevLett.119.111602
  [arXiv:1706.09330 [hep-th]].




\bibitem{Butter:2019edc}
D.~Butter, F.~Ciceri and B.~Sahoo,
``$N=4$ conformal supergravity: the complete actions,''
JHEP \textbf{01}, 029 (2020)
%doi:10.1007/JHEP01(2020)029
[arXiv:1910.11874 [hep-th]].


\bibitem{Hegde:2019ioy}
S.~Hegde and B.~Sahoo,
``New higher derivative action for tensor multiplet in $ \mathcal{N} $ = 2 conformal supergravity in four dimensions,''
JHEP \textbf{01}, 070 (2020)
%doi:10.1007/JHEP01(2020)070
[arXiv:1911.09585 [hep-th]].


\bibitem{Mishra:2020jlc}
M.~Mishra and B.~Sahoo,
``Curvature squared action in four dimensional $N = 2$ supergravity using the dilaton Weyl multiplet,''
JHEP \textbf{04}, 027 (2021)
%doi:10.1007/JHEP04(2021)027
[arXiv:2012.03760 [hep-th]].


\bibitem{Butter:2013rba}
D.~Butter, S.~M.~Kuzenko, J.~Novak and G.~Tartaglino-Mazzucchelli,
``Conformal supergravity in three dimensions: Off-shell actions,''
JHEP \textbf{10}, 073 (2013)
%doi:10.1007/JHEP10(2013)073
[arXiv:1306.1205 [hep-th]].



\bibitem{Bobev:2020egg}
N.~Bobev, A.~M.~Charles, K.~Hristov and V.~Reys,
``The Unreasonable Effectiveness of Higher-Derivative Supergravity in AdS$_4$ Holography,''
Phys. Rev. Lett. \textbf{125}, no.13, 131601 (2020)
%doi:10.1103/PhysRevLett.125.131601
[arXiv:2006.09390 [hep-th]].


\bibitem{Bobev:2021oku}
N.~Bobev, A.~M.~Charles, K.~Hristov and V.~Reys,
``Higher-derivative supergravity, AdS$_{4}$ holography, and black holes,''
JHEP \textbf{08}, 173 (2021)
%doi:10.1007/JHEP08(2021)173
[arXiv:2106.04581 [hep-th]].


\bibitem{Bobev:2021qxx}
N.~Bobev, K.~Hristov and V.~Reys,
``AdS$_5$ Holography and Higher-Derivative Supergravity,''
[arXiv:2112.06961 [hep-th]].

\bibitem{Hristov:2022lcw}
K.~Hristov,
``ABJM at finite $N$ via 4d supergravity,''
[arXiv:2204.02992 [hep-th]].


\bibitem{Bobev:2022bjm}
N.~Bobev, V.~Dimitrov, V.~Reys and A.~Vekemans,
``Higher-Derivative Corrections and AdS$_5$ Black Holes,''
[arXiv:2207.10671 [hep-th]].

\bibitem{Cassani:2022lrk}
D.~Cassani, A.~Ruip\'erez and E.~Turetta,
``Corrections to AdS$_5$ Black Hole Thermodynamics from Higher-Derivative Supergravity,''
[arXiv:2208.01007 [hep-th]].




\bibitem{Pestun:2016zxk}
V.~Pestun, M.~Zabzine, F.~Benini, T.~Dimofte, T.~T.~Dumitrescu, K.~Hosomichi, S.~Kim, K.~Lee, B.~Le Floch and M.~Marino, \textit{et al.}
``Localization techniques in quantum field theories,''
J. Phys. A \textbf{50}, no.44, 440301 (2017)
%doi:10.1088/1751-8121/aa63c1
[arXiv:1608.02952 [hep-th]].



\bibitem{Muller_hyper:1986ts}
M.~M\"uller,
``Minimal N=2 Off-Shell Supergravity,''
Phys.\ Lett.\ B {\bf 172}, 353 (1986).






\bibitem{KL} 
  S.~M.~Kuzenko and W.~D.~Linch III,
  ``On five-dimensional superspaces,''
  JHEP {\bf 0602}, 038 (2006)
  [hep-th/0507176].
  
\bibitem{KT-M5D1}
S.~M. Kuzenko and G. Tartaglino-Mazzucchelli,
``Five-dimensional N=1 AdS superspace:
Geometry,  off-shell multiplets and dynamics,''
Nucl. Phys. B {\bf 785}, 34 (2007),
0704.1185 [hep-th].


\bibitem{Howe5Dsugra}
  P.~S.~Howe,
 ``Off-shell N=2 and N=4 supergravity in five-dimensions,''
in {\it Quantum Structure of Space and Time}, 
M. J. Duff and C. J. Isham (Eds.), Cambridge University Press, Cambridge, 
1982, pp. 239--253.

\bibitem{HL}
P.~S.~Howe and U.~Lindstr\"om,
``The supercurrent in five dimensions,'' Phys.\ Lett.\ B {\bf 103}, 422 (1981).








\bibitem{Fayet:1975yi}
P.~Fayet,
``Fermi-Bose Hypersymmetry,''
Nucl. Phys. B \textbf{113}, 135 (1976).
%doi:10.1016/0550-3213(76)90458-2

\bibitem{FS2}
M.~F.~Sohnius,
``Supersymmetry and Central Charges,''
Nucl. Phys. B \textbf{138}, 109-121 (1978).


 \bibitem{N=2tensor}
J.~Wess, ``Supersymmetry and internal symmetry,''
Acta Phys.\ Austriaca {\bf 41}, 409 (1975).


\bibitem{Siegel:1978yi}
  W.~Siegel,
  ``Superfields In Higher Dimensional Space-time,''
  Phys.\ Lett.\ B {\bf 80} (1979) 220.


\bibitem{Siegel80}
W.~Siegel, ``Off-shell central charges,''
Nucl.\ Phys.\ B {\bf 173}, 51 (1980).


\bibitem{SSW}
  M.~F.~Sohnius, K.~S.~Stelle and P.~C.~West,
  ``Representations of extended supersymmetry,''
in {\it Superspace and Supergravity}, S. W. Hawking and
M. Ro\v{c}ek (Eds.) Cambridge Unieversity Press, Cambridge, 1981, p. 283.


\bibitem{deWit:1980gt}
B.~de Wit, J.~W.~van Holten and A.~Van Proeyen,
``Central Charges and Conformal Supergravity,''
Phys. Lett. B \textbf{95}, 51-55 (1980).

\bibitem{deWit:1980lyi}
B.~de Wit, J.~W.~van Holten and A.~Van Proeyen,
``Structure of N=2 Supergravity,''
Nucl. Phys. B \textbf{184}, 77 (1981)
[erratum: Nucl. Phys. B \textbf{222}, 516 (1983)].

\bibitem{deWit:1982na}
B.~de Wit, R.~Philippe and A.~Van Proeyen,
``The Improved Tensor Multiplet in $N=2$ Supergravity,''
Nucl. Phys. B \textbf{219}, 143-166 (1983).
%doi:10.1016/0550-3213(83)90432-7


\bibitem{deWit:1983xhu}
B.~de Wit, P.~G.~Lauwers, R.~Philippe, S.~Q.~Su and A.~Van Proeyen,
``Gauge and Matter Fields Coupled to N=2 Supergravity,''
Phys. Lett. B \textbf{134}, 37-43 (1984).

\bibitem{KLR}
A. Karlhede, U. Lindstr\"om and M. Ro\v cek,
``Self-interacting tensor multiplets in N = 2 superspace,''
Phys.\ Lett.\ B {\bf 147}, 297 (1984). 


\bibitem{LR3}
U.~Lindstr\"om and M.~Ro\v{c}ek,
``New hyperk\"ahler  metrics  and new supermultiplets,''
  Commun.\ Math.\ Phys.\  {\bf 115}, 21 (1988).








\bibitem{Howe:1983fr}
P.~S.~Howe, G.~Sierra and P.~K.~Townsend,
``Supersymmetry in Six-Dimensions,''
Nucl. Phys. B \textbf{221}, 331-348 (1983).


\bibitem{Koller:1982cs}
J.~Koller,
``A Six-dimensional superspace approach to extended superfields,"
Nucl. Phys. B \textbf{222}, 319-337 (1983).


\bibitem{Kugo:1982bn}
T.~Kugo and P.~K.~Townsend,
``Supersymmetry and the Division Algebras,''
Nucl. Phys. B \textbf{221}, 357-380 (1983).

\bibitem{oweAR}
P.~S.~Howe, K.~S.~Stelle and P.~C.~West,
``N=1 d = 6 Harmonic Superspace,''
Class. Quant. Grav. \textbf{2}, 815 (1985).

\bibitem{DragonNV}
N.~Dragon, E.~Ivanov, S.~Kuzenko, E.~Sokatchev and U.~Theis,
``N=2 rigid supersymmetry with gauged central charge,''
Nucl. Phys. B \textbf{538}, 411-450 (1999)
[arXiv:hep-th/9805152 [hep-th]].


\bibitem{GrundbergLindstrom_6D}
J.~Grundberg and U.~Lindstrom,
``Actions for Linear Multiplets in Six-dimensions,''
Class. Quant. Grav. \textbf{2}, L33 (1985).

\bibitem{GatesPenatiTartaglino_6D}
S.~J.~Gates, Jr., S.~Penati and G.~Tartaglino-Mazzucchelli,
``6D supersymmetry, projective superspace and 4D, N=1 superfields,''
JHEP \textbf{05}, 051 (2006)
%doi:10.1088/1126-6708/2006/05/051
[arXiv:hep-th/0508187 [hep-th]];
``6D Supersymmetric Nonlinear Sigma-Models in 4D, N=1 Superspace,''
JHEP \textbf{09}, 006 (2006)
[arXiv:hep-th/0604042 [hep-th]].

%\bibitem{GatesPenatiTartaglino_6D}
 % S.~J.~Gates, Jr., S.~Penati, G.~Tartaglino-Mazzucchelli,
%  ``6D supersymmetry, projective superspace and 4D, N=1 superfields,''
%  JHEP {\bf 0605 } (2006)  051.
 % [hep-th/0508187];
%   ``6D Supersymmetric Nonlinear Sigma-Models in 4D, N=1 Superspace,''
%  JHEP {\bf 0609 } (2006)  006.
 % [hep-th/0604042].


\bibitem{BreitenlohnerRQ}
P.~Breitenlohner and A.~Kabelschacht,
``The Auxiliary Fields of $N=2$ Extended Supergravity in 5 and 6 Space-time Dimensions,''
Nucl. Phys. B \textbf{148}, 96-106 (1979).

\bibitem{GatesQV}
S.~J.~Gates, Jr.,
``A Comment on Superspace Bianchi Identities and Six-dimensional Space-time,''
Phys. Lett. B \textbf{84}, 205 (1979).

\bibitem{GatesJV}
S.~J.~Gates, Jr. and W.~Siegel,
``Understanding Constraints in Superspace Formulations of Supergravity,''
Nucl. Phys. B \textbf{163}, 519-545 (1980).

\bibitem{SmithWAA}
A.~W.~Smith,
``N=1, D = 6 supergravity theory,''
Class. Quant. Grav. \textbf{2}, 167-177 (1985).


\bibitem{AwadaER}
M.~Awada, P.~K.~Townsend and G.~Sierra,
``Six-dimensional Simple and Extended Chiral Supergravity in Superspace,''
Class. Quant. Grav. \textbf{2}, L85 (1985).

\bibitem{BergshoeffSU}
E.~Bergshoeff, E.~Sezgin and P.~K.~Townsend,
``Superstring Actions in $D=3$, 4, 6, 10 Curved Superspace,''
Phys. Lett. B \textbf{169}, 191-196 (1986).

\bibitem{BergshoeffRB}
E.~Bergshoeff and M.~Rakowski,
``An Off-shell Superspace R(2) Action in Six-dimensions,''
Phys. Lett. B \textbf{191}, 399-403 (1987).




\end{thebibliography}
\end{document}